\documentclass[12pt,a4paper]{article}

\usepackage[utf8]{inputenc} \usepackage[T1]{fontenc}    \usepackage{newtxtext} \usepackage[margin=1in]{geometry}
\usepackage{setspace}
\usepackage[numbers,sort&compress,super]{natbib}

\usepackage{amsmath,amsfonts,bm}

\def\eqref#1{equation~\ref{#1}}

\def\1{\bm{1}}

\DeclareMathAlphabet{\mathsfit}{\encodingdefault}{\sfdefault}{m}{sl}
\SetMathAlphabet{\mathsfit}{bold}{\encodingdefault}{\sfdefault}{bx}{n}

\usepackage[hidelinks]{hyperref}
\usepackage{url}

\usepackage{booktabs}       \usepackage{amsfonts}       \usepackage{nicefrac}       \usepackage{microtype}      \usepackage{xcolor}         \usepackage{colortbl}       \usepackage{algorithm}
\usepackage{algorithmic}
\usepackage{amsmath}
\usepackage{amssymb}  \usepackage{graphicx}  
\usepackage{subcaption}
\usepackage{longtable}
\usepackage{wrapfig}
\usepackage{rotating}
\usepackage{makecell}
\usepackage{threeparttable}  \usepackage{tocloft}  \usepackage{array} 
\usepackage{subfiles}
\usepackage{etoc}
\usepackage{adjustbox}
\usepackage{listings}
\usepackage{authblk}
\usepackage{float}
\usepackage{siunitx}
\usepackage{multirow}
\usepackage{bm}
\usepackage{needspace}

\renewenvironment{abstract}{\par\noindent\bfseries}{\par\normalfont}
\date{}

\title{A Comprehensive Benchmark of Computational Models for Antibody--Antigen Binding Affinity}

\author[1]{Xinyan Zhao}
\author[1]{Yi-Ching Tang}
\author[1]{Victor J Cantu}
\author[2]{Junseok Lee\thanks{Work done while at University of Texas Health Science Center.}}
\author[1]{Adam E Stogsdill}
\author[1]{Ibraheem M Hamdi}
\author[3]{Ashwin Kumar Ramesh}
\author[3]{Zhiqiang An}
\author[1]{Xiaoqian Jiang}
\author[1]{Yejin Kim}
\affil[1]{McWilliams School of Biomedical Informatics, UTHealth Houston}
\affil[2]{Department of Industrial and Systems Engineering, Korea Advanced Institute of Science and Technology}
\affil[3]{Texas Therapeutics Institute, Brown Foundation Institute of Molecular Medicine, UTHealth Houston}

\begin{document}

\maketitle

\begin{abstract}
Computational antibody design requires benchmarks that measure functional binding rather than similarity to natural sequences. Here we present AbBiBench, which treats the antibody--antigen complex as the evaluation unit and harmonizes 216,560 experimental measurements spanning 13 antibodies and 9 antigens. We compare 20 computational approaches without task-specific training or fine-tuning.  The analyses show that global structure-conditioned inverse folding yields the strongest and most consistent affinity rankings, whereas antibody-specific fine-tuning does not consistently improve performance. Across models, rankings are more useful for prioritizing likely affinity gains than for excluding weak variants. In a subsequent case study, multi-objective evaluation of model-generated F045-092 CDR-H3 variants for H1N1 binding revealed a trade-off between predicted binding potential and evolutionary sequence plausibility; all 21 selected candidates produced H1-reactive ELISA signals. AbBiBench provides an open framework for developing and evaluating function-aware antibody design methods.
\end{abstract}

\begin{figure}[p]
  \centering
  \includegraphics[width=\textwidth]{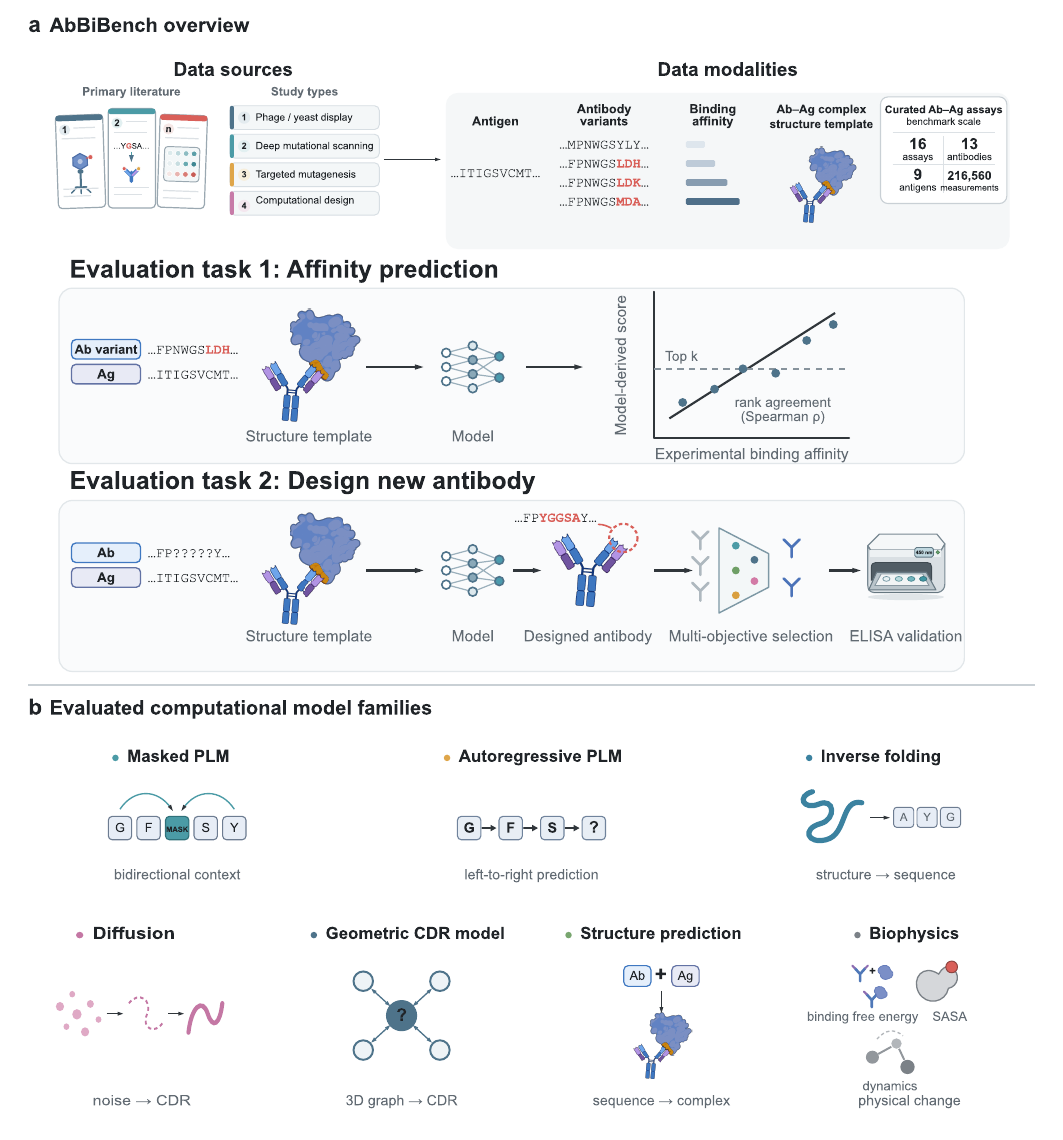}
  \caption{Overview of AbBiBench. \textbf{a}, Published studies are curated into antibody--antigen sequences, complex structures and binding measurements for zero-shot affinity prediction and antibody design, followed by multi-objective selection and ELISA validation. \textbf{b}, Evaluated protein language, inverse-folding, generative, structure-prediction and biophysical model families.}
  \label{fig:overview}
\end{figure}

Antibodies are critical components of the adaptive immune system, functioning primarily by recognizing and binding specifically to antigens such as pathogens or aberrant cells. This specific recognition is facilitated through complementary regions: the antibody's paratope and the antigen's epitope. Improving antibody–antigen affinity boosts therapeutic potency and is critical for drug discovery. Developing a therapeutic monoclonal antibody depends on multiple factors, including expression, stability, immunogenicity, aggregation, and binding affinity \mbox{\cite{chungyoun2024flab,jain2017biophysical}}. Among these, binding affinity is the most critical determinant of therapeutic efficacy, as it directly influences antibody potency. Therefore, increasing binding affinity between antibody and antigen has been a crucial process in therapeutic antibody development. 

Conventional antibody discovery approaches, for instance, phage display technology \cite{marks1992passing, mccafferty1990phage, smith1985filamentous} and animal immunization \cite{green1994antigen, kohler1975continuous}, employ iterative cycles of mutation and selection to progressively improve binding affinity. These methods have significantly advanced therapeutic antibody discovery but face challenges due to the initially limited diversity of libraries in phage display or the narrow naive B cell repertoire available in animal immunization models, restricting the comprehensive exploration of potential high-affinity antibody variants (Fig. \ref{fig:lead_antibodies_a}).  Generative-model-based antibody design approaches complement these experimental techniques by efficiently navigating the vast search space (Fig. \ref{fig:lead_antibodies_b}) and proposing high-affinity antibody variants that may not be readily accessible through experimental approaches alone. Besides, experimental affinity measurements such as surface plasmon resonance (SPR) and biolayer interferometry (BLI) remain the gold standard, but they are  difficult to scale across large antibody libraries \cite{abdiche2008determining} .

Recent advances in machine learning, especially with protein language models and structure-based generative models \cite{dauparas2022robust, hayes2025simulating, hoie2024antifold, hsu2022learning, kong2022conditional, kong2023end, li2024prosst, luo2022antigen, malherbe2024igblend, martinkus2023abdiffuser, ruffolo2021deciphering, shanker2024unsupervised, susaprot, watson2023novo, wu2023hierarchical}, have shown promise in antibody design. These models learn statistical patterns from existing amino acid sequences and protein structures, and their outputs are commonly used to assess the sequence or structural plausibility of new designs.  In addition, interface-confidence scores from structure-prediction models, including ipTM and ipSAE, have been explored as auxiliary ranking features for binder prioritization, although they are not calibrated estimators of binding affinity \cite{evansprotein,jumper2021highly,dunbrack2025ipsae}. Structure-prediction signals can even be optimized as loss functions to guide sequence design \cite{bindercraft}.  However, sequence or structural plausibility does not necessarily imply improved antibody--antigen binding affinity. In general protein design, comparing to the closest natural variant is a reasonable validation strategy because protein mutations are driven by strong evolutionary pressure. By contrast, antibody generation involves stochastic recombination and hypermutation, yielding extreme diversity. Even antibodies to the same antigen often show little sequence similarity unless clonally related. As such, evaluating designed antibodies by how closely they resemble naturally occurring ones overlooks the fundamental biology of antibody generation. This calls for new evaluation  benchmarks that better reflect the functional goals of antibody engineering, rather than assumptions borrowed from general protein design.  

From a structural biology perspective, binding affinity is determined not just by the antibody sequence, but by the quality of the interface it forms with the antigen. High-affinity binding typically arises from antibody-antigen (Ab-Ag) complexes that exhibit structural integrity \cite{shanker2024unsupervised} -- meaning they are stable, well-packed, and maintain favorable conformations with minimal strain. Structural integrity ensures optimal shape and chemical complementarity at the binding interface. Antibodies that form such stable complexes with their targets resemble naturally occurring Ab-Ag complexes such as those collected in structural databases like SAbDab \cite{dunbar2014sabdab}. Recent machine learning models \cite{dauparas2022robust, hsu2022learning, li2024prosst, luo2022antigen, malherbe2024igblend, susaprot, wu2023hierarchical} have shown success in learning Ab‑Ag complex sequence-structure patterns, enabling us to gauge whether a designed Ab-Ag complex lies within the high‑probability manifold of structurally stable, high‑affinity complexes. Therefore, incorporating the antigen into evaluation provides a more biologically grounded and functionally relevant assessment of antibody design. 

General protein benchmarks, including ProteinGym \cite{notin2023proteingym} and BindingGYM \cite{binding_gym}, provide broad mutational or protein--protein interaction coverage but are not organized around antibody binding-affinity optimization and design.  Unlike general protein benchmarks, antibody-focused benchmarks  should treat the full antibody--antigen complex as the evaluation unit and compare as broad a range of computational approaches as possible. Existing resources cover only parts of this scope: FLAb evaluates multiple fitness properties without treating the antigen complex as the scoring unit \cite{chungyoun2024flab}; AbRank emphasizes supervised pairwise ranking \cite{liu2025abrank}; AbAgym provides structurally supported mutational data, but its large-scale collection primarily comprises antigen-side mutations and uses heterogeneous DMS-derived proxies rather than equilibrium dissociation constants ($K_d$) or binding free-energy changes ($\Delta\Delta G$) \cite{cia2025abagym}; and AbDesign provides a structurally annotated, experimentally consistent antibody-mutant test set, but contains only 658 single-point mutations and reports relative ELISA binding-retainment ratios rather than direct affinity measurements \cite{janusz2025abdesign}. Therefore, our community currently lacks a comprehensive benchmark centered on antibody--antigen binding-affinity prediction and generative design.

To address the limitations discussed above, we introduce AbBiBench (\textbf{A}nti\textbf{b}ody \textbf{Bi}nding \textbf{Bench}marking),   a benchmark for evaluating computational  models of antibody binding affinity without task-specific training or fine-tuning. Rather than assessing antibodies in isolation \cite{chungyoun2024flab}, we consider the full Ab-Ag complex as the fundamental unit for evaluation. By uniting standardized experimental binding affinity datasets \footnote{\url{https://huggingface.co/datasets/AbBibench/Antibody_Binding_Benchmark_Dataset}} with a robust \textit{in silico} evaluation pipeline \footnote{\url{https://github.com/MSBMI-SAFE/AbBiBench}}, AbBiBench provides a rigorous, biologically grounded environment to accelerate computational methods that lead to clinically and diagnostically impactful discoveries.

Across 16 assays, AbBiBench standardizes 216,560 measurements spanning  13 antibodies against nine antigen targets---influenza A hemagglutinins from H1N1, H3N2, and H9N2 strains; SARS-CoV-2 Spike HR2; VEGF; hen-egg-white lysozyme; HER2; integrin-$\alpha$1; and Ang2---and compares 20 computational approaches from seven families: masked protein language models, autoregressive protein language models, inverse-folding models, diffusion-based models, geometric CDR-imputation models, structure-prediction confidence models, and biophysics-based methods. AbBiBench uses common affinity-ranking metrics and connects retrospective benchmarking with a prospective F045-092/H1N1 ELISA case study. The analyses show that global structure-conditioned inverse folding yields the strongest and most consistent affinity rankings, whereas antibody-specific fine-tuning does not consistently improve performance. Across models, rankings are more useful for prioritizing likely affinity gains than for excluding weak variants. Together, AbBiBench provides a standardized data resource and unified evaluation framework for antibody--antigen affinity prediction, and the resulting cross-model evidence offers practical guidance for selecting, evaluating and developing function-aware antibody design methods.

\section{Results}
\label{ressults}
\subsection{AbBiBench spans diverse antibody--antigen affinity landscapes}

\begin{figure}[p]
  \centering
  \includegraphics[width=\textwidth]{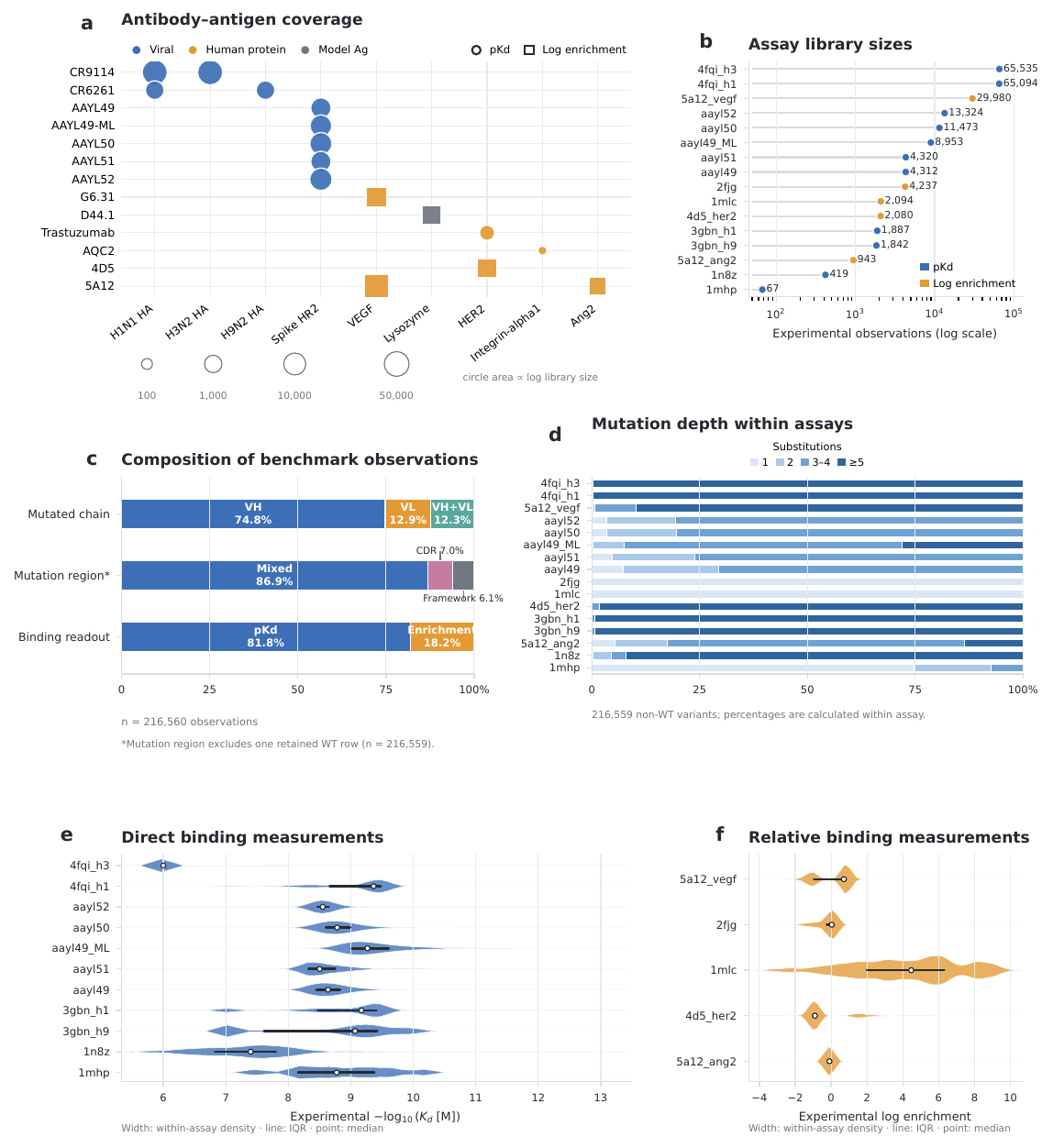}
  \caption{Composition and curation of AbBiBench. \textbf{a}, Coverage of 13 antibodies and nine antigen targets. Marker area scales with the logarithm of the assay-library size, marker color denotes antigen class and marker shape distinguishes direct $K_d$ from enrichment readouts. \textbf{b}, Number of experimental observations per assay on a logarithmic scale. \textbf{c}, Composition by reported mutated chain, Chothia-defined mutation region and experimental readout. Mutation-region percentages exclude the single retained wild-type record (216,559 non-wild-type variants). \textbf{d}, Within-assay distribution of mutation depth, calculated as the total heavy- plus light-chain Hamming distance from the corresponding wild-type pair. \textbf{e}, Distributions of direct binding measurements after conversion to $-\log_{10}(K_d\,[\mathrm{M}])$. \textbf{f}, Distributions of log-enrichment measurements, which are relative high-throughput binding proxies rather than thermodynamic affinity constants. In \textbf{e} and \textbf{f}, density width is normalized within each assay; horizontal lines and points show interquartile ranges and medians, respectively. Counts follow the 16-assay manifest in Table~\ref{tab:compact_table}.}
  \label{fig:dataset_overview}
\end{figure}

We assembled AbBiBench from eight public studies~\cite{phillips2021binding,engelhart2022dataset,li2023machine,koenig2017mutational,warszawski2020correction,Shanehsazzadeh2023.01.08.523187,Clark2006Affinity,minot2024}, yielding 216,560 experimental observations across 16 assays, 13 antibodies and nine antigen targets (Fig.~\ref{fig:dataset_overview}a). The target panel covers influenza A hemagglutinins from H1N1, H3N2 and H9N2 strains; the SARS-CoV-2 Spike HR2 region; VEGF; Ang2; HER2; integrin-$\alpha$1; and hen-egg-white lysozyme. Assay libraries span nearly three orders of magnitude, from 67 AQC2 variants measured against integrin-$\alpha$1 to 65,535 CR9114 variants measured against H3N2 hemagglutinin (Fig.~\ref{fig:dataset_overview}b). Coverage is therefore broad but unbalanced; we therefore assign equal weight to assays rather than allowing the largest libraries to dominate.

The benchmark includes 11 direct-$K_d$ assays (177,226 observations; 81.8\%) and five enrichment assays (39,334 observations; 18.2\%) (Fig.~\ref{fig:dataset_overview}c,e,f). By the reported library construction, 161,947 observations (74.8\%) contain heavy-chain mutations, 28,033 (12.9\%) contain light-chain mutations and 26,580 (12.3\%) contain mutations in both chains. Sequence comparison with the matched wild types further identified 216,559 non-wild-type variants: 7,859 (3.6\%) carried one substitution, 6,744 (3.1\%) carried two, 35,934 (16.6\%) carried three or four and 166,022 (76.7\%) carried at least five (Fig.~\ref{fig:dataset_overview}d). This aggregate depth distribution is driven by the large combinatorial influenza landscapes; the assay-level profiles range from almost entirely single-substitution deep-mutational scans to libraries dominated by higher-order variants. Mutations also span complementarity-determining and framework regions, with mixed-region variants accounting for 86.9\% of the non-wild-type collection (Fig.~\ref{fig:dataset_overview}c).

Direct-$K_d$ and enrichment readouts were kept separate because enrichment is a relative selection measure rather than a thermodynamic affinity constant, and all model comparisons were therefore performed within assay. Experimentally determined complexes were used as structural templates where available. For the five SARS-CoV-2 assays (AAYL49, AAYL49\_ML, AAYL50, AAYL51 and AAYL52), no experimentally resolved antibody--antigen complex was available; structure-conditioned results for these assays used AlphaFold 3-predicted templates and should therefore be interpreted with this additional source of structural uncertainty (Methods~\ref{sec:datasets}; Supplement~\ref{sup:data_details}).

\subsection{ Benchmarking Computational Models against Experimental Antibody--Antigen Binding Affinity }
We compared computational protein models spanning seven methodological families---masked protein language models (ESM2, ESM3, SaProt, ProSST, AntiBERTy, CurrAb), autoregressive protein language models (ProGen2, ProtGPT-2), inverse-folding models (ProteinMPNN, ESM-IF, AntiFold), diffusion-based generative models (DiffAb), geometric CDR-imputation models (MEAN, dyMEAN), structure-prediction confidence scores (AlphaFold 3, Boltz-2, ESMFold2), and biophysics-based methods (FoldX binding free energy, FreeSASA epitope SASA, ESMDance)---against experimentally measured antibody--antigen binding affinities from 16 assays. Structure-prediction models are evaluated primarily with ipSAE. Correlations obtained with iPTM and interface pLDDT (ipLDDT) as auxiliary proxies are reported only in Supplement~\ref{sup:sec:structure_prediction_correlation}.

\begin{figure}[!p]
  \centering
  \includegraphics[width=1\textwidth]{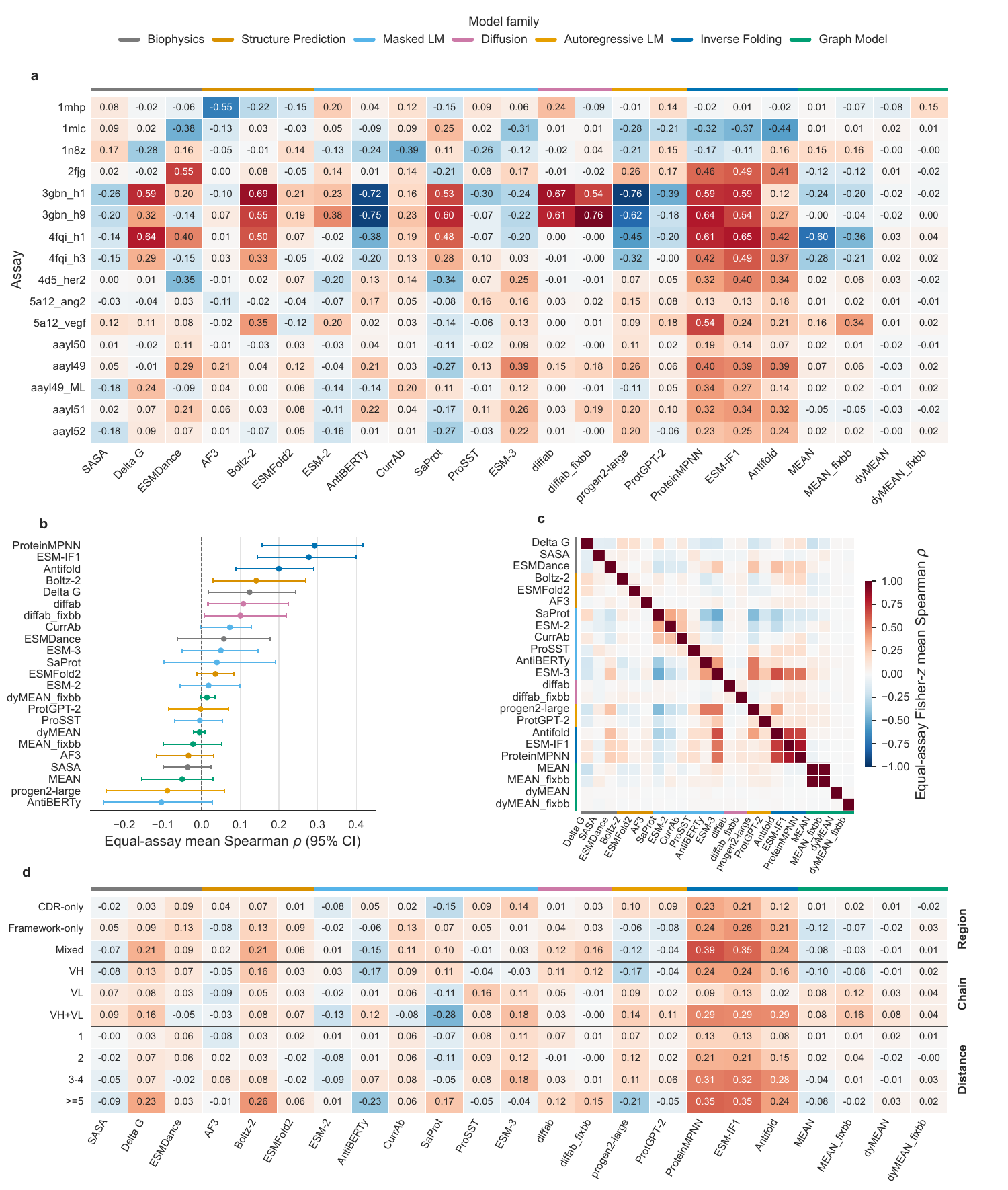}
  \caption{Zero-shot prediction of experimental binding affinity across assays and models. \textbf{a}, Spearman correlations between model scores and experimental binding affinities for each assay; blue and red denote negative and positive correlations, respectively. AlphaFold 3, Boltz-2 and ESMFold2 are scored by ipSAE on the same variants. \textbf{b}, Equal-assay mean Spearman $\rho$ with assay-bootstrap 95\% confidence intervals. \textbf{c}, Equal-assay Fisher-$z$ mean Spearman correlations between model-predicted variant rankings. \textbf{d}, Mutation-stratified equal-assay mean Spearman correlations within mutation-region, mutated-chain, and mutation-distance strata. Cells summarize assay strata with at least 20 variants after requiring at least 90\% safe-join and valid-score coverage; cross-column differences are descriptive because assay support differs. Model-family colors are used in \textbf{a}, \textbf{b}, and \textbf{c}.}
  \label{fig:correlation}
\end{figure}

\begin{figure}[H]
  \centering
  \includegraphics[width=0.87\textwidth]{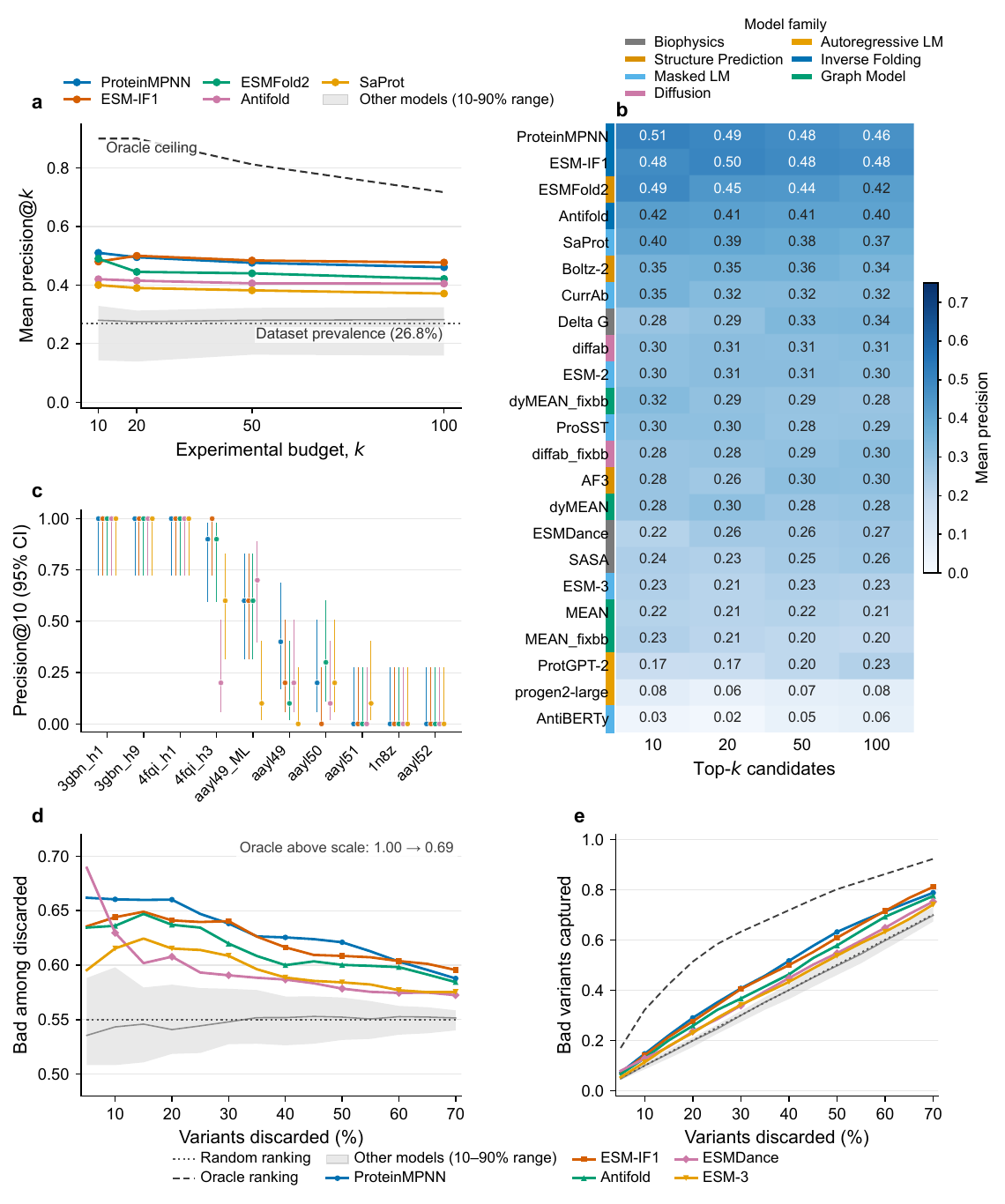}
  \caption{Candidate prioritization and negative filtering across experimental budgets. \textbf{a}, Equal-assay mean precision for the top $k=10,20,50,$ or $100$ variants. Colored lines show the five strongest methods averaged across budgets; the gray band and line show the 10th--90th percentile and median of the others. Dotted and dashed lines mark the equal-assay prevalence of variants with at least fivefold affinity improvement (26.8\%) and the finite-dataset oracle ceiling, respectively. \textbf{b}, Mean top-$k$ precision across the 10 direct-$K_d$ assays retained for this analysis; colored sidebars indicate model families. \textbf{c}, Assay-level Precision@10 for the five highlighted methods, with Wilson 95\% binomial confidence intervals. \textbf{d}, Fraction of discarded variants worse than wild type as discard budget increased from 5\% to 70\%. The displayed scale covers model curves; oracle values at 5\% and 70\% are reported. \textbf{e}, Fraction of all worse-than-wild-type variants captured at each budget. In \textbf{d} and \textbf{e}, bad variants have p$K_d$ below the assay-matched wild type; curves are equal-assay means across the eight evaluable direct-$K_d$ assays. Colored lines show the five strongest filters; the gray band and line summarize the others. Dotted lines denote random ranking, and the dashed line in \textbf{e} denotes the oracle. Boundary ties were averaged over random selection within tied score groups.}
  \label{fig:nfold}
\end{figure}

\paragraph{\textbf{Finding 1: Global structure modeling outperforms sequence-only or local structure modeling.}}
Across the 16 assays, inverse-folding models showed the strongest and most consistent zero-shot affinity prediction (Fig.~\ref{fig:correlation}a,b). ProteinMPNN achieved an equal-assay mean Spearman $\rho$ of 0.292 (bootstrap 95\% CI, 0.154--0.416), followed by ESM-IF1 at 0.278 (0.144--0.399). Across $k=10,20,50,$ and $100$, ProteinMPNN and ESM-IF1 were the two strongest methods by mean fivefold precision (Fig.~\ref{fig:nfold}). This high accuracy may stem from the model's broader structural scope. PLMs with local structure token have consistently performed best in general protein binding tasks \cite{li2024prosst, susaprot}, but the same did not hold for antibodies. Inverse-folding models, which encode the entire Ab–Ag complex as a single global representation, consistently outperform models that rely on local structural tokens—such as \texttt{SaProt}, \texttt{ProSST}, and \texttt{ESM-3}. 
Besides, purely autoregressive sequence models without structure like \texttt{ProGen2} and \texttt{ProtGPT2} likewise fail to achieve competitive accuracy on the affinity prediction task. 
We shuffled the chain order of autoregressive models, but it did not increase the accuracy either  (Table \ref{tab:autoregressive}, Supplement \ref{sup:sec:sec:chain_order}). 
From a structural biology perspective, the inverse folding strategy is effective because protein function is ultimately determined by its three-dimensional structure, which is encoded by the underlying sequence. Mutations that maintain or improve structural integrity have a higher potential to enhance functional properties \cite{shanker2024unsupervised}. On the other hand, by conditioning on the full antibody-antigen structure, inverse folding models can capture long-range residue interactions and contextual features at the binding interface—both of which are critical for affinity but often missed by models relying only on local sequence information \cite{orlandi2020antigen, wang2018local}. 
Paired analysis reinforced the family-level advantage: inverse-folding models exceeded masked language models by a mean $\Delta\rho$ of 0.245 (bootstrap 95\% CI, 0.127--0.346), whereas the remaining family contrasts and all three fixed-backbone contrasts had confidence intervals spanning zero. Model-ranking similarities further showed strong within-family agreement among ProteinMPNN, ESM-IF1, and Antifold, as well as near-identical rankings from MEAN and MEAN\_fixbb, while agreement across model families was generally weaker (Fig.~\ref{fig:correlation}c).
A mutation-stratified robustness analysis further showed that the inverse-folding advantage was not confined to one antibody region or chain (Fig.~\ref{fig:correlation}d; Supplement~\ref{sup:sec:hard_subsets}). ProteinMPNN and ESM-IF1 remained the two strongest methods for CDR-only, framework-only, and mixed-region variants and for VH and joint VH--VL variants. Performance on VL-only variants was less consistent across models. Correlations for the leading inverse-folding models were descriptively higher in the multi-mutation strata than among single mutants; however, because the available strata differed in assay composition and only two assays supported all four mutation-distance bins, we do not interpret mutation distance as an independently estimated causal difficulty effect.

We next evaluated candidate prioritization across experimental budgets (Fig.~\ref{fig:nfold}).
At $k=10$, ProteinMPNN and ESM-IF1 achieved mean precisions of 0.51 and 0.48, respectively, compared with the 0.268 random-ranking baseline and the 0.90 finite-dataset oracle ceiling. Their performance remained stable through $k=100$ (0.46 and 0.48, respectively).
Among non-inverse-folding methods, ESMFold2 scored by ipSAE achieved the strongest mean top-$k$ precision and ranked third overall across the four experimental budgets; SaProt was the next strongest non-inverse-folding method.

We additionally evaluated whether model rankings could support negative selection before experimentation by treating each model as a filter and discarding its lowest-scoring variants (Fig.~\ref{fig:nfold}d,e). At a 30\% discard budget, ProteinMPNN and ESM-IF1 achieved discard precisions of 0.638 and 0.640 and captured 0.408 and 0.405 of all experimentally worse-than-wild-type variants, respectively, compared with random-ranking expectations of 0.550 and 0.301.

\paragraph{\textbf{Finding 2: Antibody-specific finetuning can yield marginal gains or even degrade performance.}}
Interestingly, we also found that antibody-finetuned PLMs have varying impacts on zero-shot correlation to binding affinity. When we compare general PLM \texttt{ESM-2} to its Ab-finetuned model \texttt{CurrAb} \cite{Burbach2025.02.27.640641},
\texttt{CurrAb} achieved a mean Spearman correlation of 0.073, compared with 0.018 for ESM-2 (a gain of +0.055), and improved mean fivefold precision across the four budgets by +0.021 relative to ESM-2.
However, \texttt{AntiFold} \cite{hoie2024antifold}, which is an ESM-IF finetuned on structure of antibodies (OAS  \cite{Olsen2022Observed})  and Ab-Ag complex (SAbDab \cite{dunbar2014sabdab}) data, had a mean Spearman correlation 0.078 lower and mean fivefold precision across the four budgets 0.074 lower than ESM-IF1, suggesting limited gains on Ab-specific finetuning and potential catastrophic forgetting.\footnote{\texttt{AntiFold} \cite{hoie2024antifold} reports a positive \texttt{1mlc} correlation (0.427) from scoring only CDR log-likelihoods, whereas AbBiBench scores the full Ab-Ag complex and obtains a correlation of $-0.441$.}

\paragraph{\textbf{}}

\paragraph{\textbf{Finding 3: Model rankings are more useful for prioritizing likely winners than for excluding likely losers.}}
Figure~\ref{fig:nfold} reveals an asymmetric practical utility for current model rankings. When used for negative selection, even the strongest models provided only modest enrichment over random ranking. At a 30\% discard budget, ESM-IF1 and ProteinMPNN identified worse-than-wild-type variants with discard precisions of 0.640 and 0.638, only 0.090 and 0.088 above the 0.550 random expectation, and captured 0.405 and 0.408 of all such variants compared with 0.301 under random ranking. In contrast, positive prioritization of the highest-ranked variants produced substantially stronger enrichment: at $k=10$, ProteinMPNN and ESM-IF1 achieved fivefold-improvement precisions of 0.51 and 0.48, approximately 1.9-fold and 1.8-fold the 0.268 random baseline, and their precision remained 0.46 and 0.48 at $k=100$. Thus, current scores are most useful for prioritizing candidates likely to improve affinity, whereas a low model rank alone provides relatively weak evidence that a variant should be excluded from experimentation. Because the positive- and negative-selection analyses involved 10 and 8 evaluable assays, respectively, this asymmetry should be interpreted directionally rather than as a paired effect-size comparison.

\subsection{Generate Antibody Variants with H1N1 Binding Signals}

\begin{figure}[p]
  \centering
  \begin{subfigure}[t]{0.61\textwidth}
    \centering
    \includegraphics[width=\linewidth]{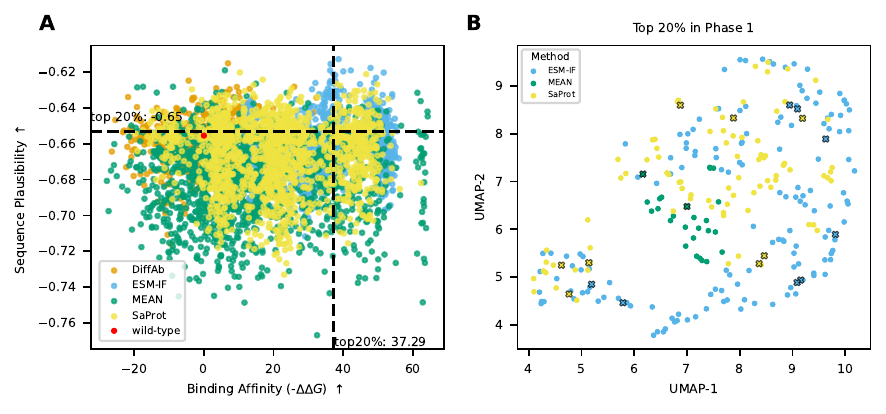}
    \caption{}
    \label{fig:phase1-screening}
  \end{subfigure}

  \vspace{0.1em}
  \begin{subfigure}[t]{0.68\textwidth}
    \centering
    \includegraphics[width=\linewidth]{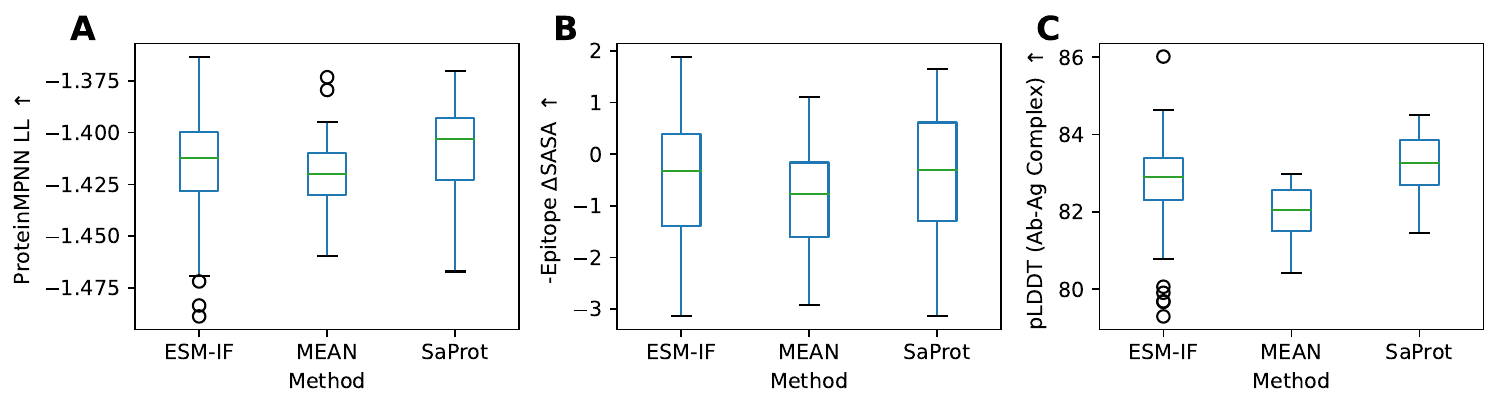}
    \caption{}
    \label{fig:phase2-1}
  \end{subfigure}

  \vspace{0.1em}
  \begin{subfigure}[t]{0.68\textwidth}
    \centering
    \includegraphics[width=\linewidth]{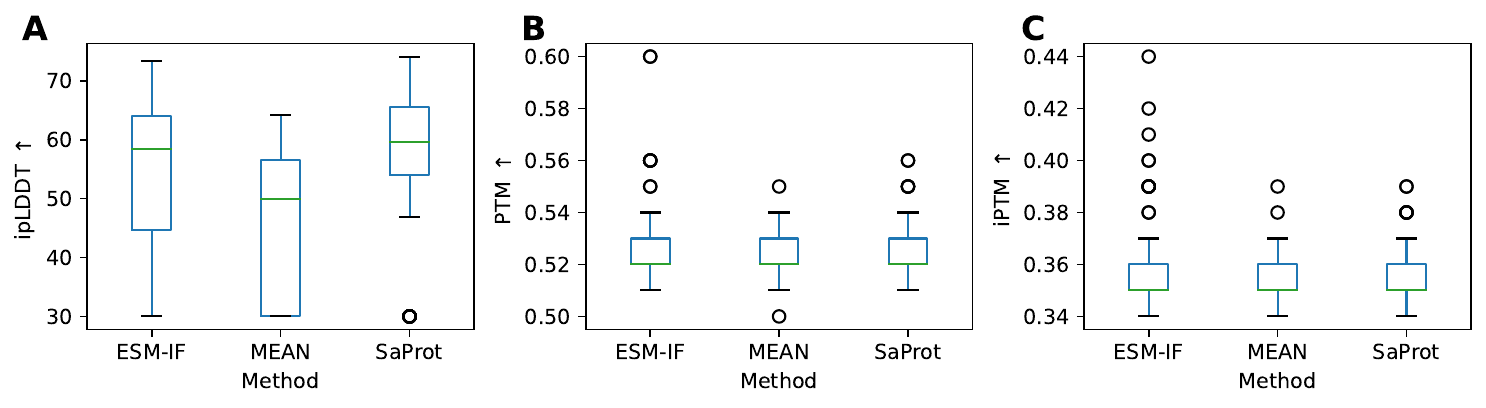}
    \caption{}
    \label{fig:phase2-2}
  \end{subfigure}

  \vspace{0.1em}
  \begin{subfigure}[t]{0.32\textwidth}
    \centering
    \includegraphics[width=\linewidth]{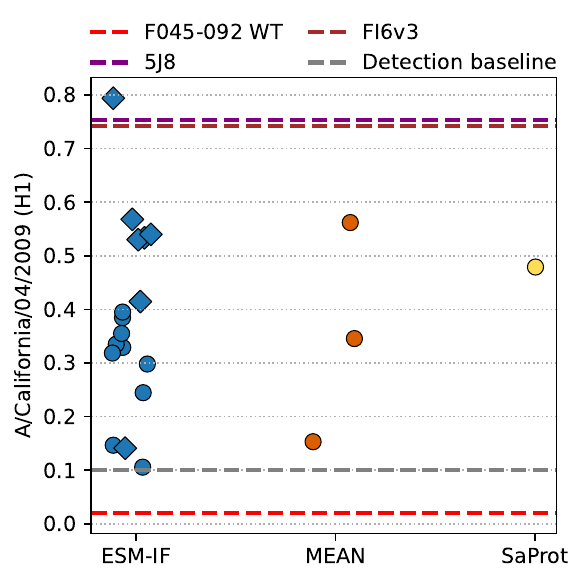}
    \caption{}
    \label{fig:elisa}
  \end{subfigure}

  \caption{Computational screening and experimental validation of generated F045-092 variants. \textbf{a}, Phase 1 evaluation of predicted binding potential, sequence plausibility and CDR-H3 sequence diversity. Predicted binding potential is reported as $-\Delta\Delta G$, where $\Delta\Delta G=\Delta G_{\mathrm{variant}}-\Delta G_{\mathrm{wild}}$ and $\Delta G_{\mathrm{wild}}=66.34$. The top 20\% of variants are shown in the upper-right region; the 18 non-dominated Pareto-optimal variants are marked with $\bm{\times}$. \textbf{b}, Phase 2 evaluation of binding potential and sequence plausibility. \textbf{c}, Phase 2 evaluation of structural integrity using side-chain orientations and confidence scores for the complex and interfacial residues. \textbf{d}, Endpoint ELISA $\text{OD}_{450}$ signals for 21 model-designed mutants and controls against H1 hemagglutinin. Diamonds indicate signals above the wild-type F045-092 control for H1; endpoint $\text{OD}_{450}$ is a relative readout of H1-reactive binding under matched assay conditions.}
  \label{fig:design-validation}
  \label{fig:phase2_merged}
\end{figure}

 To determine whether retrospective model rankings translate into practical design utility, we moved beyond scoring variants with existing affinity measurements and conducted a prospective gain-of-binding study. We evaluated whether selected models could generate antibody variants with detectable H1N1 binding signals relative to the wild-type F045-092 in an endpoint ELISA assay. F045-092 was chosen because it recognizes H3N2 but shows no detectable binding to the A/California/07/2009 H1 antigen in the source studies, providing a clear negative baseline. Each model produced 1,500 CDR-H3 variants with up to five mutations. As \texttt{DiffAb} often generated wild-type sequences with altered structures, we excluded duplicates, resulting in 467 valid \texttt{DiffAb} variants and a total of 4,967 unique variants across all models.
 We designed a two-phase filtering method: Phase 1 used predicted binding potential and AntiBERTy sequence plausibility to retain the top 20\% of variants, and Phase 2 evaluated sequence plausibility and complex-structure confidence for this reduced set (Methods~\ref{sec:multi_objective}). In Phase 1, \texttt{ESM-IF} and \texttt{SaProt} produced variants with both strong predicted binding potential and high sequence plausibility. Their average predicted $\Delta \Delta G$ values were -29.27 and -20.79, respectively, indicating more favorable computed binding energies than the wild type (Fig. \ref{fig:phase1-screening}, \ref{fig:phase1}A). These variants also retained \texttt{AntiBERTy} plausibility scores (-0.663 for \texttt{ESM-IF} and -0.666 for \texttt{SaProt}) close to that of the wild type (-0.655; Fig. \ref{fig:phase1}B).
On the other hand, \texttt{DiffAb} generated plausible sequences (-0.657) but did not improve the predicted binding energy ($\Delta \Delta G = 2.03$). Conversely, \texttt{MEAN} produced variants with more favorable predicted binding energy ($\Delta \Delta G = -13.95$), but at the cost of substantially lower sequence plausibility (-0.680) compared to the wild type.

Moving from Phase 1 to Phase 2 screening, we focused on the top 20\% of Phase 1 variants. Of these, 158, 91, and 26 candidates were selected from \texttt{ESM-IF}, \texttt{SaProt}, and \texttt{MEAN}, respectively. In Phase 2, \texttt{SaProt} achieved the highest sequence plausibility (ProteinMPNN log-likelihood = -1.406) and complex structure confidence (pLDDT = 83.22), followed by \texttt{ESM-IF} (-1.415, pLDDT = 82.82; Fig. ~\ref{fig:phase2-1},~\ref{fig:phase2-2}).

To assess the sequence diversity, we visualized CDR-H3 embeddings using AntiBERTy (Fig. \ref{fig:phase1-screening}B). \texttt{MEAN} variants formed a tight, isolated cluster, while \texttt{ESM-IF} and \texttt{SaProt} occupied broader, partially overlapping regions—indicating greater diversity.

Structural comparison revealed that \texttt{SaProt} generated CDR-H3 variants closest to wild type (mean cdrRMSD = 2.28), followed by \texttt{ESM-IF} (2.70) and \texttt{MEAN} (3.25). \texttt{SaProt} also produced variants with higher CDR-H3 structural confidence (pLDDT), suggesting better foldability (Fig. \ref{fig:diversity-rmsd},~\ref{fig:diversity-cdrh3}).
Sequence divergence analysis using cdrDist showed \texttt{SaProt} variants were most similar to wild type (0.222), compared to \texttt{ESM-IF} (0.279) and \texttt{MEAN} (0.277) (Fig. \ref{fig:diversity-rmsd}, \ref{fig:diversity-sw}). This aligns with the average number of substitutions per CDR-H3: 2.98 for \texttt{SaProt}, 3.03 for \texttt{MEAN}, and 3.75 for \texttt{ESM-IF}.

\noindent\par
\noindent\par
Across our in silico evaluations, we observed that while some models excel at one computational metric, they frequently do so at the expense of another. For instance, \texttt{MEAN} achieved more favorable predicted binding-energy scores but produced variants with substantially lower sequence plausibility and larger structural deviations from the wild type. This demonstrates a core trade-off between maximizing binding affinity and maintaining high sequence plausibility or evolutionary realism. (Fig.~\ref{fig:phase1} and Fig.~\ref{fig:diversity-cdrh3}).

\paragraph{ Experimental validation of H1 hemagglutinin binding by endpoint ELISA.} We performed ELISA binding assays on the 21 designed mutants against hemagglutinin (HA) from A/California/2009 (H1). The experimental set comprised 17 ESM-IF-derived, three MEAN-derived, and one SaProt-derived designs. Figure \ref{fig:elisa} shows that all variants produced signals above the detection threshold for H1, indicating a gain of H1 binding activity. This suggests that our in-silico evaluation pipeline enables efficient screening of variants with experimentally detectable H1-binding signals from multiple models, thereby reducing the time and cost associated with experimental validation. Notably, the highest observed $\text{OD}_{450}$ signal was produced by an ESM-IF-derived design. These ELISA results were consistent with our retrospective benchmark.

\section{Discussion}
\label{conclustions}
This study introduces AbBiBench, a biologically relevant and structurally informed benchmarking framework for antibody binding-affinity optimization and design. Recognizing the limitations of traditional computational evaluation metrics, our approach explicitly incorporates antibody-antigen complex information, thereby aligning more closely with the biological realities of antibody interactions.

Beyond establishing the strongest methods under the present protocol—where global structure-informed inverse-folding models such as \texttt{ESM-IF} and \texttt{ProteinMPNN} outperform the other evaluated implementations—AbBiBench yields critical insights for future method development. In the two matched comparisons, antibody-specific fine-tuning had model-dependent effects: \texttt{CurrAb} modestly improved over \texttt{ESM-2}, whereas \texttt{AntiFold} performed worse than \texttt{ESM-IF1} under the present protocol. These opposing outcomes indicate that antibody-specific adaptation alone is insufficient to ensure better affinity ranking.  We further highlight a fundamental trade-off between aggressively exploring novel, high-affinity conformations and preserving the evolutionary sequence plausibility required for stable proteins.

From a practical perspective, AbBiBench reveals an asymmetric use case for current zero-shot scores. ProteinMPNN and ESM-IF1 enriched fivefold affinity improvements among the highest-ranked candidates, whereas their negative-filtering performance provided only modest gains over random ranking. Current scores should therefore be used primarily to enrich experimental shortlists; a low model rank alone should not be treated as sufficient evidence to exclude a variant from experimental testing.

Bridging benchmarking to application, our case study on redesigning the F045-092 antibody successfully yielded 21 Pareto-optimal variants. Assessed in vitro via ELISA against H1N1 hemagglutinin, the highest observed $\text{OD}_{450}$ signals were directly traced back to our insights on global structure modeling. 
This study has several limitations. First, the subsequent experiment uses endpoint ELISA $\text{OD}_{450}$ as a semi-quantitative binding signal under matched assay conditions. This readout supports evaluation of candidate prioritization but does not provide an equilibrium dissociation constant or binding kinetics. Moreover, the final experimental set contained unequal numbers of candidates from ESM-IF, MEAN, and SaProt (17, 3, and 1, respectively), so the experiment alone does not constitute an unbiased head-to-head model comparison; our interpretation of ESM-IF instead relies on its concordance with the retrospective benchmark. Second, the absence of experimental neutralization readouts, such as IC$_{50}$, prevents conclusions about therapeutic potency. Finally, although AbBiBench aggregates 216,560 measurements across diverse antigens, some constituent datasets (e.g., 1mhp and 1n8z) are limited in size or derived from narrow mutational libraries, potentially reducing the statistical precision of per-dataset evaluations and the coverage of broader antibody design space. Future work should add dose-response experiments and orthogonal BLI or SPR measurements of $K_d$, $k_{\mathrm{on}}$, and $k_{\mathrm{off}}$; incorporate functional readouts and properties such as stability, immunogenicity, and developability; and increase the diversity and scale of benchmark tasks. These improvements will enable more accurate and biologically grounded evaluation of generative antibody design models. Ultimately, AbBiBench provides a roadmap for evaluating how machine-learning frameworks combine global 3D structural information with biological sequence priors.

\section{Methods}
\label{methods}

\subsection{Overview of the AbBiBench evaluation framework}
AbBiBench evaluates whether released computational methods can (i) rank experimentally measured antibody variants by binding affinity and (ii) generate antibody variants that can be prioritized for experimental testing. Throughout this study, zero-shot evaluation means using a released frozen checkpoint without AbBiBench variants or binding scores for task-specific fine-tuning, calibration, model selection or ensemble fitting. The antibody--antigen complex, rather than the antibody alone, is the common evaluation unit.

\subsection{Collection and curation of antibody binding datasets}
\label{sec:datasets}
We compiled 16 binding-affinity assays containing antigen sequences, antibody heavy- and light-chain sequences, wild-type complex structures and experimental binding measurements. The collection comprises 216,560 measurements spanning 13 antibodies and 9 antigens. Datasets were retained only when at least 20 antibody variants were available. The underlying libraries were produced by phage or yeast display, deep mutational scanning, targeted mutagenesis, machine-learning-guided design or biophysical modeling. Variant-library sizes ranged from 67 to 65,535. Variable regions of the heavy and light chains were retained because affinity-conferring mutations occur in these domains. An assay-level overview is provided in Table~\ref{tab:compact_table}; complete chain definitions, structural metadata and library-construction details are reported in Supplementary Table~\ref{tab:dataset_details}.

Heterogeneous source records were converted into complete antibody--antigen examples through a common curation procedure. We reconstructed full variable heavy- and light-chain sequences from the reported mutations and reconciled source-specific chain names and variable-region boundaries with the corresponding wild-type complexes. Experimental replicates were averaged where required by the source study. Negative controls, records with missing experimental measurements and exact duplicate records were removed before analysis. Heavy-chain, light-chain and antigen-chain identities were then recorded in a common metadata schema.

\begin{table}[H]
\centering
{
\scriptsize
\setlength{\tabcolsep}{2.5pt}
\renewcommand{\arraystretch}{1.08}
\begin{tabular}{@{}
  >{\raggedright\arraybackslash}p{0.08\textwidth}
  >{\raggedright\arraybackslash}p{0.15\textwidth}
  >{\raggedright\arraybackslash}p{0.13\textwidth}
  >{\raggedright\arraybackslash}p{0.12\textwidth}
  >{\raggedright\arraybackslash}p{0.27\textwidth}
  >{\raggedright\arraybackslash}p{0.12\textwidth}@{}}
\toprule
\textbf{Assay} & \textbf{Antibody--antigen} & \textbf{Variants} & \textbf{Mutation scope} & \textbf{Experimental source} & \textbf{Structural template} \\ \midrule
4fqi\_h1   & CR9114--H1N1 HA       & HC: 65,094                    & CDR + framework & $K_d$; germline-reverted yeast display \cite{phillips2021binding} & PDB 4FQI \\
4fqi\_h3   & CR9114--H3N2 HA       & HC: 65,535                    & CDR + framework & $K_d$; germline-reverted yeast display \cite{phillips2021binding} & PDB 4FQI \\
3gbn\_h1   & CR6261--H1N1 HA       & HC: 1,887                     & CDR + framework & $K_d$; germline-reverted yeast display \cite{phillips2021binding} & PDB 3GBN \\
3gbn\_h9   & CR6261--H9N2 HA       & HC: 1,842                     & CDR + framework & $K_d$; germline-reverted yeast display \cite{phillips2021binding} & PDB 3GBN \\
aayl49     & AAYL49--Spike HR2     & HC: 4,312                     & CDR             & $K_d$; phage-display DMS \cite{engelhart2022dataset}              & AlphaFold 3 \\
aayl49\_ML & AAYL49\_ML--Spike HR2 & HC: 8,953                    & CDR             & $K_d$; ML-generated scFv, phage display \cite{li2023machine}      & AlphaFold 3 \\
aayl50     & AAYL50--Spike HR2     & LC: 11,473                    & CDR             & $K_d$; phage-display DMS \cite{engelhart2022dataset}              & AlphaFold 3 \\
aayl51     & AAYL51--Spike HR2     & HC: 4,320                     & CDR             & $K_d$; phage-display DMS \cite{engelhart2022dataset}              & AlphaFold 3 \\
aayl52     & AAYL52--Spike HR2     & LC: 13,324                    & CDR             & $K_d$; phage-display DMS \cite{engelhart2022dataset}              & AlphaFold 3 \\
2fjg       & G6.31--VEGF           & HC: 2,223; LC: 2,014          & CDR + framework & log enrichment; phage-display DMS \cite{koenig2017mutational}    & PDB 2FJG \\
1mlc       & D44.1--lysozyme       & HC: 1,229; LC: 865            & CDR + framework & log enrichment; yeast-display DMS \cite{warszawski2020correction} & PDB 1MLC \\
1n8z       & trastuzumab--HER2     & HC: 419                       & CDR             & $K_d$; zero-shot generative design \cite{Shanehsazzadeh2023.01.08.523187} & PDB 1N8Z \\
1mhp       & AQC2--integrin-$\alpha$1 & HC: 37; LC: 25; both: 5    & CDR + framework & $K_d$; in-silico biophysical library \cite{Clark2006Affinity}    & PDB 1MHP \\
4d5\_her2  & 4D5--HER2             & Both: 2,080                   & CDR             & log enrichment; yeast-display DMS \cite{minot2024}               & Experimental PDB \\
5a12\_ang2 & 5A12--Ang2            & HC: 796; LC: 104; both: 43    & CDR             & log enrichment; yeast-display DMS \cite{minot2024}               & Experimental PDB \\
5a12\_vegf & 5A12--VEGF            & HC: 5,300; LC: 228; both: 24,452 & CDR          & log enrichment; yeast-display DMS \cite{minot2024}               & Experimental PDB \\
\bottomrule
\end{tabular}}
\caption{Experimental assays and structural templates included in AbBiBench. Variant counts are reported by mutated chain. Complete chain identities, sequence lengths and dataset-specific processing details are provided in Supplementary Table~\ref{tab:dataset_details}.}
\label{tab:compact_table}
\end{table}

\subsection{Experimental affinity labels and data harmonization}

Reported dissociation constants were first converted to molar units and then transformed as $-\log_{10}(K_d\,[\mathrm{M}])$. Enrichment measurements were log transformed, so that larger values consistently indicated stronger binding within each assay. Because enrichment is a relative high-throughput selection readout rather than a direct thermodynamic measurement, direct-$K_d$ and enrichment assays were not pooled onto a common quantitative affinity scale; all correlations and model comparisons were calculated within assay.

Where an experimentally determined antibody--antigen complex was available, the corresponding PDB structure was used as the structural template. No experimentally resolved complexes were available for the five SARS-CoV-2 assays (AAYL49, AAYL49\_ML, AAYL50, AAYL51 and AAYL52). For these assays, AlphaFold 3 was supplied with the matched antibody heavy- and light-chain sequences and the SARS-CoV-2 Spike HR2 antigen sequence to generate the antibody--antigen structural templates used by structure-conditioned methods. Dataset-specific preprocessing and structural sources are described in Supplement~\ref{sup:data_details}.

\subsection{Selection of evaluated models}
\label{sec:models}
We selected the models according to methodological diversity, availability of executable code and use of sequence, protein dynamics, local-structure or global-structure information.

The 20 base methods were ESM-2, ESM-3, SaProt, ProSST, AntiBERTy and CurrAb; ProGen2 and ProtGPT-2; ProteinMPNN, ESM-IF and AntiFold; DiffAb; MEAN and dyMEAN; AlphaFold 3, Boltz-2 and ESMFold2; and the FoldX, FreeSASA and ESMDance biophysical methods. Native and fixed-backbone implementations of DiffAb, MEAN and dyMEAN yielded 23 evaluated implementations in total.  These methods comprised masked and autoregressive protein language models, inverse-folding models, diffusion-based generators, geometric CDR-imputation models, structure-prediction models and biophysical methods.

  ESM-2 established a general masked-sequence baseline, whereas AntiBERTy and CurrAb tested antibody-specific pretraining and adaptation; SaProt and ProSST tested local-structure tokenization, and ESM-3 tested multimodal sequence--structure--function representations. ProGen2 and ProtGPT-2 supplied autoregressive, structure-agnostic sequence baselines at different model scales. ProteinMPNN and ESM-IF represented established general inverse-folding architectures, whereas AntiFold tested antibody-specific inverse-folding adaptation. DiffAb tested diffusion-based joint CDR sequence--structure generation. MEAN tested equivariant residue-level CDR imputation, and dyMEAN tested its full-atom end-to-end extension. AlphaFold 3 supplied an all-atom complex-prediction confidence baseline, Boltz-2 represented a biomolecular interaction model that jointly predicts complex structure and binding affinity, and ESMFold2 supplied an ESM-based structure-prediction comparator. FoldX, FreeSASA and ESMDance respectively tested binding-energy change, epitope exposure and learned protein-dynamics signals; ESMDance was therefore grouped under Biophysics rather than masked language models.  Detailed descriptions of the evaluated models are provided below.  

\label{sec:detailed_evaluated_models}
{
\paragraph{Masked language models.}
Masked protein language models predict masked residues from bidirectional sequence context and can capture associations between sequence motifs and higher-level functional properties~\cite{hie2024efficient,meier2021language}. ESM-2~\cite{lin2023evolutionary} is a general protein model trained on large-scale sequence collections with a masked-token objective. AntiBERTy~\cite{ruffolo2021deciphering} is a 26-million-parameter, eight-layer BERT model trained on 558 million antibody sequences obtained after clustering OAS at 95\% sequence identity. SaProt~\cite{susaprot} augments residue tokens with Foldseek-derived structural tokens~\cite{barrio2023clustering,van2024fast}, thereby encoding residue identity and local structural context jointly. ProSST~\cite{li2024prosst} uses a geometric-vector-perceptron encoder to discretize local atomic neighborhoods into structural tokens and applies disentangled attention over sequence, structure and position. ESM-3~\cite{hayes2025simulating} is a multimodal protein language model that represents sequence, structure and function through discrete token tracks. CurrAb~\cite{Burbach2025.02.27.640641} is an ESM-2 model adapted to antibodies through curriculum learning that progressively shifts from unpaired to paired OAS antibody sequences while seeking to preserve general-protein knowledge.

\paragraph{Biophysical methods.}
FoldX estimates the change in antibody--antigen binding free energy after mutation, whereas FreeSASA quantifies changes in solvent-accessible surface area at the epitope. ESMDance~\cite{hou2026esmdance} uses a frozen 35-million-parameter ESM-2 encoder with prediction heads trained on dynamic biophysical properties derived from molecular-dynamics simulations and normal-mode analyses of more than 64,000 proteins. We grouped ESMDance under Biophysics because its benchmark score measures predicted perturbation of the wild-type dynamics profile rather than masked-token likelihood; unlike FoldX and FreeSASA, it does not require an antibody--antigen complex structure at inference.

\paragraph{Autoregressive protein language models.}
Unlike masked models, autoregressive models predict each residue using the preceding sequence context. ProGen2~\cite{nijkamp2023progen2} is a decoder-only Transformer family with models of up to 6.4 billion parameters trained on approximately one billion protein sequences. ProtGPT2~\cite{ferruz2022protgpt2} uses a 738-million-parameter GPT-2 architecture trained on approximately 50 million protein sequences. Both are sequence-only models and therefore provide structure-agnostic comparators for autoregressive inverse-folding methods.

\paragraph{Inverse-folding models.}
Inverse-folding models predict amino acid sequences conditioned on protein backbone structures and have been applied to antibody design using antibody--antigen complex geometry~\cite{shanker2024unsupervised}. ProteinMPNN~\cite{dauparas2022robust} uses a message-passing neural network to encode residue interactions before autoregressive sequence prediction. ESM-IF1~\cite{hsu2022learning} combines a geometric-vector-perceptron encoder of backbone coordinates with a Transformer decoder. AntiFold~\cite{hoie2024antifold} adapts ESM-IF1 to antibodies by fine-tuning on solved and predicted antibody structures from SAbDab and OAS.

\paragraph{Diffusion-based generative models.}
Diffusion models formulate antibody design as denoising over residue identities and geometric variables. DiffAb~\cite{luo2022antigen} conditions on an antigen--antibody framework complex and jointly denoises CDR sequence and structure. We evaluated both its native co-design configuration and a fixed-backbone implementation, DiffAb\_fixbb, which predicts sequence while retaining the input backbone geometry.

\paragraph{Geometric CDR-imputation models.}
Geometry-aware models formulate antibody optimization as reconstruction of masked CDRs within the explicit three-dimensional antibody--antigen interface. MEAN~\cite{kong2022conditional} represents the complex as an E(3)-equivariant residue--atom graph and uses alternating message-passing blocks to restore CDR sequence and backbone coordinates. dyMEAN~\cite{kong2023end} extends this formulation to full-atom end-to-end design using conserved-framework initialization and a shadow-paratope representation to generate paratope sequence, side-chain geometry and binding pose jointly. For both models, we evaluated the released native setting and a fixed-backbone implementation that regenerated sequence while preserving the input structure.

\paragraph{Structure-prediction models.}
AlphaFold 3~\cite{abramson2024accurate} predicts all-atom biomolecular complexes across multiple molecular classes, whereas Boltz-2~\cite{passaro2025boltz} jointly models biomolecular complex structures and binding affinities using dedicated structure, confidence and affinity components. ESMFold2 provided an ESM-based structure-prediction comparator~\cite{candido2026language}. For all three structure predictors, ipSAE was the primary zero-shot ranking score. iPTM and interface pLDDT (ipLDDT) were evaluated only as auxiliary proxy scores in the Supplement. All confidence scores were treated as heuristic comparators rather than calibrated estimates of antibody--antigen affinity.
}

\subsection{Calculation of zero-shot model scores}
\label{sec:correlation}
For each mutant--antigen complex, we calculated a score using the native likelihood or confidence function of each frozen model. Masked models were scored with pseudo-log-likelihoods, autoregressive models with sequence log-likelihoods, inverse-folding models with structure-conditioned sequence likelihoods, and generative or geometric models with their released conditional scoring functions. Higher scores were defined to indicate greater model preference. For the biophysical baselines, lower binding free energy and lower epitope solvent-accessible surface area indicate stronger predicted binding; their signs were therefore reversed before correlation analysis.  
\label{calculation_details}
\begingroup
\paragraph{Unified scoring workflow.}
For sample $i$, let $\mathbf{s}^{\mathrm{ab}}_i$ and $\mathbf{s}^{\mathrm{ag}}_i$ denote the antibody and antigen sequences, $\mathbf{z}_i$ the available antibody--antigen structural information and $y_i$ the harmonized experimental binding score. Each frozen model $\mathcal{M}$ produced a native score
\[
\ell_i=\mathcal{M}(\mathbf{s}^{\mathrm{ab}}_i,\mathbf{s}^{\mathrm{ag}}_i,\mathbf{z}_i),
\]
with $\mathbf{z}_i$ omitted for sequence-only models. No task-specific training, calibration or ensemble fitting was performed. Scores were oriented so that larger values represented greater predicted preference and were then compared with $y_i$ within each assay.

\paragraph{Masked language-model scores.}
Let $\mathbf{s}=(s_1,\ldots,s_N)$ be the concatenated sequence presented to a model and $\mathbf{X}$ the structural representation when required.
For all masked language-model implementations, including ESM-2, CurrAb, SaProt, ProSST, ESM-3 and AntiBERTy, the complete unmasked input was passed through the model once. The score was the mean log-probability assigned to the observed token at each position,
\[
S_{\mathrm{unmasked}}(\mathbf{s},\mathbf{X})=
\frac{1}{N}\sum_{i=1}^{N}\log p_{\theta}(s_i\mid\mathbf{s},\mathbf{X}),
\]
with $\mathbf{X}$ omitted for sequence-only models and fixed assay-matched structural tokens or coordinates supplied when required.

\paragraph{Autoregressive language-model scores.}
For ProGen2 and ProtGPT2, the sequence score was the mean of left-to-right conditional log-probabilities,
\[
\overline{\log p}(\mathbf{s})=\frac{1}{N}\sum_{i=1}^{N}\log p(s_i\mid\mathbf{s}_{<i}),
\]
where $\mathbf{s}_{<i}$ denotes the residues preceding position $i$. Chain order was held fixed within each assay.

\paragraph{Inverse-folding scores.}
Inverse-folding methods additionally condition each autoregressive prediction on the complex backbone $\mathbf{X}$. The reported score was the mean structure-conditioned log-probability per residue:
\[
\overline{\log p}(\mathbf{s}\mid\mathbf{X})=
\frac{1}{N}\sum_{i=1}^{N}\log p(s_i\mid\mathbf{s}_{<i},\mathbf{X}).
\]
ProteinMPNN, ESM-IF1 and AntiFold were evaluated with their released structure-conditioned scoring functions on the same assay-matched complex template.

\paragraph{Biophysical scores.}
FoldX provided the mutation-induced binding-energy change $\Delta\Delta G=\Delta G_{\mathrm{variant}}-\Delta G_{\mathrm{wild\ type}}$, and FreeSASA provided the corresponding change in epitope solvent-accessible surface area. Because smaller values imply more favorable predicted binding, we used $-\Delta\Delta G$ and $-\Delta\mathrm{SASA}$ for ranking. For ESMDance~\cite{hou2026esmdance}, property $j$ was summarized as the mean absolute relative change between the mutant and assay-matched wild type,
\[
S_j=\operatorname{mean}\left|\frac{P_j^{\mathrm{mut}}-P_j^{\mathrm{WT}}}{P_j^{\mathrm{WT}}}\right|.
\]
The 23 property scores were quantile-normalized within each assay, and the final score was the negative geometric mean
\[
S_{\mathrm{ESMDance}}=-\left(\prod_{j=1}^{23}\widetilde{S}_j\right)^{1/23}.
\]
Thus, a larger value represents a smaller predicted perturbation of the wild-type dynamics profile.

\paragraph{Diffusion-based generative scores.}
For DiffAb~\cite{luo2022antigen}, let $\mathcal{C}$ denote the redesigned CDR residues, $\mathcal{R}_t$ their noised residue identities, coordinates and orientations at diffusion step $t$, and $\mathbf{C}$ the fixed sequence--structure context outside the CDR. The residue-type denoising objective is
\[
\mathcal{L}^{\mathrm{type}}_t=
\mathbb{E}_{\mathcal{R}_t\sim q}\left[
\frac{1}{|\mathcal{C}|}\sum_{j\in\mathcal{C}}
D_{\mathrm{KL}}\!\left(q(s^j_{t-1}\mid s^j_t,s^j_0)\,\|\,
p_{\theta}(s^j_{t-1}\mid\mathcal{R}_t,\mathbf{C})\right)
\right].
\]
Native DiffAb jointly models residue type, position and orientation, with the coordinate and orientation losses
\[
\mathcal{L}^{\mathrm{pos}}_t=
\mathbb{E}\left[\frac{1}{|\mathcal{C}|}\sum_{j\in\mathcal{C}}
\left\|\boldsymbol{\epsilon}_j-G(\mathcal{R}_t,\mathbf{C})\right\|_2^2\right],
\]
\[
\mathcal{L}^{\mathrm{ori}}_t=
\mathbb{E}\left[\frac{1}{|\mathcal{C}|}\sum_{j\in\mathcal{C}}
\left\|(\mathbf{O}_0^j)^{\top}\widehat{\mathbf{O}}_{t-1}^j-\mathbf{I}\right\|_F^2\right].
\]
The released native objective combines these three terms across diffusion steps. DiffAb\_fixbb retains all backbone coordinates and orientations and uses only the residue-type term, isolating sequence compatibility with the fixed input geometry.

\paragraph{Geometric CDR-imputation scores.}
MEAN and dyMEAN represent the antibody--antigen complex as a residue graph $\mathbf{G}=(\mathcal{V},\mathcal{E})$. Each node $i$ has an embedding $h_i$ and atomic coordinates $Z_i$, which are updated by E(3)-equivariant message passing,
\[
\{h_i^{(t+1)},Z_i^{(t+1)}\}_{i\in\mathcal{V}}=
\operatorname{GNN}_{\theta}\!\left(\{h_i^{(t)},Z_i^{(t)}\}_{i\in\mathcal{V}},\mathbf{G}\right).
\]
For a masked CDR residue $i\in\mathcal{C}$, the amino acid distribution is $p_i=\operatorname{Softmax}(Wh_i^{(T)})$. Native co-design uses sequence and coordinate losses,
\[
L=\sum_{i\in\mathcal{C}}
\left[l_{\mathrm{ce}}(p_i,\widehat{p}_i)+
\lambda l_{\mathrm{Huber}}(Z_i,\widehat{Z}_i)\right].
\]
The fixed-backbone implementations MEAN\_fixbb and dyMEAN\_fixbb preserve the input geometry and omit coordinate regression, reducing the objective to
\[
L_{\mathrm{fixbb}}=\sum_{i\in\mathcal{C}}l_{\mathrm{ce}}(p_i,\widehat{p}_i).
\]

\paragraph{Structure-prediction confidence scores.}
For AlphaFold 3, Boltz-2 and ESMFold2, ipSAE was read from each predicted antibody--antigen complex and used as the primary confidence-based ranking feature. Component CSV files corresponding to the same biological assay were concatenated before correlation, and the three predictors were compared on their intersection of valid variant rows. iPTM and interface pLDDT (ipLDDT) were retained only for the auxiliary analysis in Supplement~\ref{sup:sec:structure_prediction_correlation}; no confidence score was converted into an affinity estimate or calibrated within assay.
\endgroup

\subsection{Benchmark evaluation tasks and performance metrics}
\label{evaluation_tasks}
The primary retrospective metric was the Spearman rank correlation between the model score and harmonized experimental binding score within each assay. We summarized performance using an equal-assay mean so that large variant libraries did not dominate the analysis. We also evaluated fivefold Precision@$k$ for $k\in\{10,20,50,100\}$ on 10 of the 11 direct-$K_d$ assays; 1mhp was excluded because it contained fewer than 100 scored variants, whereas every retained assay--method pair contained at least 419 scored variants. A hit was a variant with $K_d$ at least fivefold better than the wild type. Dataset prevalence provided the expected precision under random ranking, and the finite-dataset oracle placed all observed hits first. Assay-level Precision@$k$ uncertainty was summarized with Wilson 95\% intervals.

\subsection{Generation and selection of F045-092 antibody variants}
\label{sec:case_study}
We used F045-092, an antibody that binds influenza H3N2 but does not detectably bind the A/California/07/2009 H1N1 strain in the source studies, as a subsequent model-guided design case study. The available crystal structures contain F045-092 bound to H3N2 HA (PDB IDs: 4O58 and 4O5I); no experimentally resolved F045-092--H1N1 complex was available. AlphaFold 3 was supplied with the F045-092 heavy- and light-chain sequences and the H1N1 HA sequence to predict the input complex. The predicted complex had a mean pLDDT of 83.47, iPTM of 0.39 and pTM of 0.49. ESM-IF, SaProt, DiffAb and MEAN were then used to redesign the CDR-H3 loop. Each model generated 1,500 variants, and variants with more than five substitutions were excluded to limit divergence from the parental antibody. 
The same AlphaFold 3-predicted antibody--antigen complex was used as structural input for all four sampling methods. For MEAN, alanine scanning across CDR-H3 residues was used to derive per-residue masking probabilities; masked positions were sampled from these probabilities, and sequence and structure were generated jointly through the authors' multi-round procedure. For DiffAb, CDR-H3 sequences were sampled at diffusion time points $t\in\{1,2,4,8\}$ with up to 15 random seeds to broaden sequence diversity. For ESM-IF, in-silico saturation mutagenesis provided log-likelihood scores for all 19 alternative amino acids at each CDR-H3 position. Single substitutions scoring above the wild-type residue were combined randomly into variants containing two to five mutations, with five repeats for each mutation count, following the procedure described by \cite{shanker2024unsupervised}. For SaProt, structure tokens were obtained with FoldSeek~\cite{van2024fast} and held fixed; alanine-scanning-derived masking probabilities were used to select one to five CDR-H3 positions, and only the sequence tokens were regenerated.

\par\medskip\noindent\par\smallskip
\phantomsection
\label{sec:multi_objective}
Generated variants were evaluated for sequence plausibility with AntiBERTy likelihood and for compatibility with the proposed backbone with ProteinMPNN likelihood. Binding-related criteria comprised FoldX binding-energy change, epitope solvent-accessible surface-area change and AlphaFold 3 structural confidence. In phase 1, all generated variants were scored using AntiBERTy likelihood and FoldX binding-energy change, and the top 20\% were retained. In phase 2, AlphaFold 3 complex structures were predicted for retained variants and used to calculate pLDDT, epitope solvent-accessible surface area and ProteinMPNN likelihood. Final candidates were selected from the non-dominated Pareto set across the five criteria. Design diversity was assessed with AntiBERTy embeddings, normalized Smith--Waterman CDR-H3 distance and CDR-H3 backbone RMSD relative to the wild type.
The two-phase evaluation workflow is summarized in Supplementary Fig.~\ref{fig:evaluating}.

\subsection{Endpoint ELISA experiments}
Twenty-one selected F045-092 CDR-H3 variants (17 ESM-IF-derived, 3 MEAN-derived and 1 SaProt-derived) were tested against A/California/2009 H1 hemagglutinin by endpoint ELISA. Binding was read as $\mathrm{OD}_{450}$ under matched assay conditions and compared with the wild-type F045-092 control. The endpoint signal was interpreted as a semi-quantitative H1-reactive binding readout, not as an equilibrium dissociation constant or kinetic measurement.

\noindent
\subsection{Software and computational resources}
No model training was performed. Model inference used one NVIDIA H100 80-GB GPU per model. The released benchmark repository contains evaluation scripts and model-specific instructions; the processed benchmark is distributed separately as a versioned dataset.
\section*{Data availability}
The AbBiBench dataset is available from Hugging Face at \url{https://huggingface.co/datasets/AbBibench/Antibody_Binding_Benchmark_Dataset}. The repository provides the curated antibody and antigen sequences, assay metadata and harmonized binding measurements used in this study. Primary experimental sources are cited for every constituent assay in Supplementary Table~\ref{tab:dataset_details}.

\section*{Code availability}
The evaluation code and leaderboard are available at \url{https://github.com/MSBMI-SAFE/AbBiBench}. The repository documents access to released model checkpoints and the model-specific procedures used to calculate benchmark scores.

\bibliography{references}

@article{bindercraft,
  title={One-shot design of functional protein binders with {BindCraft}},
  author={Pacesa, Martin and Nickel, Lennart and Schellhaas, Christian and Schmidt, Joseph and Pyatova, Ekaterina and Kissling, Lucas and others},
  journal={Nature},
  volume={646},
  pages={483--492},
  year={2025},
  doi={10.1038/s41586-025-09429-6}
}

@article{ruffolo2021deciphering,
  title={Deciphering antibody affinity maturation with language models and weakly supervised learning},
  author={Ruffolo, Jeffrey A and Gray, Jeffrey J and Sulam, Jeremias},
  journal={arXiv preprint arXiv:2112.07782},
  year={2021}
}

@article{chungyoun2024flab,
  title={FLAb: Benchmarking deep learning methods for antibody fitness prediction},
  author={Chungyoun, Michael and Ruffolo, Jeffrey and Gray, Jeffrey},
  journal={BioRxiv},
  pages={2024--01},
  year={2024},
  publisher={Cold Spring Harbor Laboratory}
}

@article{shanker2024unsupervised,
  title={Unsupervised evolution of protein and antibody complexes with a structure-informed language model},
  author={Shanker, Varun R and Bruun, Theodora UJ and Hie, Brian L and Kim, Peter S},
  journal={Science},
  volume={385},
  number={6704},
  pages={46--53},
  year={2024},
  publisher={American Association for the Advancement of Science}
}

@article{dunbar2014sabdab,
  title={SAbDab: the structural antibody database},
  author={Dunbar, James and Krawczyk, Konrad and Leem, Jinwoo and Baker, Terry and Fuchs, Angelika and Georges, Guy and Shi, Jiye and Deane, Charlotte M},
  journal={Nucleic acids research},
  volume={42},
  number={D1},
  pages={D1140--D1146},
  year={2014},
  publisher={Oxford University Press}
}

@article{luo2022antigen,
  title={Antigen-specific antibody design and optimization with diffusion-based generative models for protein structures},
  author={Luo, Shitong and Su, Yufeng and Peng, Xingang and Wang, Sheng and Peng, Jian and Ma, Jianzhu},
  journal={Advances in Neural Information Processing Systems},
  volume={35},
  pages={9754--9767},
  year={2022}
}

@article{dauparas2022robust,
  title={Robust deep learning--based protein sequence design using ProteinMPNN},
  author={Dauparas, Justas and Anishchenko, Ivan and Bennett, Nathaniel and Bai, Hua and Ragotte, Robert J and Milles, Lukas F and Wicky, Basile IM and Courbet, Alexis and de Haas, Rob J and Bethel, Neville and others},
  journal={Science},
  volume={378},
  number={6615},
  pages={49--56},
  year={2022},
  publisher={American Association for the Advancement of Science}
}

@inproceedings{hsu2022learning,
  title={Learning inverse folding from millions of predicted structures},
  author={Hsu, Chloe and Verkuil, Robert and Liu, Jason and Lin, Zeming and Hie, Brian and Sercu, Tom and Lerer, Adam and Rives, Alexander},
  booktitle={International conference on machine learning},
  pages={8946--8970},
  year={2022},
  organization={PMLR}
}

@article{wu2023hierarchical,
  title={A hierarchical training paradigm for antibody structure-sequence co-design},
  author={Wu, Fang and Li, Stan Z},
  journal={Advances in Neural Information Processing Systems},
  volume={36},
  pages={31140--31157},
  year={2023}
}

@article{susaprot,
  title={Saprot: Protein language modeling with structure-aware vocabulary},
  author={Su, Jin and Han, Chenchen and Zhou, Yuyang and Shan, Junjie and Zhou, Xibin and Yuan, Fajie},
  journal={BioRxiv},
  pages={2023--10},
  year={2023},
  publisher={Cold Spring Harbor Laboratory}
}

@article{li2024prosst,
  title={Prosst: Protein language modeling with quantized structure and disentangled attention},
  author={Li, Mingchen and Tan, Yang and Ma, Xinzhu and Zhong, Bozitao and Yu, Huiqun and Zhou, Ziyi and Ouyang, Wanli and Zhou, Bingxin and Tan, Pan and Hong, Liang},
  journal={Advances in Neural Information Processing Systems},
  volume={37},
  pages={35700--35726},
  year={2024}
}

@article{malherbe2024igblend,
  title={IgBlend: unifying 3D structures and sequences in antibody language models},
  author={Malherbe, Cedric and Ucar, Talip},
  journal={bioRxiv},
  pages={2024--10},
  year={2024},
  publisher={Cold Spring Harbor Laboratory}
}

@article{hayes2025simulating,
  title={Simulating 500 million years of evolution with a language model},
  author={Hayes, Thomas and Rao, Roshan and Akin, Halil and Sofroniew, Nicholas J and Oktay, Deniz and Lin, Zeming and Verkuil, Robert and Tran, Vincent Q and Deaton, Jonathan and Wiggert, Marius and others},
  journal={Science},
  volume={387},
  number={6736},
  pages={850--858},
  year={2025},
  publisher={American Association for the Advancement of Science}
}

@article{martinkus2023abdiffuser,
  title={Abdiffuser: full-atom generation of in-vitro functioning antibodies},
  author={Martinkus, Karolis and Ludwiczak, Jan and Liang, Wei-Ching and Lafrance-Vanasse, Julien and Hotzel, Isidro and Rajpal, Arvind and Wu, Yan and Cho, Kyunghyun and Bonneau, Richard and Gligorijevic, Vladimir and others},
  journal={Advances in Neural Information Processing Systems},
  volume={36},
  pages={40729--40759},
  year={2023}
}

@article{kong2023end,
  title={End-to-end full-atom antibody design},
  author={Kong, Xiangzhe and Huang, Wenbing and Liu, Yang},
  journal={arXiv preprint arXiv:2302.00203},
  year={2023}
}

@article{kong2022conditional,
  title={Conditional antibody design as 3d equivariant graph translation},
  author={Kong, Xiangzhe and Huang, Wenbing and Liu, Yang},
  journal={arXiv preprint arXiv:2208.06073},
  year={2022}
}

@article{watson2023novo,
  title={De novo design of protein structure and function with RFdiffusion},
  author={Watson, Joseph L and Juergens, David and Bennett, Nathaniel R and Trippe, Brian L and Yim, Jason and Eisenach, Helen E and Ahern, Woody and Borst, Andrew J and Ragotte, Robert J and Milles, Lukas F and others},
  journal={Nature},
  volume={620},
  number={7976},
  pages={1089--1100},
  year={2023},
  publisher={Nature Publishing Group UK London}
}

@article{lin2023evolutionary,
  title={Evolutionary-scale prediction of atomic-level protein structure with a language model},
  author={Lin, Zeming and Akin, Halil and Rao, Roshan and Hie, Brian and Zhu, Zhongkai and Lu, Wenting and Smetanin, Nikita and Verkuil, Robert and Kabeli, Ori and Shmueli, Yaniv and others},
  journal={Science},
  volume={379},
  number={6637},
  pages={1123--1130},
  year={2023},
  publisher={American Association for the Advancement of Science}
}

@article{candido2026language,
  title={Language Modeling Materializes a World Model of Protein Biology},
  author={Candido, Salvatore and Hayes, Thomas and Derry, Alexander and Rao, Roshan and Lin, Zeming and Verkuil, Robert and Wu, Bryan Z. and Lee, Jin Sub and Bruguera, Elise S. and Keval, Jehan A. and Kopylov, Mykhailo and Pak, John E. and Wu, Wesley and Thomas, Neil and Mataraso, Samson and Hsu, Alvin and Trotman-Grant, Ashton C. and Fatras, Kilian and dos Santos Costa, Allan and Badkundri, Rohil and Ak{\i}n, Halil and Oktay, Deniz and Deaton, Jonathan and Montabana, Elizabeth and Sitwala, Hrishita and Yu, Yue and Wiggert, Marius and Carlin, Dylan Alexander and Goering, Anthony W. and Blazejewski, Tomasz and Sandora, McCullen and Hla, Michael and Jia, Tina Z. and Kloker, Leon H. and Sofroniew, Nicholas J. and Uehara, Masatoshi and Pannu, Jassi and Bachas, Sharrol and Liu, Daniel S. and Sercu, Tom and Rives, Alexander},
  journal={bioRxiv},
  year={2026},
  doi={10.64898/2026.06.03.729735},
  url={https://www.biorxiv.org/content/10.64898/2026.06.03.729735},
  note={Preprint}
}

@article{notin2023proteingym,
  title={Proteingym: Large-scale benchmarks for protein fitness prediction and design},
  author={Notin, Pascal and Kollasch, Aaron and Ritter, Daniel and Van Niekerk, Lood and Paul, Steffanie and Spinner, Han and Rollins, Nathan and Shaw, Ada and Orenbuch, Rose and Weitzman, Ruben and others},
  journal={Advances in Neural Information Processing Systems},
  volume={36},
  pages={64331--64379},
  year={2023}
}

@article{abramson2024accurate,
  title={Accurate structure prediction of biomolecular interactions with AlphaFold 3},
  author={Abramson, Josh and Adler, Jonas and Dunger, Jack and Evans, Richard and Green, Tim and Pritzel, Alexander and Ronneberger, Olaf and Willmore, Lindsay and Ballard, Andrew J and Bambrick, Joshua and others},
  journal={Nature},
  volume={630},
  number={8016},
  pages={493--500},
  year={2024},
  publisher={Nature Publishing Group UK London}
}

@article{jain2017biophysical,
  title={Biophysical properties of the clinical-stage antibody landscape},
  author={Jain, Tushar and Sun, Tingwan and Durand, St{\'e}phanie and Hall, Amy and Houston, Nga Rewa and Nett, Juergen H and Sharkey, Beth and Bobrowicz, Beata and Caffry, Isabelle and Yu, Yao and others},
  journal={Proceedings of the National Academy of Sciences},
  volume={114},
  number={5},
  pages={944--949},
  year={2017},
  publisher={National Academy of Sciences}
}

@article{smith1985filamentous,
  title={Filamentous fusion phage: novel expression vectors that display cloned antigens on the virion surface},
  author={Smith, George P},
  journal={Science},
  volume={228},
  number={4705},
  pages={1315--1317},
  year={1985},
  publisher={American Association for the Advancement of Science}
}

@article{mccafferty1990phage,
  title={Phage antibodies: filamentous phage displaying antibody variable domains},
  author={McCafferty, John and Griffiths, Andrew D and Winter, Greg and Chiswell, David J},
  journal={nature},
  volume={348},
  number={6301},
  pages={552--554},
  year={1990},
  publisher={Nature Publishing Group UK London}
}

@article{marks1992passing,
  title={By--passing immunization: building high affinity human antibodies by chain shuffling},
  author={Marks, James D and Griffiths, Andrew D and Malmqvist, Magnus and Clackson, Tim P and Bye, Jacqueline M and Winter, Greg},
  journal={Bio/technology},
  volume={10},
  number={7},
  pages={779--783},
  year={1992},
  publisher={Nature Publishing Group US New York}
}

@article{kohler1975continuous,
  title={Continuous cultures of fused cells secreting antibody of predefined specificity},
  author={K{\"o}hler, Georges and Milstein, Cesar},
  journal={nature},
  volume={256},
  number={5517},
  pages={495--497},
  year={1975},
  publisher={Nature Publishing Group UK London}
}

@article{green1994antigen,
  title={Antigen--specific human monoclonal antibodies from mice engineered with human Ig heavy and light chain YACs},
  author={Green, LL and Hardy, MC and Maynard-Currie, CE and Tsuda, Hi and Louie, DM and Mendez, MJ and Abderrahim, H and Noguchi, M and Smith, DH and Zeng, Y and others},
  journal={Nature genetics},
  volume={7},
  number={1},
  pages={13--21},
  year={1994},
  publisher={Nature Publishing Group US New York}
}

@article{hoie2024antifold,
  title={AntiFold: Improved structure-based antibody design using inverse folding},
  author={H{\o}ie, Magnus Haraldson and Hummer, Alissa M and Olsen, Tobias H and Aguilar-Sanjuan, Broncio and Nielsen, Morten and Deane, Charlotte M},
  journal={Bioinformatics Advances},
  volume={5},
  number={1},
  pages={vbae202},
  year={2025},
  publisher={Oxford University Press}
}

@article{abdiche2008determining,
  title={Determining kinetics and affinities of protein interactions using a parallel real-time label-free biosensor, the Octet},
  author={Abdiche, Yasmina and Malashock, Dan and Pinkerton, Alanna and Pons, Jaume},
  journal={Analytical biochemistry},
  volume={377},
  number={2},
  pages={209--217},
  year={2008},
  publisher={Elsevier}
}

@article{jumper2021highly,
  title={Highly accurate protein structure prediction with AlphaFold},
  author={Jumper, John and Evans, Richard and Pritzel, Alexander and Green, Tim and Figurnov, Michael and Ronneberger, Olaf and Tunyasuvunakool, Kathryn and Bates, Russ and {\v{Z}}{\'\i}dek, Augustin and Potapenko, Anna and others},
  journal={nature},
  volume={596},
  number={7873},
  pages={583--589},
  year={2021},
  publisher={Nature Publishing Group UK London}
}

@article{evansprotein,
  title={Protein complex prediction with AlphaFold-Multimer},
  author={Evans, Richard and O’Neill, Michael and Pritzel, Alexander and Antropova, Natasha and Senior, Andrew and Green, Tim and {\v{Z}}{\'\i}dek, Augustin and Bates, Russ and Blackwell, Sam and Yim, Jason and others},
  journal={biorxiv},
  pages={2021--10},
  year={2021},
  publisher={Cold Spring Harbor Laboratory}
}

@article{dunbrack2025ipsae,
  title={Rēs ipSAE loquuntur: What's wrong with AlphaFold's ipTM score and how to fix it},
  author={Dunbrack Jr., Roland L.},
  journal={bioRxiv},
  pages={2025.02.10.637595},
  year={2025},
  doi={10.1101/2025.02.10.637595}
}

@article{guerois2002predicting,
  title={Predicting changes in the stability of proteins and protein complexes: a study of more than 1000 mutations},
  author={Guerois, Raphael and Nielsen, Jens Erik and Serrano, Luis},
  journal={Journal of molecular biology},
  volume={320},
  number={2},
  pages={369--387},
  year={2002},
  publisher={Elsevier}
}

@article{Mitternacht2016,
  title={FreeSASA: An open source C library for solvent accessible surface area calculations},
  author={Mitternacht, Simon},
  journal={F1000Research},
  volume={5},
  pages={189},
  year={2016}
}

@article{myung2023understanding,
  title={Understanding the complementarity and plasticity of antibody--antigen interfaces},
  author={Myung, Yoochan and Pires, Douglas EV and Ascher, David B},
  journal={Bioinformatics},
  volume={39},
  number={7},
  pages={btad392},
  year={2023},
  publisher={Oxford University Press}
}

@article{koenig2017mutational,
  title={Mutational landscape of antibody variable domains reveals a switch modulating the interdomain conformational dynamics and antigen binding},
  author={Koenig, Patrick and Lee, Chingwei V and Walters, Benjamin T and Janakiraman, Vasantharajan and Stinson, Jeremy and Patapoff, Thomas W and Fuh, Germaine},
  journal={Proceedings of the National Academy of Sciences},
  volume={114},
  number={4},
  pages={E486--E495},
  year={2017},
  publisher={National Academy of Sciences}
}

@article{phillips2021binding,
  title={Binding affinity landscapes constrain the evolution of broadly neutralizing anti-influenza antibodies},
  author={Phillips, Angela M and Lawrence, Katherine R and Moulana, Alief and Dupic, Thomas and Chang, Jeffrey and Johnson, Milo S and Cvijovic, Ivana and Mora, Thierry and Walczak, Aleksandra M and Desai, Michael M},
  journal={Elife},
  volume={10},
  pages={e71393},
  year={2021},
  publisher={eLife Sciences Publications Limited}
}

@article{warszawski2020correction,
  title={Correction: Optimizing antibody affinity and stability by the automated design of the variable light-heavy chain interfaces},
  author={Warszawski, Shira and Borenstein Katz, Aliza and Lipsh, Rosalie and Khmelnitsky, Lev and Ben Nissan, Gili and Javitt, Gabriel and Dym, Orly and Unger, Tamar and Knop, Orli and Albeck, Shira and others},
  journal={PLoS computational biology},
  volume={16},
  number={10},
  pages={e1008382},
  year={2020},
  publisher={Public Library of Science San Francisco, CA USA}
}

@article{engelhart2022dataset,
  title={A dataset comprised of binding interactions for 104,972 antibodies against a SARS-CoV-2 peptide},
  author={Engelhart, Emily and Emerson, Ryan and Shing, Leslie and Lennartz, Chelsea and Guion, Daniel and Kelley, Mary and Lin, Charles and Lopez, Randolph and Younger, David and Walsh, Matthew E},
  journal={Scientific Data},
  volume={9},
  number={1},
  pages={653},
  year={2022},
  publisher={Nature Publishing Group UK London}
}

@article{li2023machine,
  title={Machine learning optimization of candidate antibody yields highly diverse sub-nanomolar affinity antibody libraries},
  author={Li, Lin and Gupta, Esther and Spaeth, John and Shing, Leslie and Jaimes, Rafael and Engelhart, Emily and Lopez, Randolph and Caceres, Rajmonda S and Bepler, Tristan and Walsh, Matthew E},
  journal={Nature communications},
  volume={14},
  number={1},
  pages={3454},
  year={2023},
  publisher={Nature Publishing Group UK London}
}

@article{Shanehsazzadeh2023.01.08.523187,
  title={Unlocking de novo antibody design with generative artificial intelligence},
  author={Shanehsazzadeh, Amir and Bachas, Sharrol and McPartlon, Matt and Kasun, George and Sutton, John M and Steiger, Andrea K and Shuai, Richard and Kohnert, Christa and Rakocevic, Goran and Gutierrez, Jahir M and others},
  journal={BioRxiv},
  pages={2023--01},
  year={2023},
  publisher={Cold Spring Harbor Laboratory}
}

@article{clark2006Affinity,
  title={Affinity enhancement of an in vivo matured therapeutic antibody using structure-based computational design},
  author={Clark, Louis A and Boriack-Sjodin, P Ann and Eldredge, John and Fitch, Christopher and Friedman, Bethany and Hanf, Karl JM and Jarpe, Matthew and Liparoto, Stefano F and Li, You and Lugovskoy, Alexey and others},
  journal={Protein science},
  volume={15},
  number={5},
  pages={949--960},
  year={2006},
  publisher={Wiley Online Library}
}

@article{thakkar2019balancing,
  title={Balancing sensitivity and specificity in distinguishing TCR groups by CDR sequence similarity},
  author={Thakkar, Neerja and Bailey-Kellogg, Chris},
  journal={BMC bioinformatics},
  volume={20},
  number={1},
  pages={241},
  year={2019},
  publisher={Springer}
}

@article{van2024fast,
  title={Fast and accurate protein structure search with Foldseek},
  author={Van Kempen, Michel and Kim, Stephanie S and Tumescheit, Charlotte and Mirdita, Milot and Lee, Jeongjae and Gilchrist, Cameron LM and S{\"o}ding, Johannes and Steinegger, Martin},
  journal={Nature biotechnology},
  volume={42},
  number={2},
  pages={243--246},
  year={2024},
  publisher={Nature Publishing Group US New York}
}

@article{barrio2023clustering,
  title={Clustering predicted structures at the scale of the known protein universe},
  author={Barrio-Hernandez, Inigo and Yeo, Jingi and J{\"a}nes, J{\"u}rgen and Mirdita, Milot and Gilchrist, Cameron LM and Wein, Tanita and Varadi, Mihaly and Velankar, Sameer and Beltrao, Pedro and Steinegger, Martin},
  journal={Nature},
  volume={622},
  number={7983},
  pages={637--645},
  year={2023},
  publisher={Nature Publishing Group UK London}
}

@article{meier2021language,
  title={Language models enable zero-shot prediction of the effects of mutations on protein function},
  author={Meier, Joshua and Rao, Roshan and Verkuil, Robert and Liu, Jason and Sercu, Tom and Rives, Alex},
  journal={Advances in neural information processing systems},
  volume={34},
  pages={29287--29303},
  year={2021}
}

@article{hie2024efficient,
  title={Efficient evolution of human antibodies from general protein language models},
  author={Hie, Brian L and Shanker, Varun R and Xu, Duo and Bruun, Theodora UJ and Weidenbacher, Payton A and Tang, Shaogeng and Wu, Wesley and Pak, John E and Kim, Peter S},
  journal={Nature biotechnology},
  volume={42},
  number={2},
  pages={275--283},
  year={2024},
  publisher={Nature Publishing Group US New York}
}

@article{nijkamp2023progen2,
  title={Progen2: exploring the boundaries of protein language models},
  author={Nijkamp, Erik and Ruffolo, Jeffrey A and Weinstein, Eli N and Naik, Nikhil and Madani, Ali},
  journal={Cell systems},
  volume={14},
  number={11},
  pages={968--978},
  year={2023},
  publisher={Elsevier}
}

@article{ferruz2022protgpt2,
  title={ProtGPT2 is a deep unsupervised language model for protein design},
  author={Ferruz, Noelia and Schmidt, Steffen and H{\"o}cker, Birte},
  journal={Nature communications},
  volume={13},
  number={1},
  pages={4348},
  year={2022},
  publisher={Nature Publishing Group UK London}
}

@article {Burbach2025.02.27.640641,
  title={A curriculum learning approach to training antibody language models},
  author={Burbach, Sarah M and Briney, Bryan},
  journal={bioRxiv},
  year={2025}
}

@article{olsen2022Observed,
  title={Observed Antibody Space: A diverse database of cleaned, annotated, and translated unpaired and paired antibody sequences},
  author={Olsen, Tobias H and Boyles, Fergus and Deane, Charlotte M},
  journal={Protein Science},
  volume={31},
  number={1},
  pages={141--146},
  year={2022},
  publisher={Wiley Online Library}
}

@article{wang2018local,
  title={Local and global anatomy of antibody-protein antigen recognition},
  author={Wang, Meryl and Zhu, David and Zhu, Jianwei and Nussinov, Ruth and Ma, Buyong},
  journal={Journal of Molecular Recognition},
  volume={31},
  number={5},
  pages={e2693},
  year={2018},
  publisher={Wiley Online Library}
}

@article{orlandi2020antigen,
  title={Antigen-induced allosteric changes in a human IgG1 Fc increase low-affinity Fc$\gamma$ receptor binding},
  author={Orlandi, Chiara and Deredge, Daniel and Ray, Krishanu and Gohain, Neelakshi and Tolbert, William and DeVico, Anthony L and Wintrode, Patrick and Pazgier, Marzena and Lewis, George K},
  journal={Structure},
  volume={28},
  number={5},
  pages={516--527},
  year={2020},
  publisher={Elsevier}
}

@article{mackerell1998encyclopedia,
  title={The encyclopedia of computational chemistry},
  author={MacKerell, AD and Brooks, B and Brooks, CL and Nilsson, L and Roux, B and Won, Y and Karplus, M and Schleyer, P v R},
  journal={Schleyer, PVR},
  pages={271--277},
  year={1998}
}

@article{looger2001generalized,
  title={Generalized dead-end elimination algorithms make large-scale protein side-chain structure prediction tractable: implications for protein design and structural genomics},
  author={Looger, Loren L and Hellinga, Homme W},
  journal={Journal of molecular biology},
  volume={307},
  number={1},
  pages={429--445},
  year={2001},
  publisher={Elsevier}
}

@article{wisz2003empirical,
  title={An empirical model for electrostatic interactions in proteins incorporating multiple geometry-dependent dielectric constants},
  author={Wisz, Michael S and Hellinga, Homme W},
  journal={Proteins: Structure, Function, and Bioinformatics},
  volume={51},
  number={3},
  pages={360--377},
  year={2003},
  publisher={Wiley Online Library}
}

@article{kangas1998optimizing,
  title={Optimizing electrostatic affinity in ligand--receptor binding: Theory, computation, and ligand properties},
  author={Kangas, Erik and Tidor, Bruce},
  journal={The Journal of chemical physics},
  volume={109},
  number={17},
  pages={7522--7545},
  year={1998},
  publisher={American Institute of Physics}
}

@phdthesis{hanf2002protein,
  title={Protein design with hierarchical treatment of solvation and electrostatics},
  author={Hanf, Karl John Mortley},
  year={2002},
  school={Massachusetts Institute of Technology}
}

@article{binding_gym,
  title={Bindinggym: A large-scale mutational dataset toward deciphering protein-protein interactions},
  author={Lu, Wei and Zhang, Jixian and Gu, Ming and Zheng, Shuangjia},
  journal={bioRxiv},
  pages={2024--12},
  year={2024},
  publisher={Cold Spring Harbor Laboratory}
}

@article{minot2024,
  title={Meta learning addresses noisy and under-labeled data in machine learning-guided antibody engineering},
  author={Minot, Mason and Reddy, Sai T},
  journal={Cell systems},
  volume={15},
  number={1},
  pages={4--18},
  year={2024},
  publisher={Elsevier}
}

@article{liu2025abrank,
  title={Abrank: A benchmark dataset and metric-learning framework for antibody-antigen affinity ranking},
  author={Liu, Chunan and Pelissier, Aurelien and Shao, Yanjun and Denzler, Lilian and Martin, Andrew CR and Paige, Brooks and Mart{\'\i}nez, Mar{\'\i}a Rodr{\'\i}guez},
  journal={arXiv preprint arXiv:2506.17857},
  year={2025}
}

@article{cia2025abagym,
  title={{AbAgym}: a well-curated dataset for the mutational analysis of antibody--antigen complexes},
  author={Cia, Gabriel and Li, Dong and Poblete, Sim{\'o}n and Rooman, Marianne and Pucci, Fabrizio},
  journal={mAbs},
  volume={17},
  number={1},
  pages={2592421},
  year={2025},
  publisher={Taylor \& Francis},
  doi={10.1080/19420862.2025.2592421}
}

@article{janusz2025abdesign,
  title={{AbDesign}: database of point mutants of antibodies with associated structures reveals poor generalization of binding predictions from machine learning models},
  author={Janusz, Bartosz and Chomicz, Dawid and Demharter, Samuel and Arts, Marloes and de Kanter, Jurrian and Wilke, Yano and Britze, Helena and Wrobel, Sonia and Gaw{\l}owski, Tomasz and Dudzic, Pawel and others},
  journal={mAbs},
  volume={17},
  number={1},
  pages={2567319},
  year={2025},
  publisher={Taylor \& Francis},
  doi={10.1080/19420862.2025.2567319}
}

@article{passaro2025boltz,
  title={Boltz-2: Towards accurate and efficient binding affinity prediction},
  author={Passaro, Saro and Corso, Gabriele and Wohlwend, Jeremy and Reveiz, Mateo and Thaler, Stephan and Somnath, Vignesh Ram and Getz, Noah and Portnoi, Tally and Roy, Julien and Stark, Hannes and others},
  journal={bioRxiv},
  year={2025}
}

@article{hou2026esmdance,
  title={Protein language models trained on biophysical dynamics inform mutation effects},
  author={Hou, Chao and Zhao, Haiqing and Shen, Yufeng},
  journal={Proceedings of the National Academy of Sciences},
  volume={123},
  number={4},
  pages={e2530466123},
  year={2026},
  doi={10.1073/pnas.2530466123}
}
\bibliographystyle{unsrtnat}

\clearpage
\appendix
\section{Supplementary Materials}
\ifSubfilesClassLoaded{}{\renewcommand{\thefigure}{S\arabic{figure}}
  \renewcommand{\thetable}{S\arabic{table}}
  \setcounter{figure}{0}
  \setcounter{table}{0}
}

\etocsettocdepth{subsection}
\etocsetlocaltop.toc{section}
\localtableofcontents

\setcounter{secnumdepth}{2} \renewcommand\thesubsection{\arabic{subsection}} \setcounter{subsection}{0} \renewcommand\theHsubsection{supp.\arabic{subsection}} \renewcommand\theHsubsubsection{supp.\arabic{subsection}.\arabic{subsubsection}}

\subsection{Code, Dataset, and Computational Resources}
\label{sup:resources}
Code repository and leaderboard are available in \url{https://github.com/MSBMI-SAFE/AbBiBench}.
The benchmarking dataset is available in \url{https://huggingface.co/datasets/AbBibench/Antibody_Binding_Benchmark_Dataset}. Model training is not required in this study. All inference tasks are conducted on a single NVIDIA H100 80GB GPU per model.
\lstset{
  language=Python,
  basicstyle=\ttfamily\small,
  keywordstyle=\color{blue},
  stringstyle=\color{teal},
  commentstyle=\color{gray},
  showstringspaces=false,
  breaklines=true,
  frame=single,
  captionpos=b
}

\begin{lstlisting}[caption={Accessing the AbBibench dataset using Croissant}, label={lst:abbench}]
import requests
from huggingface_hub.file_download import build_hf_headers
from mlcroissant import Dataset

# Login using e.g. `huggingface-cli login` to access this dataset
headers = build_hf_headers()  # handles authentication
jsonld = requests.get(
    "https://huggingface.co/api/datasets/AbBibench/Antibody_Binding_Benchmark_Dataset/croissant",
    headers=headers
).json()
ds = Dataset(jsonld=jsonld)
records = ds.records("default")
\end{lstlisting}

\subsection{Binding Affinity Data Details}
\label{sup:data_details}

\begin{figure}[htbp] 
  \centering
  \captionsetup{font=footnotesize}
  \captionsetup[sub]{font=footnotesize}

  \begin{minipage}[t]{0.49\linewidth}
    \centering                           
    \includegraphics[
      height=24mm,
      trim=0 2mm 1mm 1mm, clip
    ]{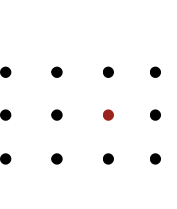}
    \subcaption{}\label{fig:lead_antibodies_a}
  \end{minipage}\hfill
  \begin{minipage}[t]{0.49\linewidth}
    \centering                           
    \raisebox{-0.5mm}[0pt][0pt]{\includegraphics[
        height=24mm,
        trim=0 1mm 0 1mm, clip
      ]{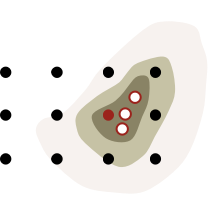}}
    \subcaption{}\label{fig:lead_antibodies_b}
  \end{minipage}

  \vspace{-3pt}        
  \caption{\footnotesize Antibody design space. (a) Traditional \textit{in vitro} screening explores a limited antibody library. (b) Protein machine learning models explore broader mutational space. \textcolor{black}{$\bullet$}: tested antibodies, \textcolor{red}{$\bullet$}: lead antibody, \textcolor{red}{$\circ$}: model-generated variants.}
  \label{fig:lead_antibodies_wrap}
\end{figure}

To obtain a robust experimental baseline for AbBiBench, we collated 16 binding-affinity assays drawn from open-source studies. Selection was guided by two criteria: (i) maximizing Ab-Ag diversity and (ii) having enough mutants per assay to ensure statistical power of our correlation analyses. For each dataset we transformed the reported binding metric to common scales and computed Spearman correlation between experimental binding scores and model log-likelihoods, providing a zero-shot test of each model’s ability to recognize affinity-improving mutations. The following sections outline, per study how raw measurements were curated and converted into the benchmark scores.
\label{supplementary:dataset}
\newcolumntype{L}[1]{>{\raggedright\arraybackslash}p{#1}}

\begin{sidewaystable}[p]
\centering
{\color{black}
\resizebox{0.98\textheight}{!}{\begin{tabular}{@{}llllllll@{}}
\toprule
\textbf{Assay} & \textbf{Antibody--antigen} & \textbf{Variants} & \textbf{Mutation scope} & \textbf{Readout} & \textbf{Structural template (chains; lengths)} & \textbf{Library construction} & \textbf{Study} \\ \midrule
4fqi\_h1   & CR9114--H1N1 HA       & HC: 65,094                    & CDR + framework & $-\!\log K_d$           & 4FQI (H/L/A,B; 121/109/324/176) & Germline-reverted yeast display & \cite{phillips2021binding} \\
4fqi\_h3   & CR9114--H3N2 HA       & HC: 65,535                    & CDR + framework & $-\!\log K_d$           & 4FQI (H/L/A,B; 121/109/324/176) & Germline-reverted yeast display & \cite{phillips2021binding} \\
3gbn\_h1   & CR6261--H1N1 HA       & HC: 1,887                     & CDR + framework & $-\!\log K_d$           & 3GBN (H/L/A,B; 121/109/328/173) & Germline-reverted yeast display & \cite{phillips2021binding} \\
3gbn\_h9   & CR6261--H9N2 HA       & HC: 1,842                     & CDR + framework & $-\!\log K_d$           & 3GBN (H/L/A,B; 121/109/328/173) & Germline-reverted yeast display & \cite{phillips2021binding} \\
aayl49     & AAYL49--Spike HR2     & HC: 4,312                     & CDR             & $-\!\log K_d$           & AlphaFold 3 (B/C/A; 118/113/14) & Phage-display DMS & \cite{engelhart2022dataset} \\
aayl49\_ML & AAYL49\_ML--Spike HR2 & HC: 8,953                    & CDR             & $-\!\log K_d$           & AlphaFold 3 (B/C/A; 118/113/14) & ML-generated scFv, phage display & \cite{li2023machine} \\
aayl50     & AAYL50--Spike HR2     & LC: 11,473                    & CDR             & $-\!\log K_d$           & AlphaFold 3 (B/C/A; 118/113/14) & Phage-display DMS & \cite{engelhart2022dataset} \\
aayl51     & AAYL51--Spike HR2     & HC: 4,320                     & CDR             & $-\!\log K_d$           & AlphaFold 3 (B/C/A; 119/115/14) & Phage-display DMS & \cite{engelhart2022dataset} \\
aayl52     & AAYL52--Spike HR2     & LC: 13,324                    & CDR             & $-\!\log K_d$           & AlphaFold 3 (B/C/A; 119/115/14) & Phage-display DMS & \cite{engelhart2022dataset} \\
2fjg       & G6.31--VEGF           & HC: 2,223; LC: 2,014          & CDR + framework & log enrichment          & 2FJG (H/L/V; 118/107/95)        & Phage-display DMS & \cite{koenig2017mutational} \\
1mlc       & D44.1--lysozyme       & HC: 1,229; LC: 865            & CDR + framework & log enrichment          & 1MLC (B/A/E; 117/109/129)       & Yeast-display DMS & \cite{warszawski2020correction} \\
1n8z       & trastuzumab--HER2     & HC: 419                       & CDR             & $-\!\log K_d$           & 1N8Z (B/A/C; 121/109/581)       & Zero-shot generative design & \cite{Shanehsazzadeh2023.01.08.523187} \\
1mhp       & AQC2--integrin-$\alpha$1 & HC: 37; LC: 25; both: 5    & CDR + framework & $-\!\log K_d$           & 1MHP (H/L/A; 118/107/184)       & In-silico biophysical library & \cite{Clark2006Affinity} \\
4d5\_her2  & 4D5--HER2             & Both: 2,080                   & CDR             & log enrichment          & PDB template (B/A/C; 121/109/581) & Yeast-display DMS & \cite{minot2024} \\
5a12\_ang2 & 5A12--Ang2            & HC: 796; LC: 104; both: 43    & CDR             & log enrichment          & PDB template (H/L/A; 215/213/220) & Yeast-display DMS & \cite{minot2024} \\
5a12\_vegf & 5A12--VEGF            & HC: 5,300; LC: 228; both: 24,452 & CDR          & log enrichment          & PDB template (H/L/D; 211/213/99)  & Yeast-display DMS & \cite{minot2024} \\
\bottomrule
\end{tabular}}
}
\caption{Complete experimental and structural metadata for the 16 AbBiBench assays. Structural-template entries list the PDB identifier or AlphaFold 3 source, followed by antibody heavy-chain, light-chain and antigen-chain identifiers and their sequence lengths.}
\label{tab:dataset_details}
\end{sidewaystable}

\paragraph{Influenza data}
For our benchmarking study, we derived data (processed by \cite{shanker2024unsupervised}) from an experiment investigating the binding affinity landscapes of two broadly neutralizing anti-influenza antibodies (bnAbs), CR6261 (PDB ID: 3GBN) and CR9114 (PDB ID: 4FQI) \cite{phillips2021binding}. Combinatorially complete libraries of all evolutionary intermediates were constructed for each bnAb’s heavy chain, spanning 11 mutations for CR6261 and 16 for CR9114. After removing entries missing dissociation constant $K_d$ values, we used $-log(K_d)$ (in M) for benchmarking. This yielded 1,887 data points for the H1N1 subtype and 1,842 for the H9H2 subtype (both against CR6261), as well as 65,094 H1N1 and 65,535 H3N2 data points (both against CR9114). 

\paragraph{VEGF data}
Mutational data for anti-VEGF antibody was sourced from two different studies. We collected data from \cite{shanker2024unsupervised}, which reported deep mutational scanning results from \cite{koenig2017mutational} to systematically analyze the impact of antibody mutations remote to the antigen-binding site. For benchmarking, we derived mutational data from positions 2–113 of the variable heavy-chain region (Kabat numbering), obtained from the G6.31 Fab phage display study for the models' input along with the structure (PDB ID: 2FJG). We collected the corresponding binding enrichment ratios (ER), defined as the log, frequency ratio of each mutation post-selection relative to pre-selection. Here, higher ER values indicate mutations having enhanced fitness, primarily reflecting increased binding affinity toward the VEGF antigen. The resulting datasets contained 2,223 heavy chain variants (subset 2fjg) and 2,014 light chain variants (subset g6\_LC). 

\paragraph{SARS-CoV-2 data} 
This dataset comprises binding assay scores of human antibodies targeting the HR2 region of the SARS-CoV-2 peptide (spike protein) developed for the purpose of benchmarking machine learning models \cite{engelhart2022dataset}. Through phage display, three antibodies were identified as binders from which the antibody library was designed by making point mutations in the CDR regions, with the curated benchmark variants containing up to 5 substitutions per sequence. Our benchmarking study focuses on two mutated variable heavy chain sequences (AAYL49 and AAYL51 assays) and two mutated variable light chain sequences (AAYL50 and AAYL52). Disassociation constant (Kd) values were provided in nM which was converted to M to derive the -log (Kd) values used for correlation (keeping consistency with \cite{shanker2024unsupervised}). After averaging triplicate results and removing negative controls, we collated 4,312 datapoints from AAYL49, 4,320 from AAYL51, 11,473 from AAYL50, and 13,324 from AAYL52.

An extension to this study involved utilizing the previously described SARS-CoV-2 data to train a Bayesian language model for scFv design. The AAYL49 assay-trained model was used to generate scFv libraries enriched for improved affinity. From these libraries, we specifically extracted mutations introduced in the heavy-chain variable region that were not present in the phage display-derived AAYL49 library, resulting in dataset AAYL49\_ML of 8,953 mutated heavy chain sequences. Due to the lack of structural information from these studies, we used AlphaFold 3 \cite{abramson2024accurate} to generate the antibody-antigen complexes for the SARS-CoV-2 data derived from this study. 

\paragraph{Lysozyme data}
Similar to the anti-VEGF antibody study, this study investigates the mutational tolerance of a variable fragment of an anti-lysozyme antibody \cite{warszawski2020correction}. A deep mutational scanning approach was applied to a yeast-display library, where mutations were introduced at 68 positions of the heavy-chain variable region (PDB ID: 1MLC), each position mutated to all 20 standard amino acids; the curated benchmark dataset contains predominantly single-point mutations with a small number of double-mutant variants (max 2 substitutions per sequence). The resulting mutation data were converted into tabular format by reconstructing the mutated heavy- and light-chain sequences and noting their corresponding log-enrichment ratios. For benchmarking, we utilized this mutational tolerance map across two benchmarking subsets: 1mlc (HC: 1,229 variants) and 1mlc\_LC (LC: 865 variants), yielding a total of 2,094 ER values. 

\paragraph{HER2 data}
The anti-HER2 data was sourced from two studies. The first aimed to develop generative AI models capable of producing antibody binders without iterative optimization (zero-shot generation) \cite{Shanehsazzadeh2023.01.08.523187}. The authors demonstrated their approach by computationally designing the heavy-chain complementarity-determining region 3 (HCDR3) of the HER2-targeting antibody trastuzumab (PDB ID: 1N8Z). For our correlation analysis, we filtered this dataset to include only mutated sequences with an HCDR3 length matching the wild-type trastuzumab sequence. The reported binding affinity (Kd) values, originally measured in nanomolar (nM), were converted to molar (M), and subsequently transformed to negative log-scale $-log(K_d)$. In total, 419 mutated HCDR3 sequences from this dataset were utilized for our benchmarking study. 

\paragraph{Integrin-$\alpha$-1 data}
In this study from 2006, researchers sought to improve the binding affinity of the AQC2 antibody fragment to the I-domain of integrin VLA1 \cite{Clark2006Affinity}. Their approach involved utilizing structure-based computational methods to propose mutants through side chain repacking \cite{hanf2002protein, looger2001generalized, wisz2003empirical} and electrostatic optimization \cite{kangas1998optimizing, mackerell1998encyclopedia}. Libraries were computationally generated by varying nearly all antigen-contacting residue positions in both antibody heavy and light chains (PDB ID: 1MHP). Successful single mutations identified experimentally were then combined to further increase affinity. Binding affinity ($K_d^{\mathrm{mut}}$) was estimated as the fold change affinity relative to wildtype using a competition ELISA. This resulted in a total of 67 variants with mutations in either or both chains.

\paragraph{Ang2}
This dataset (ID: 5a12\_ang2), sourced from \cite{minot2024} and processed by BindingGYM following the same workflow described earlier, contained 943 variants across the heavy and light chains.

\par\noindent\par

\par\noindent\par

\subsection{Computational Evaluation Metrics}
\label{sup:sec:computationalMetrics}

\begin{figure}[htbp]
  \centering
  \includegraphics[width=0.9\textwidth]{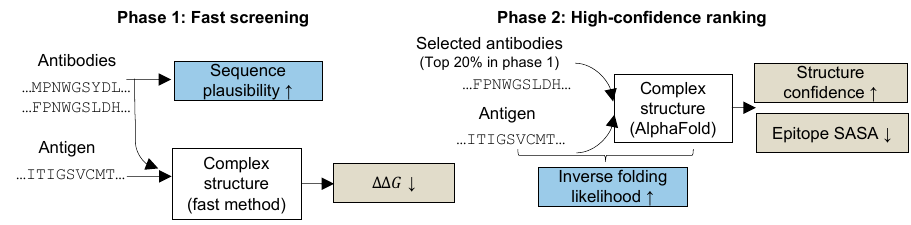}
  \caption{Evaluation of new antibody designs. Left: fast screening based on sequence plausibility and predicted binding-energy change. Right: high-confidence ranking based on AlphaFold 3-predicted complex structures, using structure confidence, epitope SASA, and inverse-folding likelihood.}
  \label{fig:evaluating}
\end{figure}

\begin{itemize}
    \item \textbf{Binding Energy} FoldX is a computational force field that calculates the free energy of protein-protein interactions by evaluating multiple physical energy terms. These terms include van der Waals forces between atoms, both inter- and intra-molecular hydrogen bonding, electrostatic interactions between charged groups, and additional contributions from solvation and entropy \cite{guerois2002predicting}. The algorithm is widely used to predict the impact of mutations on protein stability, defined as the difference in Gibbs free energy ($\Delta \Delta G$) between mutant and wild-type proteins ($\Delta G_{\text{variant}} - \Delta G_{\text{wild type}}$). In this study, we used FoldX's \texttt{analyseComplexChains} command to quantify the binding energy differences between antigen and antibody chains within their molecular complex. A lower value of $\Delta \Delta G$ indicates a stronger binding upon mutation.

    \item \textbf{Epitope SASA} \label{item:sasa} Solvent Accessible Surface Area (SASA) is a computational measure that quantifies the exposure of protein residues to the surrounding solvent. Using the FreeSASA Python module \cite{Mitternacht2016}, we calculated the surface accessibility of epitope residues in both wild-type and mutant antibody-antigen complexes. A decrease in solvent accessible surface area typically indicates tighter packing at the antibody-antigen interface, which often correlates with stronger binding affinity. We defined the epitope as antigen residues located within 5 \AA\ of the antibody chain, as this distance threshold effectively captures the antibody-antigen binding interface \cite{abramson2024accurate, myung2023understanding}. We employed relative SASA values to enable meaningful comparisons across different protein structures, representing the ratio of actual surface area to the maximum possible surface area for each residue type. The overall change in epitope accessibility was quantified as:
    \[
    \Delta \text{SASA} = \sum \left( \text{relative SASA}_{\text{variant}} \right) - \sum \left( \text{relative SASA}_{\text{wild type}} \right),
    \]
    where the summation is performed over all epitope residues.

    \item \textbf{cdrDist} Following the approach proposed by Thakkar and Bailey-Kellogg~\cite{thakkar2019balancing}, we compute the sequence distance using the normalized Smith–Waterman alignment score.  Let \( S_{\text{wild type}} \)  denote the wildtype CDR-H3 sequence and \( S_\text{variant} \)  the mutant CDR-H3 sequence. The distance between two sequences is defined as:

    \[
    \text{CDRdist}(S_{\text{wild type}}, S_\text{variant}) = 1 - \frac{\text{SW}(S_{\text{wild type}}, S_\text{variant})^2}{\text{SW}(S_{\text{wild type}}, S_\text{wild type}) \cdot \text{SW}(S_{\text{variant}}, S_\text{variant})}
    \]

    where \( \text{SW}(X, Y) \) denotes the Smith–Waterman local alignment score between sequences \( X \) and \( Y \). This formulation penalizes dissimilar alignments more heavily and ensures that the distance is normalized with respect to the self-alignment scores of the sequences being compared. The distance lies in the interval [0, 1], where 0 indicates identical sequences.
 
    \item \textbf{cdrRMSD} We used the Kabsch algorithm to superimpose the $C_\alpha$ atoms of the residues comprising each sampled CDR-H3 loop onto the corresponding CDR-H3 region of the wild-type structure, and calculated the resulting root-mean-square deviation (RMSD). 

    \item \textbf{Affinity Fold Change} The fold change was calculated by taking the difference between the pK\textsubscript{d} values of the mutant and wild-type antibodies (i.e., $-\log_{10} K_\text{d}$), and exponentiating the result as $10^{(\text{p}K_\text{d, mutant} - \text{p}K_\text{d, wild-type})}$. This yields the ratio of the dissociation constants $K_\text{d}$ between the wild-type and mutant antibodies.

    \item \textbf{Fivefold Precision@$k$, dataset prevalence, and oracle ceiling.} For assay $a$ and model $m$, we rank variants by model score and define
    \[
    P_{m,a}@k=\frac{1}{k}\sum_{i=1}^{k}\mathbb{I}\!\left(\mathrm{p}K_{d,i}\geq \mathrm{p}K_{d,\mathrm{WT}}+\log_{10}5\right),
    \]
    for $k\in\{10,20,50,100\}$.  Results are averaged with equal weight across 10 of the 11 direct-$K_d$ assays. The 1mhp assay was excluded because it contained fewer than 100 scored variants; across the retained data, every assay--method pair contained at least 419 scored variants. For each assay, Wilson intervals quantify binomial uncertainty in the observed top-$k$ hit proportion. If assay $a$ contains $H_a$ fivefold-improved variants among $N_a$ measured variants, the equal-assay random-ranking baseline and finite-dataset oracle ceiling are
    \[
    P_{\mathrm{random}}=\frac{1}{A}\sum_{a=1}^{A}\frac{H_a}{N_a},\qquad
    P_{\mathrm{oracle}}@k=\frac{1}{A}\sum_{a=1}^{A}\frac{\min(H_a,k)}{k}.
    \]
    The random baseline is therefore independent of $k$, whereas the oracle ceiling can decrease as $k$ exceeds the number of observed hits in individual assays. In the present analysis, $P_{\mathrm{random}}=0.268$ and the oracle ceilings are 0.900, 0.900, 0.812, and 0.717 for $k=10,20,50,$ and $100$, respectively.  Wild-type p$K_d$ values were available for 11 direct-$K_d$ assays, but Precision@$k$ excluded 1mhp. The values used for the retained 10 assays were 8.712198270069774 (1n8z), 7.0 (3gbn\_h1), 7.0 (3gbn\_h9), 8.424622015254007 (4fqi\_h1), 6.0 (4fqi\_h3), 8.848776 (aayl49), 8.848776 (aayl49\_ML), 8.773558 (aayl50), 8.879555 (aayl51), and 8.648002 (aayl52).

\end{itemize}

\begin{table}[p]
\centering
\tiny
\setlength{\tabcolsep}{1.8pt}
\renewcommand{\arraystretch}{1.08}
\resizebox{\textwidth}{!}{\begin{tabular}{@{}l*{22}{c}@{}}
\toprule
\textbf{Assay} & \rotatebox{65}{\textbf{ProteinMPNN}} & \rotatebox{65}{\textbf{ESM-IF1}} & \rotatebox{65}{\textbf{Antifold}} & \rotatebox{65}{\textbf{SaProt}} & \rotatebox{65}{\textbf{Boltz-2}} & \rotatebox{65}{\textbf{CurrAb}} & \rotatebox{65}{\textbf{Delta G}} & \rotatebox{65}{\textbf{diffab}} & \rotatebox{65}{\textbf{AF3}} & \rotatebox{65}{\textbf{ESM-2}} & \rotatebox{65}{\textbf{dyMEAN\_fixbb}} & \rotatebox{65}{\textbf{ProSST}} & \rotatebox{65}{\textbf{diffab\_fixbb}} & \rotatebox{65}{\textbf{dyMEAN}} & \rotatebox{65}{\textbf{ESMDance}} & \rotatebox{65}{\textbf{SASA}} & \rotatebox{65}{\textbf{ESM-3}} & \rotatebox{65}{\textbf{MEAN}} & \rotatebox{65}{\textbf{MEAN\_fixbb}} & \rotatebox{65}{\textbf{ProtGPT-2}} & \rotatebox{65}{\textbf{progen2-large}} & \rotatebox{65}{\textbf{AntiBERTy}} \\
\midrule
\texttt{1n8z} & \cellcolor{blue!8}\textcolor{black}{0.00} & \cellcolor{blue!8}\textcolor{black}{0.00} & \cellcolor{blue!8}\textcolor{black}{0.00} & \cellcolor{blue!8}\textcolor{black}{0.00} & \cellcolor{blue!8}\textcolor{black}{0.00} & \cellcolor{blue!8}\textcolor{black}{0.00} & \cellcolor{blue!8}\textcolor{black}{0.00} & \cellcolor{blue!8}\textcolor{black}{0.00} & \cellcolor{blue!8}\textcolor{black}{0.00} & \cellcolor{blue!8}\textcolor{black}{0.00} & \cellcolor{blue!8}\textcolor{black}{0.00} & \cellcolor{blue!8}\textcolor{black}{0.00} & \cellcolor{blue!8}\textcolor{black}{0.00} & \cellcolor{blue!8}\textcolor{black}{0.00} & \cellcolor{blue!8}\textcolor{black}{0.00} & \cellcolor{blue!8}\textcolor{black}{0.00} & \cellcolor{blue!8}\textcolor{black}{0.00} & \cellcolor{blue!8}\textcolor{black}{0.00} & \cellcolor{blue!8}\textcolor{black}{0.00} & \cellcolor{blue!8}\textcolor{black}{0.00} & \cellcolor{blue!8}\textcolor{black}{0.00} & \cellcolor{blue!8}\textcolor{black}{0.00} \\
\texttt{3gbn\_h1} & \cellcolor{blue!85}\textcolor{white}{1.00} & \cellcolor{blue!85}\textcolor{white}{1.00} & \cellcolor{blue!85}\textcolor{white}{1.00} & \cellcolor{blue!85}\textcolor{white}{1.00} & \cellcolor{blue!85}\textcolor{white}{1.00} & \cellcolor{blue!85}\textcolor{white}{1.00} & \cellcolor{blue!85}\textcolor{white}{1.00} & \cellcolor{blue!85}\textcolor{white}{1.00} & \cellcolor{blue!70}\textcolor{white}{0.80} & \cellcolor{blue!85}\textcolor{white}{1.00} & \cellcolor{blue!77}\textcolor{white}{0.90} & \cellcolor{blue!77}\textcolor{white}{0.90} & \cellcolor{blue!85}\textcolor{white}{1.00} & \cellcolor{blue!77}\textcolor{white}{0.90} & \cellcolor{blue!77}\textcolor{white}{0.90} & \cellcolor{blue!85}\textcolor{white}{1.00} & \cellcolor{blue!62}\textcolor{white}{0.70} & \cellcolor{blue!85}\textcolor{white}{1.00} & \cellcolor{blue!85}\textcolor{white}{1.00} & \cellcolor{blue!23}\textcolor{black}{0.20} & \cellcolor{blue!8}\textcolor{black}{0.00} & \cellcolor{blue!8}\textcolor{black}{0.00} \\
\texttt{3gbn\_h9} & \cellcolor{blue!85}\textcolor{white}{1.00} & \cellcolor{blue!85}\textcolor{white}{1.00} & \cellcolor{blue!85}\textcolor{white}{1.00} & \cellcolor{blue!85}\textcolor{white}{1.00} & \cellcolor{blue!85}\textcolor{white}{1.00} & \cellcolor{blue!77}\textcolor{white}{0.90} & \cellcolor{blue!54}\textcolor{white}{0.60} & \cellcolor{blue!85}\textcolor{white}{1.00} & \cellcolor{blue!70}\textcolor{white}{0.80} & \cellcolor{blue!85}\textcolor{white}{1.00} & \cellcolor{blue!70}\textcolor{white}{0.80} & \cellcolor{blue!77}\textcolor{white}{0.90} & \cellcolor{blue!85}\textcolor{white}{1.00} & \cellcolor{blue!70}\textcolor{white}{0.80} & \cellcolor{blue!16}\textcolor{black}{0.10} & \cellcolor{blue!62}\textcolor{white}{0.70} & \cellcolor{blue!31}\textcolor{black}{0.30} & \cellcolor{blue!85}\textcolor{white}{1.00} & \cellcolor{blue!85}\textcolor{white}{1.00} & \cellcolor{blue!46}\textcolor{white}{0.50} & \cellcolor{blue!23}\textcolor{black}{0.20} & \cellcolor{blue!8}\textcolor{black}{0.00} \\
\texttt{4fqi\_h1} & \cellcolor{blue!85}\textcolor{white}{1.00} & \cellcolor{blue!85}\textcolor{white}{1.00} & \cellcolor{blue!85}\textcolor{white}{1.00} & \cellcolor{blue!85}\textcolor{white}{1.00} & \cellcolor{blue!85}\textcolor{white}{1.00} & \cellcolor{blue!85}\textcolor{white}{1.00} & \cellcolor{blue!85}\textcolor{white}{1.00} & \cellcolor{blue!62}\textcolor{white}{0.70} & \cellcolor{blue!77}\textcolor{white}{0.90} & \cellcolor{blue!70}\textcolor{white}{0.80} & \cellcolor{blue!54}\textcolor{white}{0.60} & \cellcolor{blue!85}\textcolor{white}{1.00} & \cellcolor{blue!39}\textcolor{black}{0.40} & \cellcolor{blue!62}\textcolor{white}{0.70} & \cellcolor{blue!85}\textcolor{white}{1.00} & \cellcolor{blue!39}\textcolor{black}{0.40} & \cellcolor{blue!62}\textcolor{white}{0.70} & \cellcolor{blue!16}\textcolor{black}{0.10} & \cellcolor{blue!23}\textcolor{black}{0.20} & \cellcolor{blue!70}\textcolor{white}{0.80} & \cellcolor{blue!16}\textcolor{black}{0.10} & \cellcolor{blue!8}\textcolor{black}{0.00} \\
\texttt{4fqi\_h3} & \cellcolor{blue!77}\textcolor{white}{0.90} & \cellcolor{blue!85}\textcolor{white}{1.00} & \cellcolor{blue!23}\textcolor{black}{0.20} & \cellcolor{blue!54}\textcolor{white}{0.60} & \cellcolor{blue!31}\textcolor{black}{0.30} & \cellcolor{blue!23}\textcolor{black}{0.20} & \cellcolor{blue!8}\textcolor{black}{0.00} & \cellcolor{blue!16}\textcolor{black}{0.10} & \cellcolor{blue!8}\textcolor{black}{0.00} & \cellcolor{blue!8}\textcolor{black}{0.00} & \cellcolor{blue!31}\textcolor{black}{0.30} & \cellcolor{blue!8}\textcolor{black}{0.00} & \cellcolor{blue!8}\textcolor{black}{0.00} & \cellcolor{blue!16}\textcolor{black}{0.10} & \cellcolor{blue!8}\textcolor{black}{0.00} & \cellcolor{blue!8}\textcolor{black}{0.00} & \cellcolor{blue!8}\textcolor{black}{0.00} & \cellcolor{blue!8}\textcolor{black}{0.00} & \cellcolor{blue!8}\textcolor{black}{0.00} & \cellcolor{blue!8}\textcolor{black}{0.00} & \cellcolor{blue!8}\textcolor{black}{0.00} & \cellcolor{blue!8}\textcolor{black}{0.00} \\
\texttt{aayl49} & \cellcolor{blue!39}\textcolor{black}{0.40} & \cellcolor{blue!23}\textcolor{black}{0.20} & \cellcolor{blue!23}\textcolor{black}{0.20} & \cellcolor{blue!8}\textcolor{black}{0.00} & \cellcolor{blue!8}\textcolor{black}{0.00} & \cellcolor{blue!16}\textcolor{black}{0.10} & \cellcolor{blue!8}\textcolor{black}{0.00} & \cellcolor{blue!8}\textcolor{black}{0.00} & \cellcolor{blue!23}\textcolor{black}{0.20} & \cellcolor{blue!8}\textcolor{black}{0.00} & \cellcolor{blue!8}\textcolor{black}{0.00} & \cellcolor{blue!8}\textcolor{black}{0.00} & \cellcolor{blue!8}\textcolor{black}{0.00} & \cellcolor{blue!8}\textcolor{black}{0.00} & \cellcolor{blue!8}\textcolor{black}{0.00} & \cellcolor{blue!8}\textcolor{black}{0.00} & \cellcolor{blue!16}\textcolor{black}{0.10} & \cellcolor{blue!16}\textcolor{black}{0.10} & \cellcolor{blue!16}\textcolor{black}{0.10} & \cellcolor{blue!8}\textcolor{black}{0.00} & \cellcolor{blue!8}\textcolor{black}{0.00} & \cellcolor{blue!8}\textcolor{black}{0.00} \\
\texttt{aayl49\_ML} & \cellcolor{blue!54}\textcolor{white}{0.60} & \cellcolor{blue!54}\textcolor{white}{0.60} & \cellcolor{blue!62}\textcolor{white}{0.70} & \cellcolor{blue!16}\textcolor{black}{0.10} & \cellcolor{blue!23}\textcolor{black}{0.20} & \cellcolor{blue!23}\textcolor{black}{0.20} & \cellcolor{blue!23}\textcolor{black}{0.20} & \cellcolor{blue!23}\textcolor{black}{0.20} & \cellcolor{blue!54}\textcolor{white}{0.60} & \cellcolor{blue!23}\textcolor{black}{0.20} & \cellcolor{blue!46}\textcolor{white}{0.50} & \cellcolor{blue!23}\textcolor{black}{0.20} & \cellcolor{blue!23}\textcolor{black}{0.20} & \cellcolor{blue!23}\textcolor{black}{0.20} & \cellcolor{blue!23}\textcolor{black}{0.20} & \cellcolor{blue!31}\textcolor{black}{0.30} & \cellcolor{blue!46}\textcolor{white}{0.50} & \cellcolor{blue!8}\textcolor{black}{0.00} & \cellcolor{blue!8}\textcolor{black}{0.00} & \cellcolor{blue!23}\textcolor{black}{0.20} & \cellcolor{blue!23}\textcolor{black}{0.20} & \cellcolor{blue!23}\textcolor{black}{0.20} \\
\texttt{aayl50} & \cellcolor{blue!23}\textcolor{black}{0.20} & \cellcolor{blue!8}\textcolor{black}{0.00} & \cellcolor{blue!16}\textcolor{black}{0.10} & \cellcolor{blue!23}\textcolor{black}{0.20} & \cellcolor{blue!8}\textcolor{black}{0.00} & \cellcolor{blue!16}\textcolor{black}{0.10} & \cellcolor{blue!8}\textcolor{black}{0.00} & \cellcolor{blue!8}\textcolor{black}{0.00} & \cellcolor{blue!8}\textcolor{black}{0.00} & \cellcolor{blue!8}\textcolor{black}{0.00} & \cellcolor{blue!16}\textcolor{black}{0.10} & \cellcolor{blue!8}\textcolor{black}{0.00} & \cellcolor{blue!23}\textcolor{black}{0.20} & \cellcolor{blue!16}\textcolor{black}{0.10} & \cellcolor{blue!8}\textcolor{black}{0.00} & \cellcolor{blue!8}\textcolor{black}{0.00} & \cellcolor{blue!8}\textcolor{black}{0.00} & \cellcolor{blue!8}\textcolor{black}{0.00} & \cellcolor{blue!8}\textcolor{black}{0.00} & \cellcolor{blue!8}\textcolor{black}{0.00} & \cellcolor{blue!31}\textcolor{black}{0.30} & \cellcolor{blue!16}\textcolor{black}{0.10} \\
\texttt{aayl51} & \cellcolor{blue!8}\textcolor{black}{0.00} & \cellcolor{blue!8}\textcolor{black}{0.00} & \cellcolor{blue!8}\textcolor{black}{0.00} & \cellcolor{blue!16}\textcolor{black}{0.10} & \cellcolor{blue!8}\textcolor{black}{0.00} & \cellcolor{blue!8}\textcolor{black}{0.00} & \cellcolor{blue!8}\textcolor{black}{0.00} & \cellcolor{blue!8}\textcolor{black}{0.00} & \cellcolor{blue!8}\textcolor{black}{0.00} & \cellcolor{blue!8}\textcolor{black}{0.00} & \cellcolor{blue!8}\textcolor{black}{0.00} & \cellcolor{blue!8}\textcolor{black}{0.00} & \cellcolor{blue!8}\textcolor{black}{0.00} & \cellcolor{blue!8}\textcolor{black}{0.00} & \cellcolor{blue!8}\textcolor{black}{0.00} & \cellcolor{blue!8}\textcolor{black}{0.00} & \cellcolor{blue!8}\textcolor{black}{0.00} & \cellcolor{blue!8}\textcolor{black}{0.00} & \cellcolor{blue!8}\textcolor{black}{0.00} & \cellcolor{blue!8}\textcolor{black}{0.00} & \cellcolor{blue!8}\textcolor{black}{0.00} & \cellcolor{blue!8}\textcolor{black}{0.00} \\
\texttt{aayl52} & \cellcolor{blue!8}\textcolor{black}{0.00} & \cellcolor{blue!8}\textcolor{black}{0.00} & \cellcolor{blue!8}\textcolor{black}{0.00} & \cellcolor{blue!8}\textcolor{black}{0.00} & \cellcolor{blue!8}\textcolor{black}{0.00} & \cellcolor{blue!8}\textcolor{black}{0.00} & \cellcolor{blue!8}\textcolor{black}{0.00} & \cellcolor{blue!8}\textcolor{black}{0.00} & \cellcolor{blue!8}\textcolor{black}{0.00} & \cellcolor{blue!8}\textcolor{black}{0.00} & \cellcolor{blue!8}\textcolor{black}{0.00} & \cellcolor{blue!8}\textcolor{black}{0.00} & \cellcolor{blue!8}\textcolor{black}{0.00} & \cellcolor{blue!8}\textcolor{black}{0.00} & \cellcolor{blue!8}\textcolor{black}{0.00} & \cellcolor{blue!8}\textcolor{black}{0.00} & \cellcolor{blue!8}\textcolor{black}{0.00} & \cellcolor{blue!8}\textcolor{black}{0.00} & \cellcolor{blue!8}\textcolor{black}{0.00} & \cellcolor{blue!8}\textcolor{black}{0.00} & \cellcolor{blue!8}\textcolor{black}{0.00} & \cellcolor{blue!8}\textcolor{black}{0.00} \\
\textbf{Mean} & \cellcolor{blue!47}\textcolor{white}{0.51} & \cellcolor{blue!45}\textcolor{black}{0.48} & \cellcolor{blue!40}\textcolor{black}{0.42} & \cellcolor{blue!39}\textcolor{black}{0.40} & \cellcolor{blue!35}\textcolor{black}{0.35} & \cellcolor{blue!35}\textcolor{black}{0.35} & \cellcolor{blue!30}\textcolor{black}{0.28} & \cellcolor{blue!31}\textcolor{black}{0.30} & \cellcolor{blue!33}\textcolor{black}{0.33} & \cellcolor{blue!31}\textcolor{black}{0.30} & \cellcolor{blue!33}\textcolor{black}{0.32} & \cellcolor{blue!31}\textcolor{black}{0.30} & \cellcolor{blue!30}\textcolor{black}{0.28} & \cellcolor{blue!30}\textcolor{black}{0.28} & \cellcolor{blue!25}\textcolor{black}{0.22} & \cellcolor{blue!26}\textcolor{black}{0.24} & \cellcolor{blue!26}\textcolor{black}{0.23} & \cellcolor{blue!25}\textcolor{black}{0.22} & \cellcolor{blue!26}\textcolor{black}{0.23} & \cellcolor{blue!21}\textcolor{black}{0.17} & \cellcolor{blue!14}\textcolor{black}{0.08} & \cellcolor{blue!10}\textcolor{black}{0.03} \\
\bottomrule
\end{tabular}}
\caption{Assay-level fivefold Precision@10 for all evaluated methods. Values are the proportions of the top 10 model-ranked variants achieving at least fivefold affinity improvement over the corresponding wild type. Cell shading ranges from light to dark blue for values from 0 to 1, using the same scale in all four tables. The final row is the equal-assay-weighted mean across the 10 direct-$K_d$ assays.}
\label{tab:fivefold_precision_k10}
\end{table}

\begin{table}[p]
\centering
\tiny
\setlength{\tabcolsep}{1.8pt}
\renewcommand{\arraystretch}{1.08}
\resizebox{\textwidth}{!}{\begin{tabular}{@{}l*{22}{c}@{}}
\toprule
\textbf{Assay} & \rotatebox{65}{\textbf{ProteinMPNN}} & \rotatebox{65}{\textbf{ESM-IF1}} & \rotatebox{65}{\textbf{Antifold}} & \rotatebox{65}{\textbf{SaProt}} & \rotatebox{65}{\textbf{Boltz-2}} & \rotatebox{65}{\textbf{CurrAb}} & \rotatebox{65}{\textbf{Delta G}} & \rotatebox{65}{\textbf{diffab}} & \rotatebox{65}{\textbf{AF3}} & \rotatebox{65}{\textbf{ESM-2}} & \rotatebox{65}{\textbf{dyMEAN\_fixbb}} & \rotatebox{65}{\textbf{ProSST}} & \rotatebox{65}{\textbf{diffab\_fixbb}} & \rotatebox{65}{\textbf{dyMEAN}} & \rotatebox{65}{\textbf{ESMDance}} & \rotatebox{65}{\textbf{SASA}} & \rotatebox{65}{\textbf{ESM-3}} & \rotatebox{65}{\textbf{MEAN}} & \rotatebox{65}{\textbf{MEAN\_fixbb}} & \rotatebox{65}{\textbf{ProtGPT-2}} & \rotatebox{65}{\textbf{progen2-large}} & \rotatebox{65}{\textbf{AntiBERTy}} \\
\midrule
\texttt{1n8z} & \cellcolor{blue!8}\textcolor{black}{0.00} & \cellcolor{blue!8}\textcolor{black}{0.00} & \cellcolor{blue!8}\textcolor{black}{0.00} & \cellcolor{blue!8}\textcolor{black}{0.00} & \cellcolor{blue!8}\textcolor{black}{0.00} & \cellcolor{blue!8}\textcolor{black}{0.00} & \cellcolor{blue!8}\textcolor{black}{0.00} & \cellcolor{blue!8}\textcolor{black}{0.00} & \cellcolor{blue!8}\textcolor{black}{0.00} & \cellcolor{blue!8}\textcolor{black}{0.00} & \cellcolor{blue!8}\textcolor{black}{0.00} & \cellcolor{blue!8}\textcolor{black}{0.00} & \cellcolor{blue!8}\textcolor{black}{0.00} & \cellcolor{blue!8}\textcolor{black}{0.00} & \cellcolor{blue!8}\textcolor{black}{0.00} & \cellcolor{blue!8}\textcolor{black}{0.00} & \cellcolor{blue!8}\textcolor{black}{0.00} & \cellcolor{blue!8}\textcolor{black}{0.00} & \cellcolor{blue!8}\textcolor{black}{0.00} & \cellcolor{blue!8}\textcolor{black}{0.00} & \cellcolor{blue!8}\textcolor{black}{0.00} & \cellcolor{blue!8}\textcolor{black}{0.00} \\
\texttt{3gbn\_h1} & \cellcolor{blue!85}\textcolor{white}{1.00} & \cellcolor{blue!85}\textcolor{white}{1.00} & \cellcolor{blue!85}\textcolor{white}{1.00} & \cellcolor{blue!85}\textcolor{white}{1.00} & \cellcolor{blue!85}\textcolor{white}{1.00} & \cellcolor{blue!77}\textcolor{white}{0.90} & \cellcolor{blue!85}\textcolor{white}{1.00} & \cellcolor{blue!85}\textcolor{white}{1.00} & \cellcolor{blue!73}\textcolor{white}{0.85} & \cellcolor{blue!85}\textcolor{white}{1.00} & \cellcolor{blue!77}\textcolor{white}{0.90} & \cellcolor{blue!77}\textcolor{white}{0.90} & \cellcolor{blue!85}\textcolor{white}{1.00} & \cellcolor{blue!81}\textcolor{white}{0.95} & \cellcolor{blue!81}\textcolor{white}{0.95} & \cellcolor{blue!85}\textcolor{white}{1.00} & \cellcolor{blue!54}\textcolor{white}{0.60} & \cellcolor{blue!85}\textcolor{white}{1.00} & \cellcolor{blue!81}\textcolor{white}{0.95} & \cellcolor{blue!43}\textcolor{black}{0.45} & \cellcolor{blue!8}\textcolor{black}{0.00} & \cellcolor{blue!8}\textcolor{black}{0.00} \\
\texttt{3gbn\_h9} & \cellcolor{blue!85}\textcolor{white}{1.00} & \cellcolor{blue!85}\textcolor{white}{1.00} & \cellcolor{blue!85}\textcolor{white}{1.00} & \cellcolor{blue!85}\textcolor{white}{1.00} & \cellcolor{blue!85}\textcolor{white}{1.00} & \cellcolor{blue!70}\textcolor{white}{0.80} & \cellcolor{blue!58}\textcolor{white}{0.65} & \cellcolor{blue!85}\textcolor{white}{1.00} & \cellcolor{blue!66}\textcolor{white}{0.75} & \cellcolor{blue!85}\textcolor{white}{1.00} & \cellcolor{blue!70}\textcolor{white}{0.80} & \cellcolor{blue!73}\textcolor{white}{0.85} & \cellcolor{blue!85}\textcolor{white}{1.00} & \cellcolor{blue!70}\textcolor{white}{0.80} & \cellcolor{blue!31}\textcolor{black}{0.30} & \cellcolor{blue!54}\textcolor{white}{0.60} & \cellcolor{blue!39}\textcolor{black}{0.40} & \cellcolor{blue!85}\textcolor{white}{1.00} & \cellcolor{blue!77}\textcolor{white}{0.90} & \cellcolor{blue!43}\textcolor{black}{0.45} & \cellcolor{blue!23}\textcolor{black}{0.20} & \cellcolor{blue!8}\textcolor{black}{0.00} \\
\texttt{4fqi\_h1} & \cellcolor{blue!85}\textcolor{white}{1.00} & \cellcolor{blue!85}\textcolor{white}{1.00} & \cellcolor{blue!85}\textcolor{white}{1.00} & \cellcolor{blue!85}\textcolor{white}{1.00} & \cellcolor{blue!85}\textcolor{white}{1.00} & \cellcolor{blue!85}\textcolor{white}{1.00} & \cellcolor{blue!85}\textcolor{white}{1.00} & \cellcolor{blue!58}\textcolor{white}{0.65} & \cellcolor{blue!73}\textcolor{white}{0.85} & \cellcolor{blue!77}\textcolor{white}{0.90} & \cellcolor{blue!58}\textcolor{white}{0.65} & \cellcolor{blue!85}\textcolor{white}{1.00} & \cellcolor{blue!43}\textcolor{black}{0.45} & \cellcolor{blue!70}\textcolor{white}{0.80} & \cellcolor{blue!85}\textcolor{white}{1.00} & \cellcolor{blue!39}\textcolor{black}{0.40} & \cellcolor{blue!58}\textcolor{white}{0.65} & \cellcolor{blue!12}\textcolor{black}{0.05} & \cellcolor{blue!16}\textcolor{black}{0.10} & \cellcolor{blue!62}\textcolor{white}{0.70} & \cellcolor{blue!12}\textcolor{black}{0.05} & \cellcolor{blue!8}\textcolor{black}{0.00} \\
\texttt{4fqi\_h3} & \cellcolor{blue!73}\textcolor{white}{0.85} & \cellcolor{blue!85}\textcolor{white}{1.00} & \cellcolor{blue!39}\textcolor{black}{0.40} & \cellcolor{blue!54}\textcolor{white}{0.60} & \cellcolor{blue!31}\textcolor{black}{0.30} & \cellcolor{blue!16}\textcolor{black}{0.10} & \cellcolor{blue!8}\textcolor{black}{0.00} & \cellcolor{blue!16}\textcolor{black}{0.10} & \cellcolor{blue!12}\textcolor{black}{0.05} & \cellcolor{blue!8}\textcolor{black}{0.00} & \cellcolor{blue!20}\textcolor{black}{0.15} & \cellcolor{blue!8}\textcolor{black}{0.00} & \cellcolor{blue!8}\textcolor{black}{0.00} & \cellcolor{blue!20}\textcolor{black}{0.15} & \cellcolor{blue!8}\textcolor{black}{0.00} & \cellcolor{blue!8}\textcolor{black}{0.00} & \cellcolor{blue!8}\textcolor{black}{0.00} & \cellcolor{blue!8}\textcolor{black}{0.00} & \cellcolor{blue!8}\textcolor{black}{0.00} & \cellcolor{blue!8}\textcolor{black}{0.00} & \cellcolor{blue!8}\textcolor{black}{0.00} & \cellcolor{blue!8}\textcolor{black}{0.00} \\
\texttt{aayl49} & \cellcolor{blue!23}\textcolor{black}{0.20} & \cellcolor{blue!23}\textcolor{black}{0.20} & \cellcolor{blue!20}\textcolor{black}{0.15} & \cellcolor{blue!8}\textcolor{black}{0.00} & \cellcolor{blue!8}\textcolor{black}{0.00} & \cellcolor{blue!12}\textcolor{black}{0.05} & \cellcolor{blue!8}\textcolor{black}{0.00} & \cellcolor{blue!8}\textcolor{black}{0.00} & \cellcolor{blue!16}\textcolor{black}{0.10} & \cellcolor{blue!8}\textcolor{black}{0.00} & \cellcolor{blue!8}\textcolor{black}{0.00} & \cellcolor{blue!8}\textcolor{black}{0.00} & \cellcolor{blue!8}\textcolor{black}{0.00} & \cellcolor{blue!8}\textcolor{black}{0.00} & \cellcolor{blue!8}\textcolor{black}{0.00} & \cellcolor{blue!8}\textcolor{black}{0.00} & \cellcolor{blue!12}\textcolor{black}{0.05} & \cellcolor{blue!12}\textcolor{black}{0.05} & \cellcolor{blue!12}\textcolor{black}{0.05} & \cellcolor{blue!8}\textcolor{black}{0.00} & \cellcolor{blue!8}\textcolor{black}{0.00} & \cellcolor{blue!8}\textcolor{black}{0.00} \\
\texttt{aayl49\_ML} & \cellcolor{blue!62}\textcolor{white}{0.70} & \cellcolor{blue!62}\textcolor{white}{0.70} & \cellcolor{blue!46}\textcolor{white}{0.50} & \cellcolor{blue!20}\textcolor{black}{0.15} & \cellcolor{blue!27}\textcolor{black}{0.25} & \cellcolor{blue!31}\textcolor{black}{0.30} & \cellcolor{blue!31}\textcolor{black}{0.30} & \cellcolor{blue!35}\textcolor{black}{0.35} & \cellcolor{blue!46}\textcolor{white}{0.50} & \cellcolor{blue!23}\textcolor{black}{0.20} & \cellcolor{blue!35}\textcolor{black}{0.35} & \cellcolor{blue!27}\textcolor{black}{0.25} & \cellcolor{blue!27}\textcolor{black}{0.25} & \cellcolor{blue!27}\textcolor{black}{0.25} & \cellcolor{blue!35}\textcolor{black}{0.35} & \cellcolor{blue!27}\textcolor{black}{0.25} & \cellcolor{blue!43}\textcolor{black}{0.45} & \cellcolor{blue!12}\textcolor{black}{0.05} & \cellcolor{blue!16}\textcolor{black}{0.10} & \cellcolor{blue!20}\textcolor{black}{0.15} & \cellcolor{blue!20}\textcolor{black}{0.15} & \cellcolor{blue!16}\textcolor{black}{0.10} \\
\texttt{aayl50} & \cellcolor{blue!23}\textcolor{black}{0.20} & \cellcolor{blue!12}\textcolor{black}{0.05} & \cellcolor{blue!12}\textcolor{black}{0.05} & \cellcolor{blue!16}\textcolor{black}{0.10} & \cellcolor{blue!8}\textcolor{black}{0.00} & \cellcolor{blue!12}\textcolor{black}{0.05} & \cellcolor{blue!8}\textcolor{black}{0.00} & \cellcolor{blue!8}\textcolor{black}{0.00} & \cellcolor{blue!8}\textcolor{black}{0.00} & \cellcolor{blue!8}\textcolor{black}{0.00} & \cellcolor{blue!16}\textcolor{black}{0.10} & \cellcolor{blue!8}\textcolor{black}{0.00} & \cellcolor{blue!16}\textcolor{black}{0.10} & \cellcolor{blue!12}\textcolor{black}{0.05} & \cellcolor{blue!8}\textcolor{black}{0.00} & \cellcolor{blue!12}\textcolor{black}{0.05} & \cellcolor{blue!8}\textcolor{black}{0.00} & \cellcolor{blue!8}\textcolor{black}{0.00} & \cellcolor{blue!8}\textcolor{black}{0.00} & \cellcolor{blue!8}\textcolor{black}{0.00} & \cellcolor{blue!20}\textcolor{black}{0.15} & \cellcolor{blue!16}\textcolor{black}{0.10} \\
\texttt{aayl51} & \cellcolor{blue!8}\textcolor{black}{0.00} & \cellcolor{blue!12}\textcolor{black}{0.05} & \cellcolor{blue!12}\textcolor{black}{0.05} & \cellcolor{blue!12}\textcolor{black}{0.05} & \cellcolor{blue!8}\textcolor{black}{0.00} & \cellcolor{blue!8}\textcolor{black}{0.00} & \cellcolor{blue!8}\textcolor{black}{0.00} & \cellcolor{blue!8}\textcolor{black}{0.00} & \cellcolor{blue!8}\textcolor{black}{0.00} & \cellcolor{blue!8}\textcolor{black}{0.00} & \cellcolor{blue!8}\textcolor{black}{0.00} & \cellcolor{blue!8}\textcolor{black}{0.00} & \cellcolor{blue!8}\textcolor{black}{0.00} & \cellcolor{blue!8}\textcolor{black}{0.00} & \cellcolor{blue!8}\textcolor{black}{0.00} & \cellcolor{blue!8}\textcolor{black}{0.00} & \cellcolor{blue!8}\textcolor{black}{0.00} & \cellcolor{blue!8}\textcolor{black}{0.00} & \cellcolor{blue!8}\textcolor{black}{0.00} & \cellcolor{blue!8}\textcolor{black}{0.00} & \cellcolor{blue!8}\textcolor{black}{0.00} & \cellcolor{blue!8}\textcolor{black}{0.00} \\
\texttt{aayl52} & \cellcolor{blue!8}\textcolor{black}{0.00} & \cellcolor{blue!8}\textcolor{black}{0.00} & \cellcolor{blue!8}\textcolor{black}{0.00} & \cellcolor{blue!8}\textcolor{black}{0.00} & \cellcolor{blue!8}\textcolor{black}{0.00} & \cellcolor{blue!8}\textcolor{black}{0.00} & \cellcolor{blue!8}\textcolor{black}{0.00} & \cellcolor{blue!8}\textcolor{black}{0.00} & \cellcolor{blue!8}\textcolor{black}{0.00} & \cellcolor{blue!8}\textcolor{black}{0.00} & \cellcolor{blue!8}\textcolor{black}{0.00} & \cellcolor{blue!8}\textcolor{black}{0.00} & \cellcolor{blue!12}\textcolor{black}{0.05} & \cellcolor{blue!8}\textcolor{black}{0.00} & \cellcolor{blue!8}\textcolor{black}{0.00} & \cellcolor{blue!8}\textcolor{black}{0.00} & \cellcolor{blue!8}\textcolor{black}{0.00} & \cellcolor{blue!8}\textcolor{black}{0.00} & \cellcolor{blue!8}\textcolor{black}{0.00} & \cellcolor{blue!8}\textcolor{black}{0.00} & \cellcolor{blue!8}\textcolor{black}{0.00} & \cellcolor{blue!8}\textcolor{black}{0.00} \\
\textbf{Mean} & \cellcolor{blue!46}\textcolor{black}{0.49} & \cellcolor{blue!46}\textcolor{white}{0.50} & \cellcolor{blue!40}\textcolor{black}{0.41} & \cellcolor{blue!38}\textcolor{black}{0.39} & \cellcolor{blue!35}\textcolor{black}{0.35} & \cellcolor{blue!33}\textcolor{black}{0.32} & \cellcolor{blue!31}\textcolor{black}{0.29} & \cellcolor{blue!32}\textcolor{black}{0.31} & \cellcolor{blue!32}\textcolor{black}{0.31} & \cellcolor{blue!32}\textcolor{black}{0.31} & \cellcolor{blue!31}\textcolor{black}{0.30} & \cellcolor{blue!31}\textcolor{black}{0.30} & \cellcolor{blue!30}\textcolor{black}{0.29} & \cellcolor{blue!31}\textcolor{black}{0.30} & \cellcolor{blue!28}\textcolor{black}{0.26} & \cellcolor{blue!26}\textcolor{black}{0.23} & \cellcolor{blue!25}\textcolor{black}{0.21} & \cellcolor{blue!25}\textcolor{black}{0.21} & \cellcolor{blue!24}\textcolor{black}{0.21} & \cellcolor{blue!21}\textcolor{black}{0.17} & \cellcolor{blue!12}\textcolor{black}{0.06} & \cellcolor{blue!10}\textcolor{black}{0.02} \\
\bottomrule
\end{tabular}}
\caption{Assay-level fivefold Precision@20 for all evaluated methods. Values are the proportions of the top 20 model-ranked variants achieving at least fivefold affinity improvement over the corresponding wild type. Cell shading ranges from light to dark blue for values from 0 to 1, using the same scale in all four tables. The final row is the equal-assay-weighted mean across the 10 direct-$K_d$ assays.}
\label{tab:fivefold_precision_k20}
\end{table}

\begin{table}[p]
\centering
\tiny
\setlength{\tabcolsep}{1.8pt}
\renewcommand{\arraystretch}{1.08}
\resizebox{\textwidth}{!}{\begin{tabular}{@{}l*{22}{c}@{}}
\toprule
\textbf{Assay} & \rotatebox{65}{\textbf{ProteinMPNN}} & \rotatebox{65}{\textbf{ESM-IF1}} & \rotatebox{65}{\textbf{Antifold}} & \rotatebox{65}{\textbf{SaProt}} & \rotatebox{65}{\textbf{Boltz-2}} & \rotatebox{65}{\textbf{CurrAb}} & \rotatebox{65}{\textbf{Delta G}} & \rotatebox{65}{\textbf{diffab}} & \rotatebox{65}{\textbf{AF3}} & \rotatebox{65}{\textbf{ESM-2}} & \rotatebox{65}{\textbf{dyMEAN\_fixbb}} & \rotatebox{65}{\textbf{ProSST}} & \rotatebox{65}{\textbf{diffab\_fixbb}} & \rotatebox{65}{\textbf{dyMEAN}} & \rotatebox{65}{\textbf{ESMDance}} & \rotatebox{65}{\textbf{SASA}} & \rotatebox{65}{\textbf{ESM-3}} & \rotatebox{65}{\textbf{MEAN}} & \rotatebox{65}{\textbf{MEAN\_fixbb}} & \rotatebox{65}{\textbf{ProtGPT-2}} & \rotatebox{65}{\textbf{progen2-large}} & \rotatebox{65}{\textbf{AntiBERTy}} \\
\midrule
\texttt{1n8z} & \cellcolor{blue!8}\textcolor{black}{0.00} & \cellcolor{blue!8}\textcolor{black}{0.00} & \cellcolor{blue!8}\textcolor{black}{0.00} & \cellcolor{blue!8}\textcolor{black}{0.00} & \cellcolor{blue!8}\textcolor{black}{0.00} & \cellcolor{blue!8}\textcolor{black}{0.00} & \cellcolor{blue!8}\textcolor{black}{0.00} & \cellcolor{blue!8}\textcolor{black}{0.00} & \cellcolor{blue!8}\textcolor{black}{0.00} & \cellcolor{blue!8}\textcolor{black}{0.00} & \cellcolor{blue!8}\textcolor{black}{0.00} & \cellcolor{blue!8}\textcolor{black}{0.00} & \cellcolor{blue!8}\textcolor{black}{0.00} & \cellcolor{blue!8}\textcolor{black}{0.00} & \cellcolor{blue!8}\textcolor{black}{0.00} & \cellcolor{blue!8}\textcolor{black}{0.00} & \cellcolor{blue!8}\textcolor{black}{0.00} & \cellcolor{blue!8}\textcolor{black}{0.00} & \cellcolor{blue!8}\textcolor{black}{0.00} & \cellcolor{blue!8}\textcolor{black}{0.00} & \cellcolor{blue!8}\textcolor{black}{0.00} & \cellcolor{blue!8}\textcolor{black}{0.00} \\
\texttt{3gbn\_h1} & \cellcolor{blue!85}\textcolor{white}{1.00} & \cellcolor{blue!85}\textcolor{white}{1.00} & \cellcolor{blue!85}\textcolor{white}{1.00} & \cellcolor{blue!85}\textcolor{white}{1.00} & \cellcolor{blue!85}\textcolor{white}{1.00} & \cellcolor{blue!79}\textcolor{white}{0.92} & \cellcolor{blue!85}\textcolor{white}{1.00} & \cellcolor{blue!85}\textcolor{white}{1.00} & \cellcolor{blue!73}\textcolor{white}{0.84} & \cellcolor{blue!85}\textcolor{white}{1.00} & \cellcolor{blue!76}\textcolor{white}{0.88} & \cellcolor{blue!70}\textcolor{white}{0.80} & \cellcolor{blue!85}\textcolor{white}{1.00} & \cellcolor{blue!76}\textcolor{white}{0.88} & \cellcolor{blue!77}\textcolor{white}{0.90} & \cellcolor{blue!85}\textcolor{white}{1.00} & \cellcolor{blue!63}\textcolor{white}{0.72} & \cellcolor{blue!83}\textcolor{white}{0.98} & \cellcolor{blue!79}\textcolor{white}{0.92} & \cellcolor{blue!50}\textcolor{white}{0.54} & \cellcolor{blue!20}\textcolor{black}{0.16} & \cellcolor{blue!17}\textcolor{black}{0.12} \\
\texttt{3gbn\_h9} & \cellcolor{blue!85}\textcolor{white}{1.00} & \cellcolor{blue!85}\textcolor{white}{1.00} & \cellcolor{blue!85}\textcolor{white}{1.00} & \cellcolor{blue!85}\textcolor{white}{1.00} & \cellcolor{blue!85}\textcolor{white}{1.00} & \cellcolor{blue!74}\textcolor{white}{0.86} & \cellcolor{blue!65}\textcolor{white}{0.74} & \cellcolor{blue!85}\textcolor{white}{1.00} & \cellcolor{blue!62}\textcolor{white}{0.70} & \cellcolor{blue!85}\textcolor{white}{1.00} & \cellcolor{blue!67}\textcolor{white}{0.76} & \cellcolor{blue!70}\textcolor{white}{0.80} & \cellcolor{blue!85}\textcolor{white}{1.00} & \cellcolor{blue!62}\textcolor{white}{0.70} & \cellcolor{blue!42}\textcolor{black}{0.44} & \cellcolor{blue!71}\textcolor{white}{0.82} & \cellcolor{blue!39}\textcolor{black}{0.40} & \cellcolor{blue!83}\textcolor{white}{0.98} & \cellcolor{blue!74}\textcolor{white}{0.86} & \cellcolor{blue!57}\textcolor{white}{0.64} & \cellcolor{blue!25}\textcolor{black}{0.22} & \cellcolor{blue!13}\textcolor{black}{0.06} \\
\texttt{4fqi\_h1} & \cellcolor{blue!85}\textcolor{white}{1.00} & \cellcolor{blue!85}\textcolor{white}{1.00} & \cellcolor{blue!85}\textcolor{white}{1.00} & \cellcolor{blue!85}\textcolor{white}{1.00} & \cellcolor{blue!85}\textcolor{white}{1.00} & \cellcolor{blue!74}\textcolor{white}{0.86} & \cellcolor{blue!85}\textcolor{white}{1.00} & \cellcolor{blue!57}\textcolor{white}{0.64} & \cellcolor{blue!70}\textcolor{white}{0.80} & \cellcolor{blue!71}\textcolor{white}{0.82} & \cellcolor{blue!63}\textcolor{white}{0.72} & \cellcolor{blue!85}\textcolor{white}{1.00} & \cellcolor{blue!46}\textcolor{white}{0.50} & \cellcolor{blue!70}\textcolor{white}{0.80} & \cellcolor{blue!85}\textcolor{white}{1.00} & \cellcolor{blue!45}\textcolor{black}{0.48} & \cellcolor{blue!57}\textcolor{white}{0.64} & \cellcolor{blue!10}\textcolor{black}{0.02} & \cellcolor{blue!11}\textcolor{black}{0.04} & \cellcolor{blue!53}\textcolor{white}{0.58} & \cellcolor{blue!13}\textcolor{black}{0.06} & \cellcolor{blue!10}\textcolor{black}{0.02} \\
\texttt{4fqi\_h3} & \cellcolor{blue!70}\textcolor{white}{0.80} & \cellcolor{blue!82}\textcolor{white}{0.96} & \cellcolor{blue!51}\textcolor{white}{0.56} & \cellcolor{blue!54}\textcolor{white}{0.60} & \cellcolor{blue!33}\textcolor{black}{0.32} & \cellcolor{blue!14}\textcolor{black}{0.08} & \cellcolor{blue!10}\textcolor{black}{0.02} & \cellcolor{blue!17}\textcolor{black}{0.12} & \cellcolor{blue!16}\textcolor{black}{0.10} & \cellcolor{blue!8}\textcolor{black}{0.00} & \cellcolor{blue!19}\textcolor{black}{0.14} & \cellcolor{blue!8}\textcolor{black}{0.00} & \cellcolor{blue!11}\textcolor{black}{0.04} & \cellcolor{blue!20}\textcolor{black}{0.16} & \cellcolor{blue!8}\textcolor{black}{0.00} & \cellcolor{blue!8}\textcolor{black}{0.00} & \cellcolor{blue!8}\textcolor{black}{0.00} & \cellcolor{blue!8}\textcolor{black}{0.00} & \cellcolor{blue!8}\textcolor{black}{0.00} & \cellcolor{blue!8}\textcolor{black}{0.00} & \cellcolor{blue!8}\textcolor{black}{0.00} & \cellcolor{blue!8}\textcolor{black}{0.00} \\
\texttt{aayl49} & \cellcolor{blue!20}\textcolor{black}{0.16} & \cellcolor{blue!19}\textcolor{black}{0.14} & \cellcolor{blue!14}\textcolor{black}{0.08} & \cellcolor{blue!8}\textcolor{black}{0.00} & \cellcolor{blue!8}\textcolor{black}{0.00} & \cellcolor{blue!13}\textcolor{black}{0.06} & \cellcolor{blue!8}\textcolor{black}{0.00} & \cellcolor{blue!8}\textcolor{black}{0.00} & \cellcolor{blue!14}\textcolor{black}{0.08} & \cellcolor{blue!10}\textcolor{black}{0.02} & \cellcolor{blue!11}\textcolor{black}{0.04} & \cellcolor{blue!8}\textcolor{black}{0.00} & \cellcolor{blue!11}\textcolor{black}{0.04} & \cellcolor{blue!8}\textcolor{black}{0.00} & \cellcolor{blue!8}\textcolor{black}{0.00} & \cellcolor{blue!8}\textcolor{black}{0.00} & \cellcolor{blue!13}\textcolor{black}{0.06} & \cellcolor{blue!10}\textcolor{black}{0.02} & \cellcolor{blue!10}\textcolor{black}{0.02} & \cellcolor{blue!8}\textcolor{black}{0.00} & \cellcolor{blue!8}\textcolor{black}{0.00} & \cellcolor{blue!11}\textcolor{black}{0.04} \\
\texttt{aayl49\_ML} & \cellcolor{blue!54}\textcolor{white}{0.60} & \cellcolor{blue!54}\textcolor{white}{0.60} & \cellcolor{blue!36}\textcolor{black}{0.36} & \cellcolor{blue!20}\textcolor{black}{0.16} & \cellcolor{blue!30}\textcolor{black}{0.28} & \cellcolor{blue!36}\textcolor{black}{0.36} & \cellcolor{blue!48}\textcolor{white}{0.52} & \cellcolor{blue!28}\textcolor{black}{0.26} & \cellcolor{blue!37}\textcolor{black}{0.38} & \cellcolor{blue!26}\textcolor{black}{0.24} & \cellcolor{blue!28}\textcolor{black}{0.26} & \cellcolor{blue!23}\textcolor{black}{0.20} & \cellcolor{blue!28}\textcolor{black}{0.26} & \cellcolor{blue!22}\textcolor{black}{0.18} & \cellcolor{blue!26}\textcolor{black}{0.24} & \cellcolor{blue!23}\textcolor{black}{0.20} & \cellcolor{blue!43}\textcolor{black}{0.46} & \cellcolor{blue!22}\textcolor{black}{0.18} & \cellcolor{blue!20}\textcolor{black}{0.16} & \cellcolor{blue!26}\textcolor{black}{0.24} & \cellcolor{blue!19}\textcolor{black}{0.14} & \cellcolor{blue!20}\textcolor{black}{0.16} \\
\texttt{aayl50} & \cellcolor{blue!20}\textcolor{black}{0.16} & \cellcolor{blue!17}\textcolor{black}{0.12} & \cellcolor{blue!11}\textcolor{black}{0.04} & \cellcolor{blue!11}\textcolor{black}{0.04} & \cellcolor{blue!10}\textcolor{black}{0.02} & \cellcolor{blue!11}\textcolor{black}{0.04} & \cellcolor{blue!10}\textcolor{black}{0.02} & \cellcolor{blue!11}\textcolor{black}{0.04} & \cellcolor{blue!10}\textcolor{black}{0.02} & \cellcolor{blue!10}\textcolor{black}{0.02} & \cellcolor{blue!13}\textcolor{black}{0.06} & \cellcolor{blue!8}\textcolor{black}{0.00} & \cellcolor{blue!11}\textcolor{black}{0.04} & \cellcolor{blue!13}\textcolor{black}{0.06} & \cellcolor{blue!8}\textcolor{black}{0.00} & \cellcolor{blue!10}\textcolor{black}{0.02} & \cellcolor{blue!10}\textcolor{black}{0.02} & \cellcolor{blue!10}\textcolor{black}{0.02} & \cellcolor{blue!10}\textcolor{black}{0.02} & \cellcolor{blue!10}\textcolor{black}{0.02} & \cellcolor{blue!16}\textcolor{black}{0.10} & \cellcolor{blue!14}\textcolor{black}{0.08} \\
\texttt{aayl51} & \cellcolor{blue!11}\textcolor{black}{0.04} & \cellcolor{blue!10}\textcolor{black}{0.02} & \cellcolor{blue!10}\textcolor{black}{0.02} & \cellcolor{blue!10}\textcolor{black}{0.02} & \cellcolor{blue!10}\textcolor{black}{0.02} & \cellcolor{blue!8}\textcolor{black}{0.00} & \cellcolor{blue!10}\textcolor{black}{0.02} & \cellcolor{blue!8}\textcolor{black}{0.00} & \cellcolor{blue!8}\textcolor{black}{0.00} & \cellcolor{blue!8}\textcolor{black}{0.00} & \cellcolor{blue!8}\textcolor{black}{0.00} & \cellcolor{blue!10}\textcolor{black}{0.02} & \cellcolor{blue!10}\textcolor{black}{0.02} & \cellcolor{blue!8}\textcolor{black}{0.00} & \cellcolor{blue!8}\textcolor{black}{0.00} & \cellcolor{blue!10}\textcolor{black}{0.02} & \cellcolor{blue!10}\textcolor{black}{0.02} & \cellcolor{blue!8}\textcolor{black}{0.00} & \cellcolor{blue!8}\textcolor{black}{0.00} & \cellcolor{blue!8}\textcolor{black}{0.00} & \cellcolor{blue!10}\textcolor{black}{0.02} & \cellcolor{blue!10}\textcolor{black}{0.02} \\
\texttt{aayl52} & \cellcolor{blue!8}\textcolor{black}{0.00} & \cellcolor{blue!8}\textcolor{black}{0.00} & \cellcolor{blue!8}\textcolor{black}{0.00} & \cellcolor{blue!8}\textcolor{black}{0.00} & \cellcolor{blue!8}\textcolor{black}{0.00} & \cellcolor{blue!8}\textcolor{black}{0.00} & \cellcolor{blue!8}\textcolor{black}{0.00} & \cellcolor{blue!8}\textcolor{black}{0.00} & \cellcolor{blue!8}\textcolor{black}{0.00} & \cellcolor{blue!8}\textcolor{black}{0.00} & \cellcolor{blue!8}\textcolor{black}{0.00} & \cellcolor{blue!8}\textcolor{black}{0.00} & \cellcolor{blue!10}\textcolor{black}{0.02} & \cellcolor{blue!8}\textcolor{black}{0.00} & \cellcolor{blue!8}\textcolor{black}{0.00} & \cellcolor{blue!8}\textcolor{black}{0.00} & \cellcolor{blue!8}\textcolor{black}{0.00} & \cellcolor{blue!8}\textcolor{black}{0.00} & \cellcolor{blue!8}\textcolor{black}{0.00} & \cellcolor{blue!8}\textcolor{black}{0.00} & \cellcolor{blue!8}\textcolor{black}{0.00} & \cellcolor{blue!8}\textcolor{black}{0.00} \\
\textbf{Mean} & \cellcolor{blue!45}\textcolor{black}{0.48} & \cellcolor{blue!45}\textcolor{black}{0.48} & \cellcolor{blue!39}\textcolor{black}{0.41} & \cellcolor{blue!37}\textcolor{black}{0.38} & \cellcolor{blue!36}\textcolor{black}{0.36} & \cellcolor{blue!32}\textcolor{black}{0.32} & \cellcolor{blue!34}\textcolor{black}{0.33} & \cellcolor{blue!32}\textcolor{black}{0.31} & \cellcolor{blue!30}\textcolor{black}{0.29} & \cellcolor{blue!32}\textcolor{black}{0.31} & \cellcolor{blue!30}\textcolor{black}{0.29} & \cellcolor{blue!30}\textcolor{black}{0.28} & \cellcolor{blue!30}\textcolor{black}{0.29} & \cellcolor{blue!29}\textcolor{black}{0.28} & \cellcolor{blue!28}\textcolor{black}{0.26} & \cellcolor{blue!28}\textcolor{black}{0.25} & \cellcolor{blue!26}\textcolor{black}{0.23} & \cellcolor{blue!25}\textcolor{black}{0.22} & \cellcolor{blue!24}\textcolor{black}{0.20} & \cellcolor{blue!24}\textcolor{black}{0.20} & \cellcolor{blue!13}\textcolor{black}{0.07} & \cellcolor{blue!12}\textcolor{black}{0.05} \\
\bottomrule
\end{tabular}}
\caption{Assay-level fivefold Precision@50 for all evaluated methods. Values are the proportions of the top 50 model-ranked variants achieving at least fivefold affinity improvement over the corresponding wild type. Cell shading ranges from light to dark blue for values from 0 to 1, using the same scale in all four tables. The final row is the equal-assay-weighted mean across the 10 direct-$K_d$ assays.}
\label{tab:fivefold_precision_k50}
\end{table}

\begin{table}[p]
\centering
\tiny
\setlength{\tabcolsep}{1.8pt}
\renewcommand{\arraystretch}{1.08}
\resizebox{\textwidth}{!}{\begin{tabular}{@{}l*{22}{c}@{}}
\toprule
\textbf{Assay} & \rotatebox{65}{\textbf{ProteinMPNN}} & \rotatebox{65}{\textbf{ESM-IF1}} & \rotatebox{65}{\textbf{Antifold}} & \rotatebox{65}{\textbf{SaProt}} & \rotatebox{65}{\textbf{Boltz-2}} & \rotatebox{65}{\textbf{CurrAb}} & \rotatebox{65}{\textbf{Delta G}} & \rotatebox{65}{\textbf{diffab}} & \rotatebox{65}{\textbf{AF3}} & \rotatebox{65}{\textbf{ESM-2}} & \rotatebox{65}{\textbf{dyMEAN\_fixbb}} & \rotatebox{65}{\textbf{ProSST}} & \rotatebox{65}{\textbf{diffab\_fixbb}} & \rotatebox{65}{\textbf{dyMEAN}} & \rotatebox{65}{\textbf{ESMDance}} & \rotatebox{65}{\textbf{SASA}} & \rotatebox{65}{\textbf{ESM-3}} & \rotatebox{65}{\textbf{MEAN}} & \rotatebox{65}{\textbf{MEAN\_fixbb}} & \rotatebox{65}{\textbf{ProtGPT-2}} & \rotatebox{65}{\textbf{progen2-large}} & \rotatebox{65}{\textbf{AntiBERTy}} \\
\midrule
\texttt{1n8z} & \cellcolor{blue!8}\textcolor{black}{0.00} & \cellcolor{blue!8}\textcolor{black}{0.00} & \cellcolor{blue!8}\textcolor{black}{0.00} & \cellcolor{blue!8}\textcolor{black}{0.00} & \cellcolor{blue!8}\textcolor{black}{0.00} & \cellcolor{blue!8}\textcolor{black}{0.00} & \cellcolor{blue!8}\textcolor{black}{0.00} & \cellcolor{blue!8}\textcolor{black}{0.00} & \cellcolor{blue!8}\textcolor{black}{0.00} & \cellcolor{blue!8}\textcolor{black}{0.00} & \cellcolor{blue!8}\textcolor{black}{0.00} & \cellcolor{blue!8}\textcolor{black}{0.00} & \cellcolor{blue!8}\textcolor{black}{0.00} & \cellcolor{blue!8}\textcolor{black}{0.00} & \cellcolor{blue!8}\textcolor{black}{0.00} & \cellcolor{blue!8}\textcolor{black}{0.00} & \cellcolor{blue!8}\textcolor{black}{0.00} & \cellcolor{blue!8}\textcolor{black}{0.00} & \cellcolor{blue!8}\textcolor{black}{0.00} & \cellcolor{blue!8}\textcolor{black}{0.00} & \cellcolor{blue!8}\textcolor{black}{0.00} & \cellcolor{blue!8}\textcolor{black}{0.00} \\
\texttt{3gbn\_h1} & \cellcolor{blue!85}\textcolor{white}{1.00} & \cellcolor{blue!85}\textcolor{white}{1.00} & \cellcolor{blue!85}\textcolor{white}{1.00} & \cellcolor{blue!85}\textcolor{white}{1.00} & \cellcolor{blue!85}\textcolor{white}{1.00} & \cellcolor{blue!79}\textcolor{white}{0.92} & \cellcolor{blue!85}\textcolor{white}{1.00} & \cellcolor{blue!85}\textcolor{white}{1.00} & \cellcolor{blue!72}\textcolor{white}{0.83} & \cellcolor{blue!84}\textcolor{white}{0.99} & \cellcolor{blue!76}\textcolor{white}{0.88} & \cellcolor{blue!70}\textcolor{white}{0.81} & \cellcolor{blue!85}\textcolor{white}{1.00} & \cellcolor{blue!77}\textcolor{white}{0.90} & \cellcolor{blue!78}\textcolor{white}{0.91} & \cellcolor{blue!85}\textcolor{white}{1.00} & \cellcolor{blue!62}\textcolor{white}{0.70} & \cellcolor{blue!83}\textcolor{white}{0.97} & \cellcolor{blue!77}\textcolor{white}{0.90} & \cellcolor{blue!63}\textcolor{white}{0.71} & \cellcolor{blue!23}\textcolor{black}{0.19} & \cellcolor{blue!23}\textcolor{black}{0.20} \\
\texttt{3gbn\_h9} & \cellcolor{blue!85}\textcolor{white}{1.00} & \cellcolor{blue!85}\textcolor{white}{1.00} & \cellcolor{blue!81}\textcolor{white}{0.95} & \cellcolor{blue!85}\textcolor{white}{1.00} & \cellcolor{blue!85}\textcolor{white}{1.00} & \cellcolor{blue!73}\textcolor{white}{0.84} & \cellcolor{blue!73}\textcolor{white}{0.85} & \cellcolor{blue!85}\textcolor{white}{1.00} & \cellcolor{blue!66}\textcolor{white}{0.75} & \cellcolor{blue!83}\textcolor{white}{0.98} & \cellcolor{blue!64}\textcolor{white}{0.73} & \cellcolor{blue!68}\textcolor{white}{0.78} & \cellcolor{blue!85}\textcolor{white}{1.00} & \cellcolor{blue!65}\textcolor{white}{0.74} & \cellcolor{blue!50}\textcolor{white}{0.55} & \cellcolor{blue!70}\textcolor{white}{0.80} & \cellcolor{blue!46}\textcolor{black}{0.49} & \cellcolor{blue!80}\textcolor{white}{0.93} & \cellcolor{blue!70}\textcolor{white}{0.81} & \cellcolor{blue!61}\textcolor{white}{0.69} & \cellcolor{blue!29}\textcolor{black}{0.27} & \cellcolor{blue!13}\textcolor{black}{0.06} \\
\texttt{4fqi\_h1} & \cellcolor{blue!85}\textcolor{white}{1.00} & \cellcolor{blue!85}\textcolor{white}{1.00} & \cellcolor{blue!85}\textcolor{white}{1.00} & \cellcolor{blue!85}\textcolor{white}{1.00} & \cellcolor{blue!85}\textcolor{white}{1.00} & \cellcolor{blue!77}\textcolor{white}{0.90} & \cellcolor{blue!83}\textcolor{white}{0.97} & \cellcolor{blue!62}\textcolor{white}{0.70} & \cellcolor{blue!70}\textcolor{white}{0.80} & \cellcolor{blue!70}\textcolor{white}{0.81} & \cellcolor{blue!66}\textcolor{white}{0.75} & \cellcolor{blue!85}\textcolor{white}{1.00} & \cellcolor{blue!53}\textcolor{white}{0.59} & \cellcolor{blue!65}\textcolor{white}{0.74} & \cellcolor{blue!85}\textcolor{white}{1.00} & \cellcolor{blue!50}\textcolor{white}{0.54} & \cellcolor{blue!55}\textcolor{white}{0.61} & \cellcolor{blue!10}\textcolor{black}{0.02} & \cellcolor{blue!10}\textcolor{black}{0.02} & \cellcolor{blue!53}\textcolor{white}{0.58} & \cellcolor{blue!13}\textcolor{black}{0.06} & \cellcolor{blue!13}\textcolor{black}{0.06} \\
\texttt{4fqi\_h3} & \cellcolor{blue!71}\textcolor{white}{0.82} & \cellcolor{blue!80}\textcolor{white}{0.93} & \cellcolor{blue!57}\textcolor{white}{0.63} & \cellcolor{blue!46}\textcolor{black}{0.49} & \cellcolor{blue!30}\textcolor{black}{0.29} & \cellcolor{blue!18}\textcolor{black}{0.13} & \cellcolor{blue!9}\textcolor{black}{0.01} & \cellcolor{blue!16}\textcolor{black}{0.11} & \cellcolor{blue!17}\textcolor{black}{0.12} & \cellcolor{blue!8}\textcolor{black}{0.00} & \cellcolor{blue!16}\textcolor{black}{0.10} & \cellcolor{blue!8}\textcolor{black}{0.00} & \cellcolor{blue!11}\textcolor{black}{0.04} & \cellcolor{blue!17}\textcolor{black}{0.12} & \cellcolor{blue!8}\textcolor{black}{0.00} & \cellcolor{blue!8}\textcolor{black}{0.00} & \cellcolor{blue!8}\textcolor{black}{0.00} & \cellcolor{blue!8}\textcolor{black}{0.00} & \cellcolor{blue!8}\textcolor{black}{0.00} & \cellcolor{blue!8}\textcolor{black}{0.00} & \cellcolor{blue!8}\textcolor{black}{0.00} & \cellcolor{blue!8}\textcolor{black}{0.00} \\
\texttt{aayl49} & \cellcolor{blue!16}\textcolor{black}{0.11} & \cellcolor{blue!20}\textcolor{black}{0.15} & \cellcolor{blue!13}\textcolor{black}{0.06} & \cellcolor{blue!8}\textcolor{black}{0.00} & \cellcolor{blue!8}\textcolor{black}{0.00} & \cellcolor{blue!10}\textcolor{black}{0.03} & \cellcolor{blue!8}\textcolor{black}{0.00} & \cellcolor{blue!9}\textcolor{black}{0.01} & \cellcolor{blue!13}\textcolor{black}{0.06} & \cellcolor{blue!9}\textcolor{black}{0.01} & \cellcolor{blue!10}\textcolor{black}{0.03} & \cellcolor{blue!9}\textcolor{black}{0.01} & \cellcolor{blue!10}\textcolor{black}{0.02} & \cellcolor{blue!8}\textcolor{black}{0.00} & \cellcolor{blue!8}\textcolor{black}{0.00} & \cellcolor{blue!8}\textcolor{black}{0.00} & \cellcolor{blue!12}\textcolor{black}{0.05} & \cellcolor{blue!10}\textcolor{black}{0.02} & \cellcolor{blue!10}\textcolor{black}{0.02} & \cellcolor{blue!8}\textcolor{black}{0.00} & \cellcolor{blue!9}\textcolor{black}{0.01} & \cellcolor{blue!10}\textcolor{black}{0.03} \\
\texttt{aayl49\_ML} & \cellcolor{blue!48}\textcolor{white}{0.52} & \cellcolor{blue!54}\textcolor{white}{0.60} & \cellcolor{blue!36}\textcolor{black}{0.36} & \cellcolor{blue!23}\textcolor{black}{0.19} & \cellcolor{blue!26}\textcolor{black}{0.23} & \cellcolor{blue!34}\textcolor{black}{0.34} & \cellcolor{blue!45}\textcolor{black}{0.48} & \cellcolor{blue!27}\textcolor{black}{0.25} & \cellcolor{blue!37}\textcolor{black}{0.38} & \cellcolor{blue!25}\textcolor{black}{0.22} & \cellcolor{blue!28}\textcolor{black}{0.26} & \cellcolor{blue!27}\textcolor{black}{0.25} & \cellcolor{blue!33}\textcolor{black}{0.32} & \cellcolor{blue!29}\textcolor{black}{0.27} & \cellcolor{blue!24}\textcolor{black}{0.21} & \cellcolor{blue!27}\textcolor{black}{0.25} & \cellcolor{blue!36}\textcolor{black}{0.37} & \cellcolor{blue!20}\textcolor{black}{0.15} & \cellcolor{blue!22}\textcolor{black}{0.18} & \cellcolor{blue!29}\textcolor{black}{0.27} & \cellcolor{blue!19}\textcolor{black}{0.14} & \cellcolor{blue!24}\textcolor{black}{0.21} \\
\texttt{aayl50} & \cellcolor{blue!19}\textcolor{black}{0.14} & \cellcolor{blue!14}\textcolor{black}{0.08} & \cellcolor{blue!11}\textcolor{black}{0.04} & \cellcolor{blue!10}\textcolor{black}{0.02} & \cellcolor{blue!10}\textcolor{black}{0.02} & \cellcolor{blue!10}\textcolor{black}{0.03} & \cellcolor{blue!10}\textcolor{black}{0.02} & \cellcolor{blue!11}\textcolor{black}{0.04} & \cellcolor{blue!9}\textcolor{black}{0.01} & \cellcolor{blue!10}\textcolor{black}{0.02} & \cellcolor{blue!13}\textcolor{black}{0.06} & \cellcolor{blue!8}\textcolor{black}{0.00} & \cellcolor{blue!10}\textcolor{black}{0.03} & \cellcolor{blue!13}\textcolor{black}{0.06} & \cellcolor{blue!10}\textcolor{black}{0.02} & \cellcolor{blue!10}\textcolor{black}{0.03} & \cellcolor{blue!11}\textcolor{black}{0.04} & \cellcolor{blue!10}\textcolor{black}{0.02} & \cellcolor{blue!10}\textcolor{black}{0.02} & \cellcolor{blue!9}\textcolor{black}{0.01} & \cellcolor{blue!13}\textcolor{black}{0.07} & \cellcolor{blue!13}\textcolor{black}{0.06} \\
\texttt{aayl51} & \cellcolor{blue!10}\textcolor{black}{0.02} & \cellcolor{blue!9}\textcolor{black}{0.01} & \cellcolor{blue!9}\textcolor{black}{0.01} & \cellcolor{blue!9}\textcolor{black}{0.01} & \cellcolor{blue!9}\textcolor{black}{0.01} & \cellcolor{blue!9}\textcolor{black}{0.01} & \cellcolor{blue!10}\textcolor{black}{0.02} & \cellcolor{blue!8}\textcolor{black}{0.00} & \cellcolor{blue!8}\textcolor{black}{0.00} & \cellcolor{blue!9}\textcolor{black}{0.01} & \cellcolor{blue!8}\textcolor{black}{0.00} & \cellcolor{blue!9}\textcolor{black}{0.01} & \cellcolor{blue!9}\textcolor{black}{0.01} & \cellcolor{blue!8}\textcolor{black}{0.00} & \cellcolor{blue!8}\textcolor{black}{0.00} & \cellcolor{blue!9}\textcolor{black}{0.01} & \cellcolor{blue!10}\textcolor{black}{0.02} & \cellcolor{blue!8}\textcolor{black}{0.00} & \cellcolor{blue!8}\textcolor{black}{0.00} & \cellcolor{blue!8}\textcolor{black}{0.00} & \cellcolor{blue!10}\textcolor{black}{0.02} & \cellcolor{blue!10}\textcolor{black}{0.02} \\
\texttt{aayl52} & \cellcolor{blue!8}\textcolor{black}{0.00} & \cellcolor{blue!8}\textcolor{black}{0.00} & \cellcolor{blue!8}\textcolor{black}{0.00} & \cellcolor{blue!8}\textcolor{black}{0.00} & \cellcolor{blue!8}\textcolor{black}{0.00} & \cellcolor{blue!8}\textcolor{black}{0.00} & \cellcolor{blue!8}\textcolor{black}{0.00} & \cellcolor{blue!8}\textcolor{black}{0.00} & \cellcolor{blue!8}\textcolor{black}{0.00} & \cellcolor{blue!8}\textcolor{black}{0.00} & \cellcolor{blue!8}\textcolor{black}{0.00} & \cellcolor{blue!8}\textcolor{black}{0.00} & \cellcolor{blue!9}\textcolor{black}{0.01} & \cellcolor{blue!8}\textcolor{black}{0.00} & \cellcolor{blue!8}\textcolor{black}{0.00} & \cellcolor{blue!8}\textcolor{black}{0.00} & \cellcolor{blue!8}\textcolor{black}{0.00} & \cellcolor{blue!8}\textcolor{black}{0.00} & \cellcolor{blue!8}\textcolor{black}{0.00} & \cellcolor{blue!8}\textcolor{black}{0.00} & \cellcolor{blue!8}\textcolor{black}{0.00} & \cellcolor{blue!8}\textcolor{black}{0.00} \\
\textbf{Mean} & \cellcolor{blue!43}\textcolor{black}{0.46} & \cellcolor{blue!45}\textcolor{black}{0.48} & \cellcolor{blue!39}\textcolor{black}{0.40} & \cellcolor{blue!37}\textcolor{black}{0.37} & \cellcolor{blue!35}\textcolor{black}{0.35} & \cellcolor{blue!33}\textcolor{black}{0.32} & \cellcolor{blue!34}\textcolor{black}{0.34} & \cellcolor{blue!32}\textcolor{black}{0.31} & \cellcolor{blue!31}\textcolor{black}{0.29} & \cellcolor{blue!31}\textcolor{black}{0.30} & \cellcolor{blue!30}\textcolor{black}{0.28} & \cellcolor{blue!30}\textcolor{black}{0.29} & \cellcolor{blue!31}\textcolor{black}{0.30} & \cellcolor{blue!30}\textcolor{black}{0.28} & \cellcolor{blue!29}\textcolor{black}{0.27} & \cellcolor{blue!28}\textcolor{black}{0.26} & \cellcolor{blue!26}\textcolor{black}{0.23} & \cellcolor{blue!24}\textcolor{black}{0.21} & \cellcolor{blue!23}\textcolor{black}{0.20} & \cellcolor{blue!25}\textcolor{black}{0.23} & \cellcolor{blue!14}\textcolor{black}{0.08} & \cellcolor{blue!13}\textcolor{black}{0.06} \\
\bottomrule
\end{tabular}}
\caption{Assay-level fivefold Precision@100 for all evaluated methods. Values are the proportions of the top 100 model-ranked variants achieving at least fivefold affinity improvement over the corresponding wild type. Cell shading ranges from light to dark blue for values from 0 to 1, using the same scale in all four tables. The final row is the equal-assay-weighted mean across the 10 direct-$K_d$ assays.}
\label{tab:fivefold_precision_k100}
\end{table}

\subsection{Mutation-Stratified Zero-Shot Analysis}
\label{sup:sec:hard_subsets}

We assessed whether the zero-shot results were robust to three mutation-level strata. Mutated chains were classified as VH, VL, or joint VH--VL. Mutation distance was the total amino-acid Hamming distance from the wild-type heavy--light pair and was grouped as 1, 2, 3--4, or $\geq 5$ substitutions. Mutation regions were assigned from the residue numbering in each benchmark wild-type structure using Chothia CDR ranges (VH: 26--32, 52--56, and 95--102; VL: 24--34, 50--56, and 89--97). Variants were then classified as CDR-only, framework-only, or mixed. Wild-type rows were excluded. Within each assay, model, and stratum containing at least 20 variants, we calculated Spearman correlation between model score and experimental binding score. We report the equal-assay mean correlations in Fig.~\ref{fig:correlation}d.

ProteinMPNN and ESM-IF1 ranked first or second across all mutation-region strata and across VH and joint VH--VL variants (Fig.~\ref{fig:correlation}d). Their equal-assay mean correlations increased descriptively with mutation count, from 0.131 and 0.133 for single mutants to 0.355 and 0.346 for variants with at least five substitutions. This comparison is not fully paired across assays: only two assays contained at least 20 variants in every distance bin. A common-assay sensitivity analysis retained the general direction for ProteinMPNN and Antifold but was underpowered; the distance trend should therefore be interpreted as descriptive rather than as an isolated effect of distance.

\subsection{Impact of Chain Order in Autoregressive PLMs in Correlation Studies}

\label{sup:sec:sec:chain_order}
Since autoregressive models (ESM-IF1 \cite{hsu2022learning}, ProGen \cite{nijkamp2023progen2}, ProGPT2 \cite{ferruz2022protgpt2}) use previous tokens as context to predict the next, it was hypothesized that providing antigen (mimicking SHM) and light chain context before the mutated heavy chain could improve zero-shot correlation of general autoregressive PLMs to experimental binding affinity. The benchmarking results show otherwise, with sporadic and minimal impact of chain ordering seen across both structure-based (ESM-IF1 \cite{hsu2022learning}) and sequence-based (ProGen2 \cite{nijkamp2023progen2}, ProGPT 2 \cite{ferruz2022protgpt2}) correlation studies across the eleven Ab-Ag datasets. Therefore, demonstrating that autoregressive PLMs trained on single chain protein data remain largely insensitive to multimer chain ordering.

\begin{table}[htbp]\centering
\tiny
\setlength\tabcolsep{2pt}\renewcommand\arraystretch{0.6}
\begin{tabular}{L{2.2cm} c c c c c c c c c c c}
\toprule
Datasets&anti-VEGF&Influenza&Influenza&Influenza&Influenza&SARS-CoV-2&SARS-CoV-2&SARS-CoV-2&anti-lysozyme&anti-HER2&anti-integrin\\
Models&2fjg&3gbn\_h1&3gbn\_h9&4fqi\_h1&4fqi\_h3&AAYL49&AAYL51&AAYL49(ML)&1mlc&1n8z&1mhp\\
\midrule
\textbf{ESM-IF}&\textit{0.5504}&\textit{0.5950}&\textbf{\textit{0.5399}}&\textbf{\textit{0.6459}}&\textit{0.4938}&\textit{0.3871}&\textbf{\textit{0.3439}}&\textit{0.2662}&\textit{-0.3574}&\textit{-0.1083}&\textit{-0.3564}\\
\textbf{ESM-IF (A/H/L)}&0.5586&\textbf{0.6015}&\textbf{0.5399}&0.5306&0.4745&\textbf{0.3958}&0.3335&\textbf{0.2880}&-0.3569&-0.2180&-0.3053*\\
\textbf{ESM-IF (L/A/H)}&\textbf{0.5600}&0.2408&0.2606&0.6286&\textbf{0.4981}&0.3855&0.3283&0.2694&-0.3578&-0.0931*&-0.3904\\
\textbf{ProtGPT2}&0.0372&-0.3913&-0.1763&-0.2017&-0.0014*&0.0634&0.1028&0.0634&-0.2120*&0.1463&-0.0962*\\
\textbf{ProtGPT2 (A/H/L)}&0.0114*&-0.4254&-0.4622&-0.2507&-0.0125&0.0246*&0.0783&0.0992&\textbf{-0.1617}&\textbf{0.1489}&\textbf{0.1834*}\\
\textbf{ProtGPT2 (L/A/H)}&-0.0536&-0.5087&-0.3812&-0.4656&-0.2340&0.0858&0.0903&0.0883&-0.2142&-0.0325*&-0.2441*\\
\textbf{ProGen2 (Base)}&\textit{0.2861}&\textit{-0.6707}&\textit{-0.5972}&\textit{-0.4643}&\textit{-0.3127}&\textit{0.2668}&\textit{0.1934}&\textit{-0.1124}&\textit{-0.3851}&\textit{-0.1932}&\textit{-0.3532}\\
\textbf{ProGen2 (Base) (A/H/L)}&0.3554&-0.5716&-0.5594&-0.4156&-0.2504&0.2665&0.2001&-0.0865&-0.2985&-0.1209&-0.1503*\\
\textbf{ProGen2 (Base) (L/A/H)}&0.4324&-0.5417&-0.5498&-0.3890&-0.2339&0.2537&0.2501&-0.1122&-0.3165&-0.0324*&-0.1746*\\
\textbf{ProGen2 (Small)}&\textit{0.2039}&\textit{-0.6800}&\textit{0.6498}&\textit{-0.6321} & \textit{-0.3592}&\textit{0.2814}&\textit{0.2294} & \textit{-0.0412}&\textit{-0.2222}&\textit{-0.2232} & -0.1045*\\
\textbf{ProGen2 (Small) (A/H/L)}&0.2914&-0.4463&-0.4975&-0.5841&-0.3278&0.2812&0.2103&-0.0417&-0.2340&-0.1180&-0.3592\\
\textbf{ProGen2 (Small) (L/A/H)}&0.3851&-0.4762&-0.4762&-0.5632&-0.3318&0.3014&0.2243&-0.0899&-0.2646&-0.2036&-0.0534*\\
\textbf{ProGen2 (Medium)}&\textit{0.2537}&\textit{-0.6887}&\textit{-0.6204}&\textit{-0.5284}&\textit{-0.3729}&\textit{0.2865}&\textit{0.2117}&\textit{-0.0702}&\textit{-0.2780}&\textit{-0.2669}&-0.0221\\
\textbf{ProGen2 (Medium) (A/H/L)}&0.3023&-0.4679&-0.5396&-0.6204&-0.4178&0.2980&0.2215&-0.0996&-0.2847&-0.0780*&0.1231*\\
\textbf{ProGen2 (Medium) (L/A/H)}&0.3710&-0.4543&-0.5496&-0.6400&-0.4292&0.3069&0.2005&-0.0565&-0.2962&0.0046*&0.1162*\\
\textbf{ProGen2 (Large)}&\textit{0.2704}&\textit{-0.7555}&\textit{-0.6245}&\textit{-0.4478}&\textit{-0.3203}&\textit{0.2578}&\textit{0.2017}&\textit{-0.1124}&\textit{-0.2869}&\textit{-0.2086}&\textit{-0.3777}\\
\textbf{ProGen2 (Large) (A/H/L)}&0.3229&-0.6290&-0.6516&-0.4859&-0.3789&0.2764&0.2239&-0.1038&-0.2693&0.0270*&-0.1325*\\
\textbf{ProGen2 (Large) (L/A/H)}&0.3097&-0.6518&-0.6661&-0.4533&-0.3625&0.2750&0.2109&-0.0512&-0.3391&0.0499*&-0.0344*\\
\bottomrule
\end{tabular}
\caption{Performance of autoregressive models across diverse Ab-Ag binding assays. Highest correlation per dataset is marked in bold and baseline model values are italicized. Asterisk (*) denotes correlation values are insignificant ($p value > 0.05$).}
\label{tab:autoregressive}
\end{table}

\subsection{Structure Prediction Confidence Scores as Binding Affinity Proxies}
\label{sup:sec:structure_prediction_correlation}

The main analysis uses ipSAE for AF3, Boltz-2 and ESMFold2. Table~\ref{tab:af3_boltz_spearman} reports the corresponding auxiliary iPTM and ipLDDT correlations on the same ipSAE common-case variant population. The equal-assay mean correlations were $-0.023$ and $-0.064$ for AF3, 0.132 and 0.108 for Boltz-2, and 0.065 and 0.009 for ESMFold2, respectively. Their variation in sign and magnitude across assays reinforces that these confidence measures are heuristic affinity proxies rather than calibrated affinity estimators.

\begin{table}[htbp]
\centering
\small
\setlength\tabcolsep{4pt}
\renewcommand\arraystretch{0.85}
\color{black}
\resizebox{\textwidth}{!}{\begin{tabular}{l cc cc cc}
\toprule
 & \multicolumn{2}{c}{\textbf{AlphaFold3}} & \multicolumn{2}{c}{\textbf{Boltz-2}} & \multicolumn{2}{c}{\textbf{ESMFold2}} \\
\cmidrule(lr){2-3}\cmidrule(lr){4-5}\cmidrule(lr){6-7}
\textbf{Assay} & \textbf{iPTM} & \textbf{ipLDDT} & \textbf{iPTM} & \textbf{ipLDDT} & \textbf{iPTM} & \textbf{ipLDDT} \\
\midrule
1mhp & $-0.532$ & $-0.540$ & $-0.216$ & $-0.441$ & $-0.165$ & $-0.277$ \\
1mlc & $-0.174$ & $-0.151$ & 0.016 & $-0.139$ & $-0.083$ & $-0.109$ \\
1n8z & $-0.157$ & $-0.033$ & $-0.063$ & 0.159 & 0.194 & $-0.171$ \\
2fjg & 0.091 & 0.073 & 0.086 & 0.074 & 0.030 & 0.084 \\
3gbn\_h1 & $-0.053$ & $-0.411$ & 0.708 & 0.572 & 0.275 & $-0.003$ \\
3gbn\_h9 & 0.128 & $-0.158$ & 0.558 & 0.453 & 0.230 & 0.014 \\
4fqi\_h1 & 0.047 & $-0.013$ & 0.396 & 0.318 & 0.164 & 0.027 \\
4fqi\_h3 & 0.051 & 0.012 & 0.273 & 0.327 & 0.007 & $-0.049$ \\
4d5\_her2 & 0.023 & $-0.017$ & 0.029 & 0.018 & 0.083 & 0.026 \\
5a12\_ang2 & $-0.110$ & $-0.102$ & $-0.017$ & $-0.040$ & $-0.043$ & $-0.072$ \\
5a12\_vegf & $-0.009$ & 0.002 & 0.312 & 0.201 & $-0.145$ & 0.117 \\
aayl49 & 0.234 & 0.245 & 0.017 & 0.046 & 0.189 & 0.157 \\
aayl49\_ML & 0.031 & 0.003 & 0.020 & 0.144 & 0.078 & 0.064 \\
aayl50 & $-0.007$ & $-0.008$ & $-0.016$ & 0.052 & 0.045 & 0.060 \\
aayl51 & 0.065 & 0.075 & 0.048 & $-0.001$ & 0.152 & 0.145 \\
aayl52 & 0.006 & $-0.001$ & $-0.042$ & $-0.020$ & 0.027 & 0.129 \\
\midrule
\textbf{Average} & $-0.023$ & $-0.064$ & 0.132 & 0.108 & 0.065 & 0.009 \\
\bottomrule
\end{tabular}}
\caption{Auxiliary Spearman correlations between iPTM or interface pLDDT (ipLDDT) and experimental binding affinity for AlphaFold3, Boltz-2 and ESMFold2. ipSAE is the primary structure-prediction score in the main analysis; the proxy correlations shown here use the same common-case variant population and the same assay-component merging rules.}
\label{tab:af3_boltz_spearman}
\end{table}

\subsection{ Supplementary analyses of generated F045-092 variants}
\label{sup:sec:case_study}

\subsubsection{} \label{sup:sec:sec:input_data} 

\subsubsection{} \label{sup:sec:sec:design_cdrh3} 

\begin{itemize}

    \item 

    \item 
    
    \item  

    \item ~
\end{itemize}

\begin{figure}[!ht]
   \centering
   \includegraphics[width=1\textwidth]{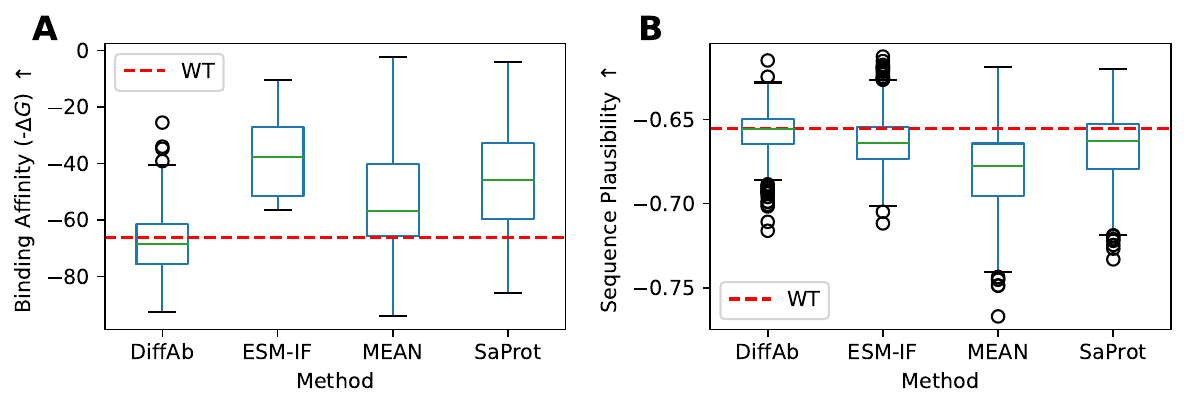} 
   \caption{
        \textbf{A.} Boxplots of the binding energy (-$\Delta G$). The wild type is -66.340. 
        \textbf{B.} Boxplots of biological plausibility of model-predicted antibody sequences. The wild type is -0.655}
   \label{fig:phase1}
 \end{figure}
 \vspace{-0.1em}  

\begin{figure}[!ht]
   \centering
   \includegraphics[width=1\textwidth]{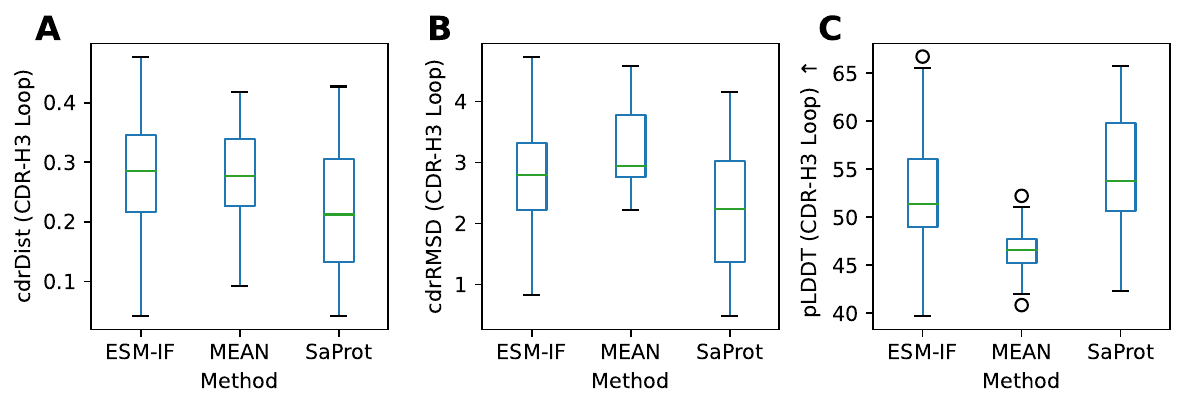} 
   \caption{
     \textbf{A.} Comparing sequence similarity. 
     \textbf{B.} Comparing loop conformation similarity. 
     \textbf{C.} Comparing against the accuracies of side-chain orientations of the CDRH3 residues.}
   \label{fig:diversity-cdrh3}
 \end{figure}
 \vspace{-0.9em}  

\begin{figure}[ht]
    \centering
    \includegraphics[width=\textwidth]{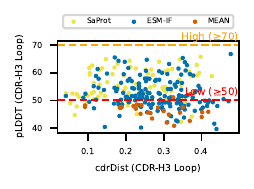} 
   \caption{
        A scatter plot of pLDDT and sequence divergence indicated by cdrDist. Phase 1-screened variants were used for this analysis.
        }
    \label{fig:diversity-sw}
\end{figure}

\begin{figure}[ht]
    \centering
    \includegraphics[width=0.99\linewidth]{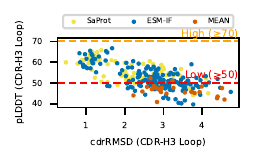}
    \caption{
        Relationship between structure confidence (pLDDT) and structural deviation from wild type (cdrRMSD) of CDR-H3 generated by models.  
    }
    \label{fig:diversity-rmsd}
\end{figure}

\newpage

\end{document}